\documentclass[11pt,a4paper]{article}

\usepackage{graphicx}
\usepackage{afterpage}
\usepackage{epsfig,cite}
\usepackage{amssymb}
\usepackage{amsmath}
\usepackage{dsfont}
\usepackage{multirow}
\usepackage{url,hyperref}
\usepackage{feynarts}

\usepackage{url}
\usepackage{hyperref}

\usepackage{tikz}
\usetikzlibrary{shapes,arrows}

\newcommand{\be}{\begin{equation}}
\newcommand{\ee}{\end{equation}}
\newcommand{\bea}{\begin{eqnarray}}
\newcommand{\eea}{\end{eqnarray}}
\newcommand{\bi}{\begin{itemize}}
\newcommand{\ei}{\end{itemize}}
\newcommand{\ben}{\begin{enumerate}}
\newcommand{\een}{\end{enumerate}}
\newcommand{\la}{\left\langle}
\newcommand{\ra}{\right\rangle}
\newcommand{\lc}{\left[}
\newcommand{\rc}{\right]}
\newcommand{\lp}{\left(}
\newcommand{\rp}{\right)}

\def\frac#1#2{{{#1}\over {#2}}}
\def\gsim{\mathrel{\rlap{\lower4pt\hbox{\hskip1pt$\sim$}}
    \raise1pt\hbox{$>$}}}         
\def\lsim{\mathrel{\rlap{\lower4pt\hbox{\hskip1pt$\sim$}}
    \raise1pt\hbox{$<$}}}         

\newcommand{\draft}[1]{}

\def\beq{\begin{equation}}  
\def\eeq{\end{equation}}  

\def \n0{N_j^{(0)}}

\def\lapprox{\lower .7ex\hbox{$\;\stackrel{\textstyle <}{\sim}\;$}}
\def\gapprox{\lower .7ex\hbox{$\;\stackrel{\textstyle >}{\sim}\;$}}

\begin{document}
\begin{figure}[h]
\epsfig{width=0.32\textwidth,figure=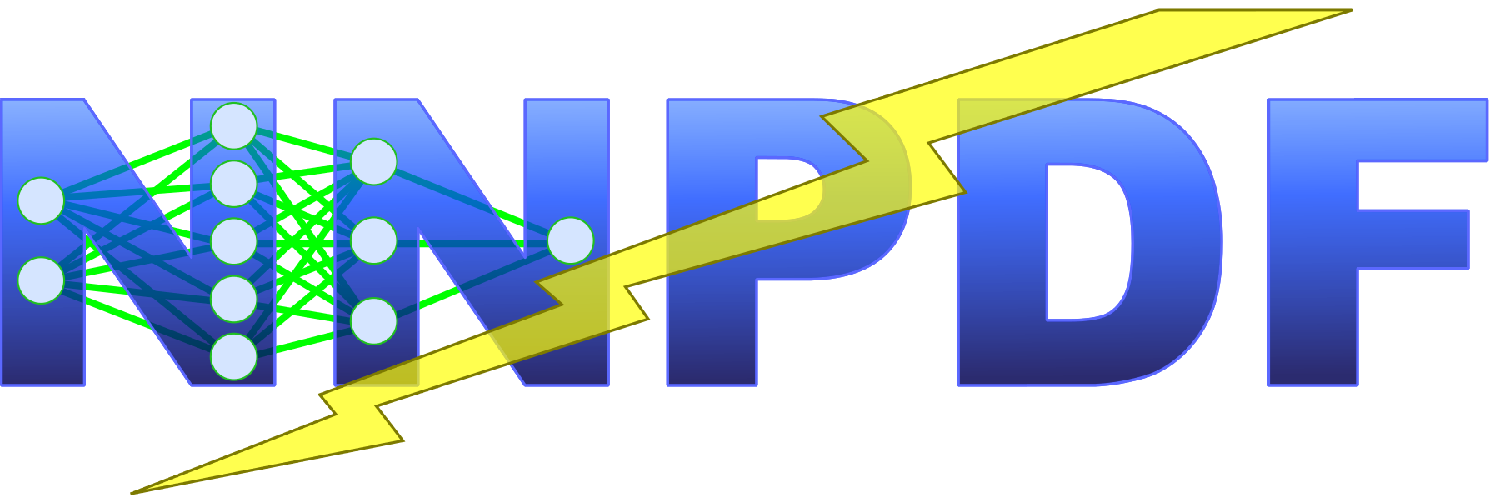}
\end{figure}
\vspace{-2.0cm}
\begin{flushright}
Edinburgh 2013/20\\
IFUM-1014-FT\\
FR-PHENO-2013-008\\
CERN-PH-TH/2013-075\\
\end{flushright}

\vspace{1.0cm}

\begin{center}
{\Large \bf Parton distributions with QED corrections}
\vspace{1.0cm}

{\bf  The NNPDF Collaboration:}\\
 Richard~D.~Ball$^{1}$,  Valerio~Bertone$^{2}$,
Stefano~Carrazza$^{3}$, Luigi~Del~Debbio$^{1}$,\\
Stefano~Forte$^3$, Alberto~Guffanti$^{4}$, 
Nathan~P.~Hartland$^1$, Juan~Rojo$^5$.

\vspace{.3cm}
{\it ~$^1$ Higgs Centre for Theoretical Physics, University of Edinburgh,\\
JCMB, KB, Mayfield Rd, Edinburgh EH9 3JZ, Scotland\\
~$^2$  Physikalisches Institut, Albert-Ludwigs-Universit\"at Freiburg,\\ 
Hermann-Herder-Stra\ss e 3, D-79104 Freiburg i. B., Germany  \\
~$^3$ Dipartimento di Fisica, Universit\`a di Milano and
INFN, Sezione di Milano,\\ Via Celoria 16, I-20133 Milano, Italy\\
~$^4$ The Niels Bohr International Academy and Discovery Center, \\
The Niels Bohr Institute, Blegdamsvej 17, DK-2100 Copenhagen, Denmark\\
~$^5$ PH Department, TH Unit, CERN, CH-1211 Geneva 23, Switzerland \\}
\end{center}

\vspace{1.0cm}

\begin{center}
{\bf \large Abstract}
\end{center}

We present a set of parton distribution functions (PDFs), based on the
NNPDF2.3 set, which includes a photon PDF, and QED
contributions to parton evolution. We describe the implementation of the combined
QCD+QED evolution in the NNPDF framework.  We then 
provide a first determination of
the full set of PDFs based on deep-inelastic
scattering data and LHC data for $W$ and $Z/\gamma^*$ Drell-Yan
production, using leading-order
QED and  NLO or NNLO QCD. We compare the ensuing NNPDF2.3QED PDF set to the older 
MRST2004QED set.  
We perform a preliminary investigation of the phenomenological implications of
NNPDF2.3QED: specifically,
photon-induced corrections to direct photon production at HERA,
and high-mass dilepton and $W$ pair production at the LHC.

\clearpage

\tableofcontents

\clearpage

\section{Introduction}

\label{sec-intro}


Because of the need for precision phenomenology at the
LHC~\cite{Forte:2010dt,Forte:2013wc}  the
parton distributions (PDFs) of the nucleon are currently determined using
next-to-next-to leading order (NNLO) QCD. At this level of
accuracy, however, electroweak (EW) corrections also become relevant, and indeed
EW corrections to various hadron collider processes have been studied
in detail, such as for instance
inclusive $W$ and $Z$ production~\cite{Baur:1998kt,Zykunov:2001mn,Dittmaier:2001ay,Baur:2001ze,Baur:2004ig,Arbuzov:2007db,Arbuzov:2005dd,Brensing:2007qm,Balossini:2009sa,CarloniCalame:2007cd,Dittmaier:2009cr}, $W$ and
$Z$ boson production in association with jets~\cite{Denner:2009gj,Denner:2011vu,Denner:2012ts},
dijet production~\cite{Moretti:2005ut,Dittmaier:2012kx} and top quark pair production~\cite{Bernreuther:2005is,Kuhn:2005it,Hollik:2007sw,Hollik:2011ps,Kuhn:2013zoa}.

A consistent inclusion of EW corrections, however, requires the use
of PDFs which incorporate QED effects. This in particular
implies the inclusion of QED corrections to perturbative evolution,
and thus a photon PDF.
There is currently only one PDF set with
QED corrections available, the MRST2004QED set~\cite{Martin:2004dh}. In this
pioneering work, the photon PDF was determined based on a model
inspired 
by photon radiation off constituent quarks (though consistency with
some HERA data was checked a posteriori), and therefore not provided
with a PDF uncertainty. 

In this work we will construct a PDF set including QED corrections, with a
photon PDF parametrized in the same way as all the other PDFs, and determined from a fit
to hard-scattering experimental data. Our goal is to construct a PDF set with the
following features:
\begin{itemize}
\item QCD corrections included up to NLO or NNLO;
\item QED corrections included to LO;
\item a photon PDF obtained from a fit to deep-inelastic scattering
  (DIS) and Drell-Yan (both low mass, on-shell $W$ and $Z$ production,
  and high mass) data;
\item all other PDFs constrained by the same data included in the
  NNPDF2.3 PDF determination~\cite{Ball:2012cx}.
\end{itemize}
The lepton PDF, as well as weak contributions
to evolution equations~\cite{Ciafaloni:2001mu, Ciafaloni:2005fm} are
negligible and will not be considered here.

In principle, this goal could be achieved by simply performing a 
global fit including 
QED and QCD corrections both to perturbative evolution and to hard 
matrix elements, and with data which constrain the photon PDF. In
practice, this would require the availability of
a fast interface, like {\tt APPLgrid}~\cite{Carli:2010rw} or
{\tt FastNLO}~\cite{Wobisch:2011ij}, to codes which include
QED corrections to processes  which are
sensitive to the photon PDF, such as single or double gauge boson
production. Because such interfaces are not
available, we adopt instead a reweighting procedure, which turns out
to be sufficiently accurate to accommodate all relevant existing data. 

The reweighting procedure we use works as follows (see
Fig.~\ref{flowchart}). First, we construct a set of PDFs (NNPDF2.3QED
DIS-only), including a photon PDF, by performing a fit to
deep-inelastic scattering (DIS) data
only, based on the same DIS data  used for NNPDF2.3, and using
either NLO or NNLO QCD and LO QED theory. To leading order in QED, the
photon PDF only contributes to DIS through perturbative evolution
(just like the gluon PDF to leading order in QCD). Therefore, the
photon PDF is only weakly constrained by DIS data, and thus the photon
PDF in the NNPDF2.3QED DIS-only set is affected by large
uncertainties. The result is a pair of PDF sets: NNPDF2.3QED
DIS-only, NLO or NNLO, according to how QCD evolution has been treated.

In the next step, each replica of the photon PDF from the NNPDF2.3QED 
DIS-only set is combined with a random PDF replica 
of a set of the default NNPDF2.3 PDFs,
fitted to the global dataset. This works
because of the small correlation between the photon PDF and other
PDFs, as we shall explicitly check. Also, the
violation of the momentum sum rule that this procedure entails is  
not larger than the uncertainty on the momentum sum rule in the global
QCD fit. The procedure is performed using NLO or NNLO NNPDF2.3 PDFs,
for three values of $\alpha_s(M_z)=0.117,\>0.118,\>0.119$. 
The photon PDF
determined in the NNPDF2.3QED 
DIS-only fit is in fact almost independent of the value of
$\alpha_s$ within this range.
This leads to several sets of PDF replicas, which we call NNPDF2.3QED 
prior, at the scale $Q_0$. 
The NNPDF2.3QED prior PDFs  are  then evolved to all $Q^2$
using combined QCD+QED evolution equations, to LO in QED and either to
NLO or NNLO in QCD and with the appropriate value of $\alpha_s$.

The LHC $W$ and $Z/\gamma^*$ production data are now included in the
fit by Bayesian reweighting~\cite{Ball:2010gb} of the NNPDF2.3QED
prior PDF set.  The set of
reweighted replicas is then unweighted~\cite{Ball:2011gg} in order to obtain 
a standard set of 100 replicas of our final NNPDF2.3QED set. 

\tikzstyle{block} = [rectangle, draw, fill=blue!20, 
    text width=26em, text centered, rounded corners, minimum height=4em]
\tikzstyle{line} = [draw, -latex']
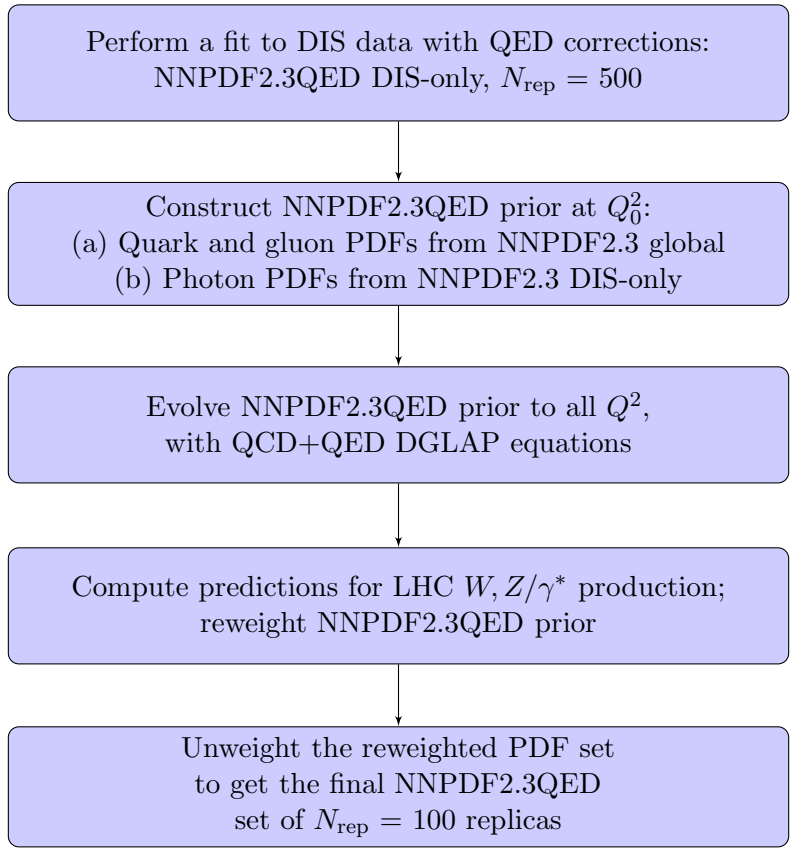
\begin{figure} [hb!]
  \begin{center}
    \begin{tikzpicture}[node distance = 2.4cm, auto]
      \node [block] (f1) {Perform a fit to DIS data with QED
        corrections:\\
        NNPDF2.3QED DIS-only, $N_{\rm rep}$ = 500};
      \node [block, below of=f1] (f2) {Construct NNPDF2.3QED prior at $Q_0^2$: \\
        (a) Quark and gluon PDFs from NNPDF2.3 global \\
        (b) Photon PDFs from NNPDF2.3 DIS-only };
      \node [block,below of=f2] (f3) {Evolve NNPDF2.3QED prior to all $Q^2$, \\
        with QCD+QED DGLAP equations};
      \node [block,below of=f3] (f4) {Compute predictions for LHC $W,Z/\gamma^{*}$ production;\\
        reweight NNPDF2.3QED prior }; \node [block,below of=f4] (f5)
      {Unweight the reweighted
        PDF set to get the final NNPDF2.3QED\\
        set of $N_{\rm rep}$ = 100 replicas}; \path [line] (f1) --
      (f2); \path [line] (f2) -- (f3); \path [line] (f3) -- (f4);
      \path [line] (f4) -- (f5);
    \end{tikzpicture}
  \end{center}
  \caption{\label{flowchart} \small Flow-chart for the construction of
    the NNPDF2.3QED set. }
\end{figure}

The photon PDF in our final set turns out to be in good agreement with
that from the MRST2004QED set at medium large $x\gsim0.03$, while for
smaller $x$ values it is substantially smaller (by about a factor
three for $x\sim10^{-3}$), though everywhere 
affected by sizable uncertainties, typically of order 50\%.
We will perform some first illustrative phenomenological studies using the
NNPDF2.3QED set, and in particular discuss deep-inelastic
direct photon production at  HERA,
photon-induced corrections to backgrounds in $W'$ and $Z'$ searches and
electroweak corrections to vector boson pair production.

The paper is organized as follows. In Sect.~\ref{sec:disfit} we
present the implementation in the NNPDF framework of 
combined QCD and QED evolution
equations, and its comparison with publicly available QED evolution codes.
In Sect.~\ref{sec:disfit} we also discuss the first step of our procedure,
namely, the determination of NNPDF2.3QED DIS-only PDF set. The subsequent steps, namely the
construction of the NNPDF2.3QED prior set, and its reweighting and unweighting leading to the
final NNPDF2.3QED set are presented in
Sect.~\ref{sec:lhcwz}. Finally, our phenomenological
investigations are presented 
in Sect.~\ref{sec:searches}.

\section{Deep-inelastic scattering with QED corrections}
\label{sec:disfit}

The first step of our procedure is to perform a PDF fit to DIS data in which QED
corrections are included to leading order. 
This requires  solving the
perturbative evolution equations which combine QED and QCD
collinear radiation.
 We first review our implementation of combined QED+QCD
evolution in the NNPDF perturbative
evolution framework, in particular using the {\tt FastKernel} method first
presented in Ref.~\cite{Ball:2008by}, and then turn to the PDF determination.

\subsection{QED corrections to PDF evolution}

The resummation of collinear singularities related to QED radiation
through the solution of
QED evolution equations, and its combination with QCD evolution
equations has been understood for many years~\cite{DeRujula:1979jj,Kripfganz:1988bd,Blumlein:1989gk}: collinear photon radiation from charged
leptons or quarks leads to a scale dependence which has the same
structure as that of QCD evolution. At leading order the $P_{qq}$,
$P_{q\gamma}$ and $P_{\gamma q}$ splitting functions coincide with
their  $P_{qq}$,
$P_{qg}$ and $P_{gq}$ counterparts, up to the value of the coupling,
which in QED is proportional to the square of the electric charge. 
Because there is no photon self-coupling, the  $P_{\gamma \gamma}$ splitting
function only receives contributions from self-energy virtual
corrections, and is thus proportional to a $\delta(1-x)$.
The combined QCD+QED evolution equations thus take the form
\begin{equation}
Q^{2}\frac{\partial}{\partial Q^{2}}f(x,Q^{2})=
\left[\frac{\alpha(Q^{2})}{2\pi} P^{\rm QED}+\frac{\alpha_s(Q^{2})}{2\pi}
  P^{\rm QCD}\right] \otimes f(x,Q^2),\end{equation}
where $f(x,Q^{2})$ is a vector which includes all parton
distributions, $P^{\rm QED}(x)$ and $P^{\rm QCD}(x)$ are respectively QED and
QCD matrices of splitting functions, which admit respectively 
an expansion in powers of the
fine structure constant $\alpha$ and the strong coupling $\alpha_s$,
and $\otimes$ denotes the standard convolution. 

A priori, the vector $f(x,Q^{2})$ includes quark, gluon, lepton and
photon PDFs. In practice, however, the lepton PDFs of the nucleon are
negligibly small, and may be safely neglected: thus we will take
$f(x,Q^{2})$ to include the usual parton PDFs and a photon PDF
$\gamma(x,Q^2)$. 
The combined QED+QCD evolution equations can be solved as usual by
taking a Mellin transform, whereby they reduce to coupled  ordinary
differential equations. It is easy to check that the 
QCD and QED splitting function matrices do not commute. However, their
commutator is of order $\alpha \alpha_s$, which is subleading. 
It follows that, up to $O(\alpha \alpha_s)$ corrections, the
solution can be written in factorized form as
\begin{equation}
  f_i(N,Q^{2})=\Gamma^{\text{\rm QCD}}_{ik}(N,Q^{2},Q^{2}_{0}) 
  \Gamma^{\text{\rm QED}}_{kj}(N,Q^{2},Q^{2}_{0}) f_j(N,Q^{2}_{0}),
\label{qedqcdev}
\end{equation}
where $f(N,Q^2)$ is the Mellin transform of the vector of parton
distributions $f(x,Q^2)$, and
$\Gamma^{\text{\rm QCD}}_{ik}(N,Q^{2},Q^{2}_{0}) $ and
$\Gamma^{\text{\rm QED}}_{ik}(N,Q^{2},Q^{2}_{0}) $ are respectively the
evolution kernels which solve the QCD and QED evolution equations. 

We have implemented a combined solution of the QED+QCD evolution
equations based on Eq.~(\ref{qedqcdev}) in the NNPDF code, using the
{\tt FastKernel} method of Ref.~\cite{Ball:2008by}, and with
leading-order running of the fine structure constant. Details of the
implementation will be given in a separate
publication~\cite{nnpdfqedth}.  

Previous 
numerical implementations of combined QED and QCD evolution were
presented in
Refs.~\cite{Spiesberger:1994dm,Roth:2004ti,Martin:2004dh}, and
specifically in Ref.~\cite{Roth:2004ti} a public code ({\tt
  partonevolution}) for the solution of combined evolution equations to 
leading order in the QED coupling and up to next-to-leading order in the QCD
coupling was made
available. 

First of all, we study the effect of the inclusion of QED corrections
to perturbative 
evolution by comparing results obtained from
our solution to the QED+QCD combined
evolution equation as implemented in {\tt FastKernel}, with those
found when QED effects are  switched off. For this first comparison,
we  assume that the photon PDF vanishes 
at a reference  scale, $\gamma \lp x,Q_0^2\rp=0$ for $Q^2_0=2$ GeV$^2$,
and it is generated radiatively by evolution, and all other PDFs at
the same scale are those from the Les Houches PDF
benchmarks~\cite{Dittmar:2009ii}.

\begin{figure}[t]
  \begin{centering}
    \includegraphics[scale=0.37]{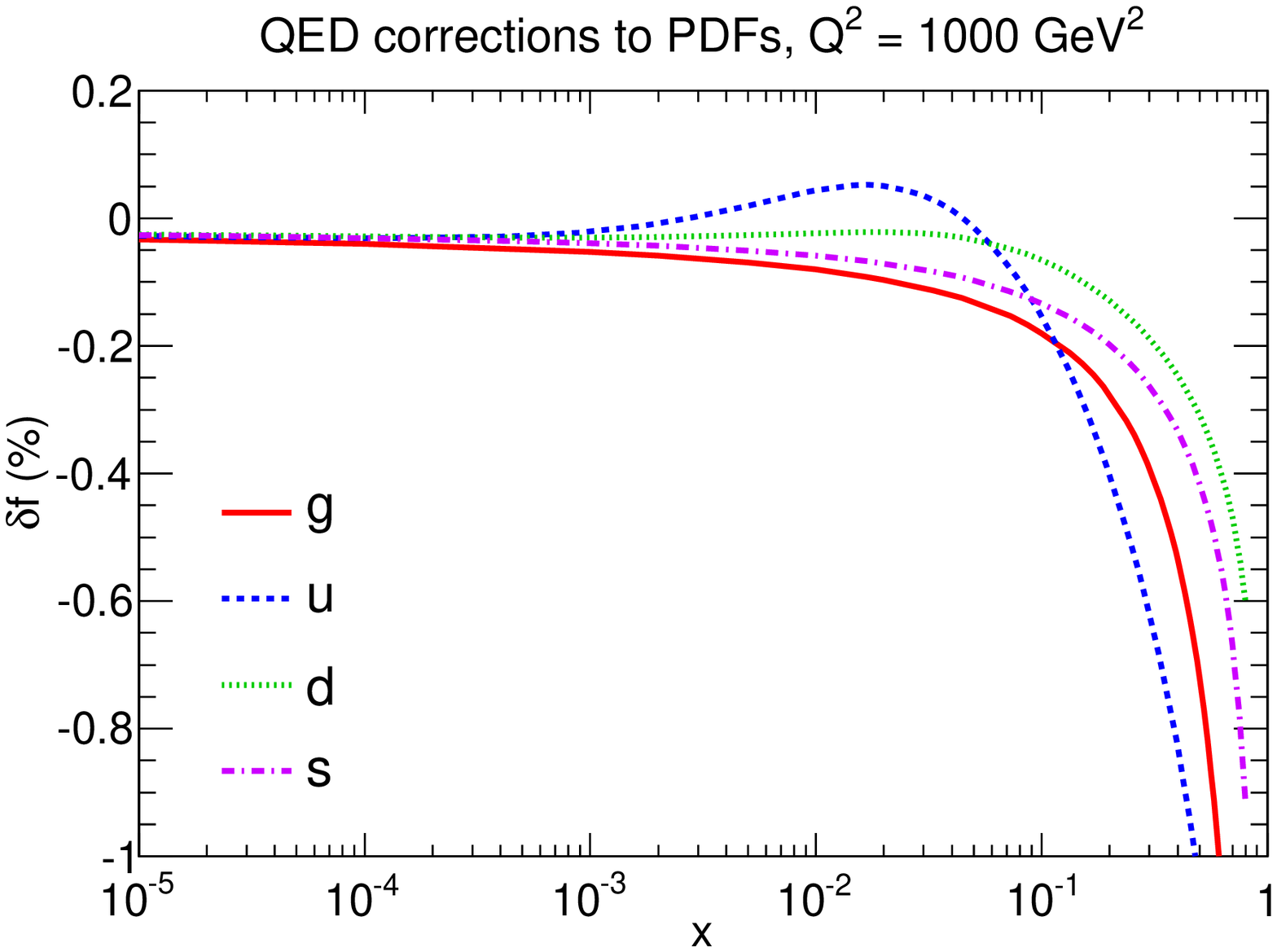}
    \includegraphics[scale=0.37]{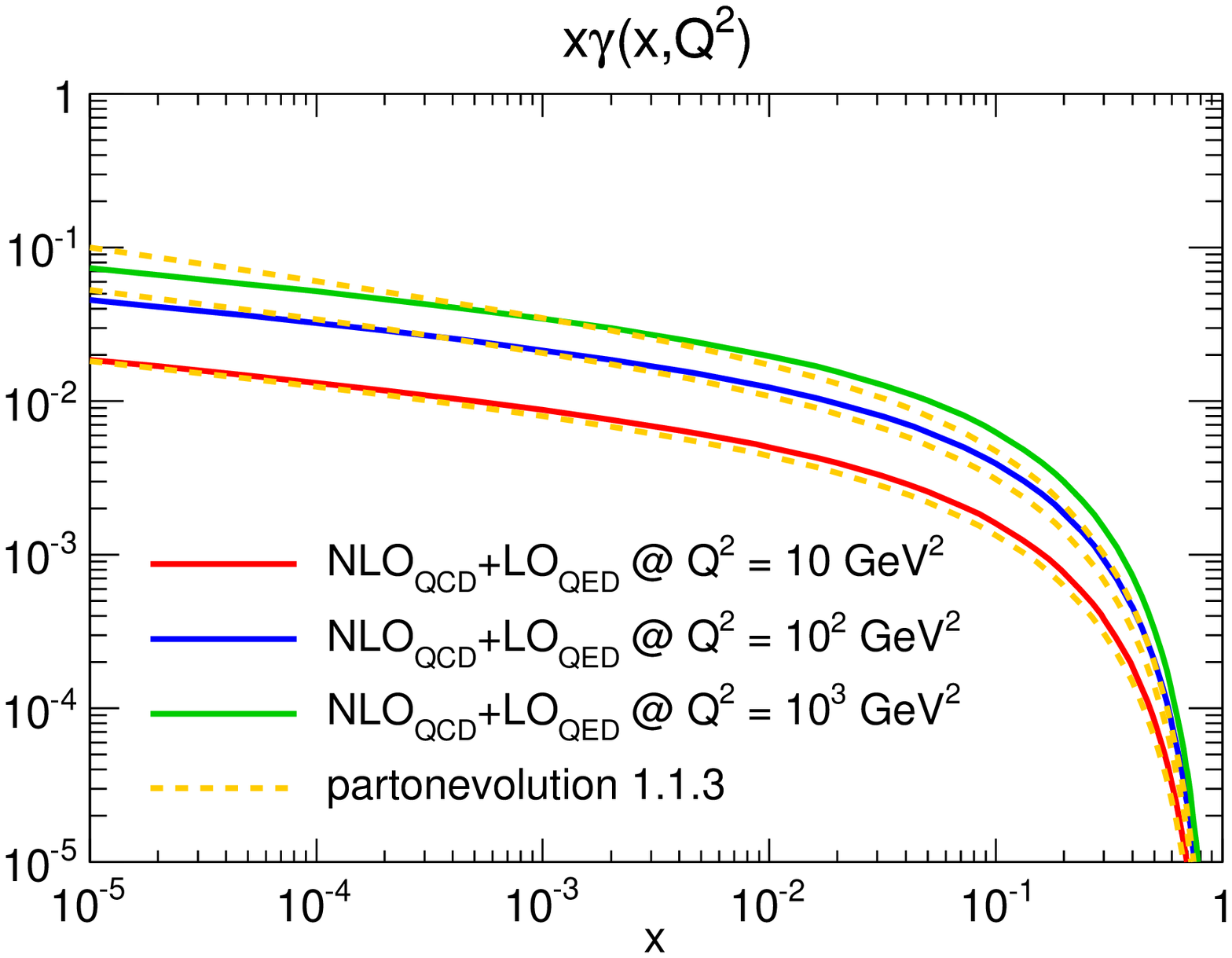}
    \par\end{centering}
  \caption{\label{fig:qedfig1} Left: relative difference between 
    PDFs obtained from pure NLO QCD evolution and  QCD NLO + QED LO
    evolution at $Q^2=10^3$~GeV$^2$. Right:  dynamically generated
    photon PDF  at $Q^2=10,\>10^2$ and $10^3$~GeV$^2$; the result
    obtained using the  {\tt
  partonevolution} code of Ref.~\cite{Weinzierl:2002mv,Roth:2004ti} is
    also shown
(dashed curves).  }
\end{figure}

In Fig.~\ref{fig:qedfig1} we plot the relative difference between PDFs
evolved with and without QED corrections as a function of $x$ at
$Q^2=10^3$~GeV$^2$ for
various PDFs, and  the photon PDF which has been generated
dynamically  by perturbative evolution at three different scales. As
expected, the QED corrections to all PDFs are small, below 1\%, and
concentrated at large $x\gsim 0.1$, while the dynamically generated
photon PDF is very small at large $x$ and then
grows monotonically as $x$ decreases. The correction to the
up quark is larger than that to the down quark, because of larger
absolute value of the former's electric charge.

We also compare our results for combined  QED+QCD evolution to 
those obtained using
the {\tt partonevolution} code~\cite{Roth:2004ti}, 
version {\tt v1.1.3}, bearing in mind that the two codes differ by
terms of $O\lp
\alpha \alpha_s \rp$. Indeed, the solution of Ref.~\cite{Roth:2004ti}
is based on the diagonalization of the
full anomalous dimension matrix, rather than its factorization into QCD and
QED components according to Eq.~(\ref{qedqcdev}), although 
terms of $O\lp\alpha \alpha_s \rp$ are also neglected. Our calculation
 should thus be 
equivalent up to subleading terms.

\begin{figure}[t]
\begin{centering}
\includegraphics[scale=0.37]{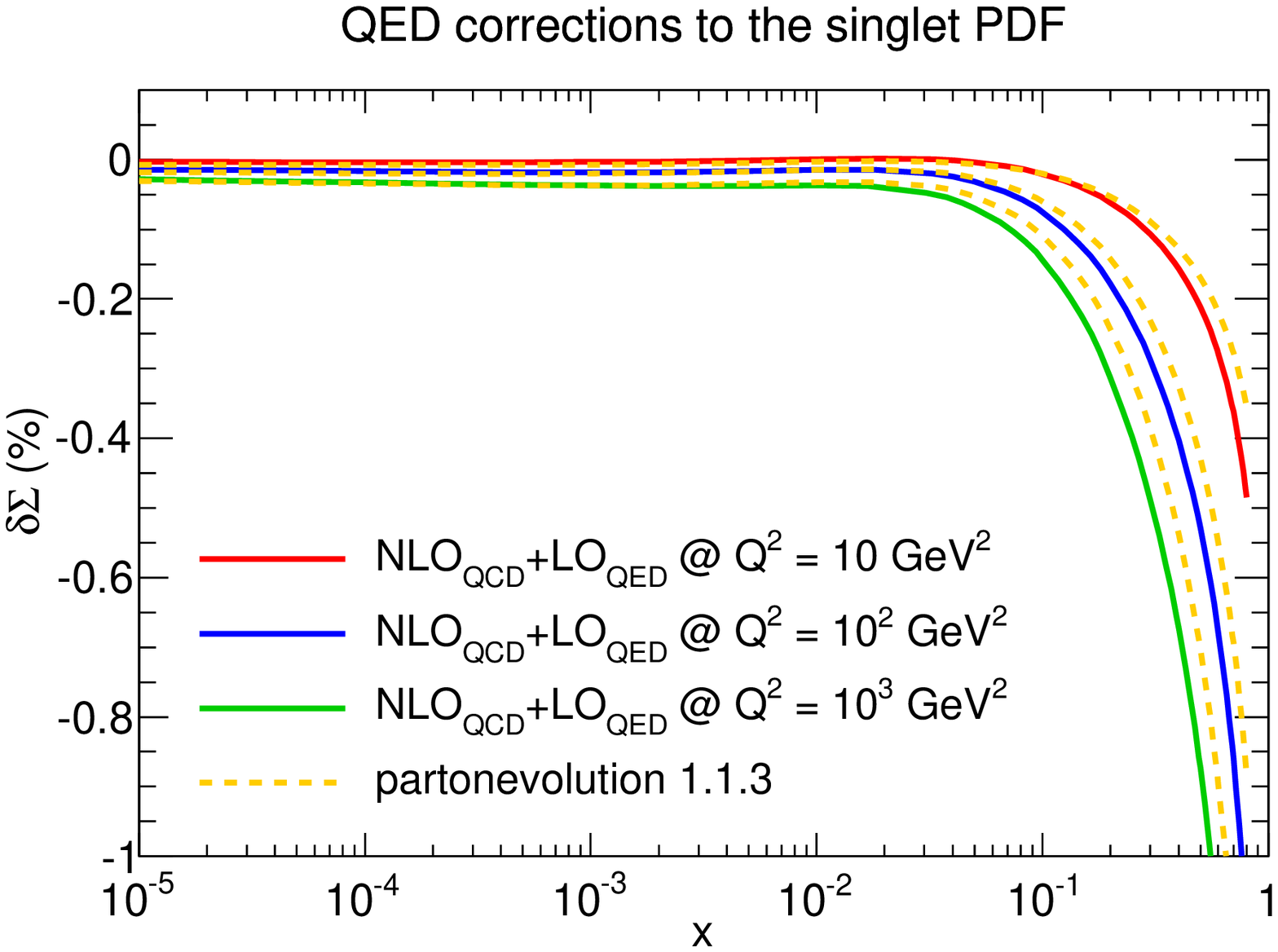}
\includegraphics[scale=0.37]{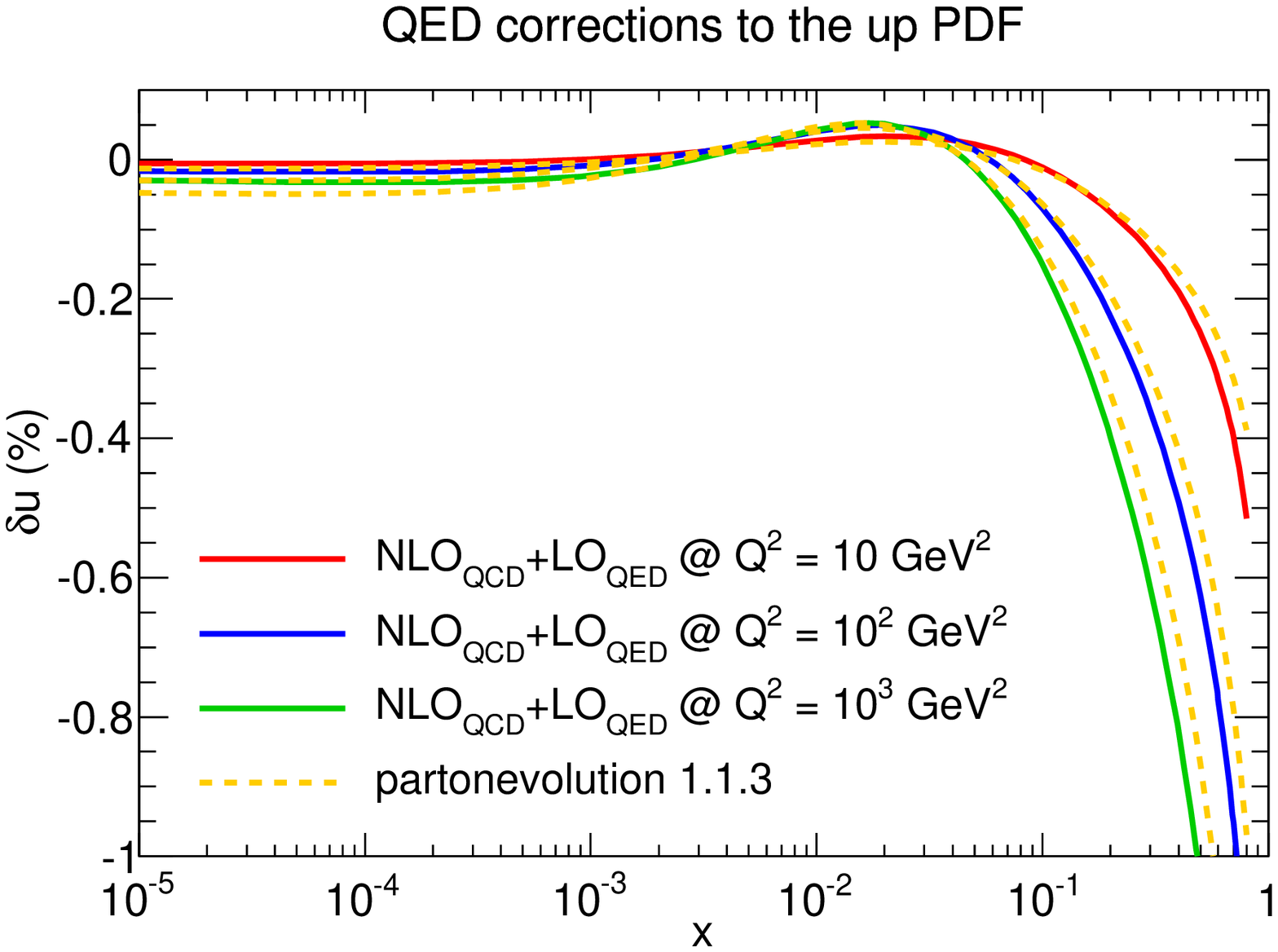}
\par\end{centering}
\caption{\label{fig:qedfig2} The relative differences of
  Fig.~\ref{fig:qedfig1} for the singlet (left) and up (right) PDFs
at $Q^2=10,\>10^2$ and $10^3$~GeV$^2$, computed using  the NNPDF  
{\tt FastKernel} implementation (solid curves) and the  {\tt
  partonevolution} code of Ref.~\cite{Weinzierl:2002mv,Roth:2004ti}
(dashed curves). }
\end{figure}

The comparison is shown in Fig.~\ref{fig:qedfig1} for the photon PDF, and
in Fig.~\ref{fig:qedfig2} for 
 the percentage differences of 
Fig.~\ref{fig:qedfig1}, here shown for two
different PDF combinations at various scales. Excellent agreement is
found, with differences between the two codes much smaller than the
effect of QED corrections to the photon PDF or PDF evolution. We have
also checked that the scale dependence obtained using our code is 
in good agreement with that of the MRST2004QED PDFs. These comparisons
will be discussed in detail in Ref.~\cite{nnpdfqedth}.

\subsection{Fitting PDFs with QED corrections}
\label{sec:disqed}

We now wish to obtain a first determination of the  photon PDF from a
fit to deep-inelastic data.
We want to include QED corrections to DIS at LO, i.e., more
 accurately, the 
leading log level. This means that the splitting functions $P^{\rm QED}$
are computed to  $O(\alpha)$, while all partonic cross sections
(coefficient functions) are determined to lowest order in $\alpha$. 
Because the photon is electrically neutral, the photon deep-inelastic
 coefficient function only starts at $O(\alpha^2)$, while quark
 coefficient functions start  at $O(\alpha)$. This means that at LO
 the photon coefficient function vanishes, and the photon only
contributes to  DIS  through its mixing 
with quarks due to perturbative evolution. This is fully analogous to
the role of the gluon in the standard LO QCD description of DIS: the
gluon coefficient function only starts at $O(\alpha_s)$ while the
quark coefficient function starts  at $O(1)$, so at LO the gluon
only contributes to deep-inelastic scattering through its mixing with
quarks upon perturbative evolution. 

An important  issue when including QED corrections is the  choice of
factorization scheme in the subtraction of QED collinear
singularities~\cite{Diener:2005me,Dittmaier:2009cr}. Different
factorization schemes differ by next-to-leading log terms.
Because  our treatment
of QED evolution is at the leading log level, our results  do not depend on
the choice of factorization scheme. This means that if our photon PDF
is used in conjunction with a next-to-leading log computation of QED
cross-sections, the latter can be taken in any (reasonable)
factorization scheme. The difference in results found when changing
the QED factorization scheme should be considered to be part of the
theoretical uncertainty. 
However, in practice, in some schemes 
the perturbative expansion  may show faster
convergence (so, for example,  next-to-leading log results are closer
to leading-log ones in some schemes than others).
We will indeed see in the next section that when DIS data are combined
with Drell-Yan data it is advantageous to use the 
DIS factorization scheme, which is defined by requiring that the
deep-inelastic structure function $F_2$ is given to all orders by its
leading-order expression~\cite{Diener:2005me,Dittmaier:2009cr}.

The starting point of our fit to DIS data including QED corrections is 
the NNPDF2.3 PDF
determination,
in terms of experimental data, theory settings and methodology.
We will perform fits at NLO and NNLO in QCD, for three different values of 
$\alpha_s\lp M_Z\rp=0.117,\>0.118$ and 0.119, all with LO QED evolution.
Unless otherwise stated, in the following all results, tables, and plots 
will use the $\alpha_s=0.119$ PDF sets. 

We add to the NNPDF default set
of seven independent PDF combinations a new, independently
parametrized PDF for the photon, in a completely analogous 
way to all other PDFs
(see~\cite{Ball:2012cx} and references therein), with a small
modification related to positivity to be discussed below:
\be
\gamma(x,Q_0^2) = \lp 1-x\rp^{m_{\gamma}}x^{-n_{\gamma}}
 {\rm NN}_{\gamma}(x),
 \label{eq:photon}
   \ee
 where ${\rm NN}_{\gamma}(x)$ is a
 multi-layer feed-forward neural network 
 with 2-5-3-1 architecture, with a total of 37 parameters to be determined by 
 experimental data, and the prefactor is a preprocessing function
 used to speed up minimization, and on which the final result should not depend.
 The preprocessing function is parametrized by the exponents $m_\gamma$
 and $n_\gamma$, whose values are chosen at random for each replica, with
 uniform distribution in the range
 \be\label{prerange}
 1 \le m_{\gamma} \le 20,  \qquad
 -1.5 \le n_{\gamma} \le 1.5.  
 \ee
 We have explicitly checked that the results are independent 
 on the preprocessing range, by computing for each replica the
 effective small- and large-$x$ exponents~\cite{Ball:2013lla}, defined as 
\be
n_\gamma[ \gamma(x,Q^2)]=\frac{\ln  \gamma(x,Q^2)}{\ln\frac{1}{x}}\mbox{ ,}
\qquad m_\gamma[  \gamma(x,Q^2)]=\frac{ \ln  \gamma(x,Q^2) }{\ln(1-x)}\mbox{ ,}
\label{eq:exp}
\ee
and verifying that the range of the effective exponents at small- and 
large-$x$ respectively is well within
the range of variation of the preprocessing exponents, thus showing
that the small- and large-$x$ behaviour of the best-fit PDFs is not
constrained by the choice of preprocessing but rather
determined by experimental data. 
 
Parton distributions must satisfy positivity conditions which follow
from the requirement that, even though PDFs are not directly
physically observable, they must lead to positive-definite physical
cross sections~\cite{Altarelli:1998gn}. Leading-order PDFs are
directly observable, and thus they must be positive-definite: indeed,
they admit a probabilistic interpretation. Because we treat QED
effects at LO, the photon PDF must be positive definite. This is
achieved, as in the construction of the 
NNPDF2.1 LO PDF sets~\cite{Ball:2011uy}, by squaring the
 output of the neuron in the last (linear) layer of the neural network 
${\rm NN}_{\gamma}(x)$, so that 
${\rm NN}_{\gamma}(x)$ is a positive semi-definite function.
  
Once QED evolution is switched on, isospin is no longer a good
symmetry, and thus it can no longer be used to relate  
the PDFs of the proton and neutron. Because deuteron
deep-inelastic scattering data are used in the fit, in principle this
requires an independent parametrization for proton and neutron
PDFs. Experimental data for 
the neutron PDFs would then no
longer provide a useful constraint, and in particular they would no
longer constrain the isospin triplet PDF. Whereas future
PDF fits including substantially more LHC data might allow for
an accurate PDF determination without using deuteron data, this does
not seem to be possible at present. 

\begin{table}[h]
\centering
\small
\begin{tabular}{c||c|c||c|c}
\hline 
& \multicolumn{2}{c||}{\bf NLO} & \multicolumn{2}{|c}{\bf NNLO}  \\
\hline 
Experiment  & QCD & QCD+QED  & QCD  & QCD+QED  \\
\hline 
\hline 
Total  &  1.10 &  1.10  & 1.10 & 1.10 \\  
 \hline 
NMC-pd              & 0.88  & 0.87      & 0.88   &0.88     \\
 NMC                 & 1.68  &  1.70  & 1.67  &1.69    \\
SLAC                &  1.36   & 1.40   & 1.08 &  1.10     \\
BCDMS               &  1.17  &  1.16  &  1.24 & 1.23     \\
CHORUS              &  1.01  &  1.01  &  0.98  & 0.99    \\
NTVDMN              &  0.54   & 0.54   & 0.56   & 0.54    \\
HERAI-AV            & 1.01    & 1.01   & 1.04  & 1.03     \\
FLH108              &  1.34  &  1.34  &  1.25  &1.24    \\
ZEUS-H2             &  1.26   & 1.25   & 1.24  & 1.25    \\
ZEUS $F_2^c$        &  0.75   &  0.75  & 0.76  &0.78    \\
H1 $F_2^c$          &  1.55  &  1.50  &  1.41  &1.39    \\
 \hline
\hline
\end{tabular}
\caption{\small \label{tab:chi2} The $\chi^2$ values
per data point for individual experiments computed in the
  NNPDF2.3 DIS-only NLO and NNLO PDF sets, in the QCD-only fits
compared to the results with combined QCD+QED evolution.
 All  $\chi^2$ values have been
obtained using  $N_{\rm rep}$=100 replicas with
  $\alpha_s(M_{\rm Z})=0.119$.
Normalization uncertainties have been included using 
the experimental definition of the covariance matrix,
see App. A of Ref.~\cite{Ball:2012wy}, while in the actual
fitting the $t_0$ definition was used~\cite{Ball:2009qv}. 
}
\end{table}

There are two separate issues here: one, is the amount of isospin
violation in the quark and gluon PDFs, and the second is the amount of
isospin violation in the photon PDF. At the scale
at which PDFs are parametrized, which is of the order of the nucleon
mass, we expect isospin violating effects in the quark and gluon PDFs
to be of the same order as that displayed in baryon spectroscopy,
which is at the per mille level, much below the current PDF
uncertainties (isospin violations of this order have  been predicted,
among others,
on the basis of bag model estimates~\cite{Londergan:2003pq}). The second is
the amount of isospin violation in the photon distribution itself:
this could be somewhat larger (perhaps at the percent level), however
any reasonable amount of isospin violation in the photon is way below
the uncertainty on the photon PDF. Therefore, we will assume that no
isospin violation is present at the initial scale.

Of course, even with isospin conserving PDFs at the starting scale,
isospin violation is then generated by QED evolution: this is
consistently accounted for when solving the evolution equations, by
determining separate solutions for the proton and neutron so that at
any scale $Q\not =Q_0$, $u^p(x,Q^2)\not=d^n(x,Q^2)$ and
$d^p(x,Q^2)\not=u^n(x,Q^2)$. Because of the larger
electric charge of the  up quark, the dynamically generated
photon PDF ends up being larger for the proton than it is
for the neutron. 
 
In Ref.~\cite{Martin:2004dh} isospin violation was parametrized on the
basis of model assumptions. 
We will compare our results for isospin
violation to those of this reference in Sect.~\ref{sec:nnpdf23qed}
below: we will see that while indeed the amount of isospin violation
in the photon PDF from that reference is somewhat larger than our own,
it is much smaller than the relavant uncertainty.

\subsection{The photon PDF from DIS data}
\label{sec:disfitgam}

Using the standard NNPDF PDF parametrization supplemented with 
Eq.~(\ref{eq:photon}), we have performed two fits at NLO and NNLO
to DIS data only, with the same settings used for NNPDF2.3, but with 
QED corrections in the PDF evolution now included, as discussed in the
previous section.

The $\chi^2$ for the fit to the  total dataset and the
individual DIS experiments are shown in Table~\ref{tab:chi2},
with and without QED corrections, and with QCD corrections
included either at NLO or at NNLO. The $\chi^2$ listed in the table
use the so-called experimental definition of the $\chi^2$, in
which normalization uncertainties are included in the covariance
matrix (see App. A of Ref.~\cite{Ball:2012wy}): this definition is
most suitable for benchmarking purposes, as it is independent of the
fit results, but it is unsuitable for minimization as it would lead to
biased fit results.
It is clear that there is essentially no difference in  fit quality between
the QCD and QED+QCD fits. 
Indeed, a direct comparison of the PDFs
 obtained in the pairs of fits with and without QED corrections show
that they differ very little. 
\begin{figure}[ht]
  \begin{centering}
    \includegraphics[scale=0.70]{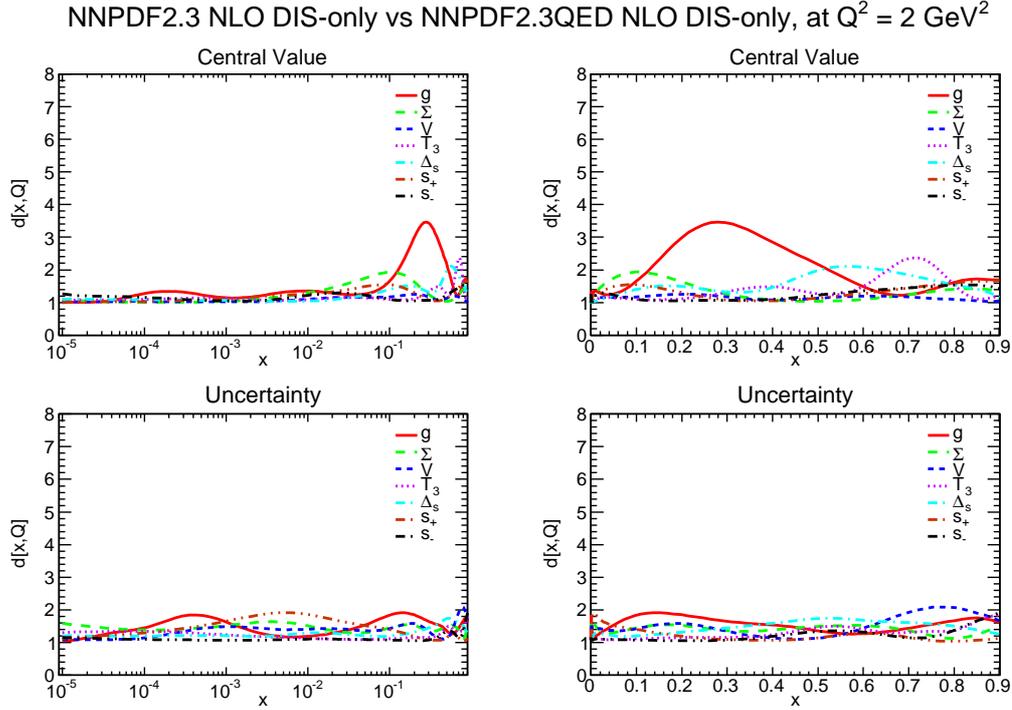}
    \par\end{centering}
  \caption{\label{fig:distances-nlo} \small Distances
    between PDFs in the NNPDF2.3 NLO DIS-only fit and the fit
    including QED corrections, at the input scale of $Q_0^2$=2
    GeV$^2$. Distances between central values (top) and uncertainties
    (bottom) are shown, on a logarithmic (left) or linear (right)
    scale in $x$.}
\end{figure}

\begin{figure}[ht]
  \begin{centering}
    \includegraphics[scale=0.37]{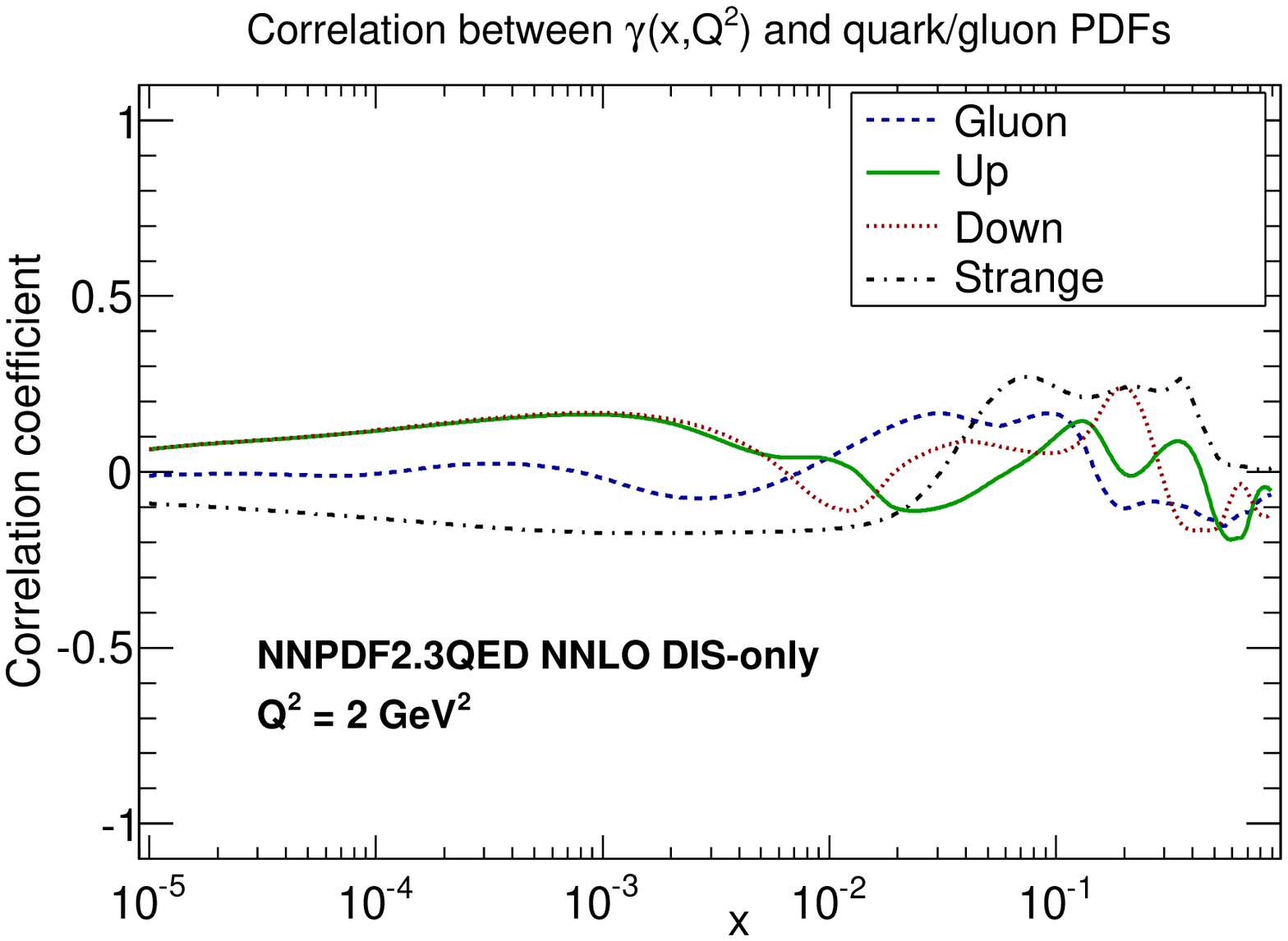}
    \includegraphics[scale=0.37]{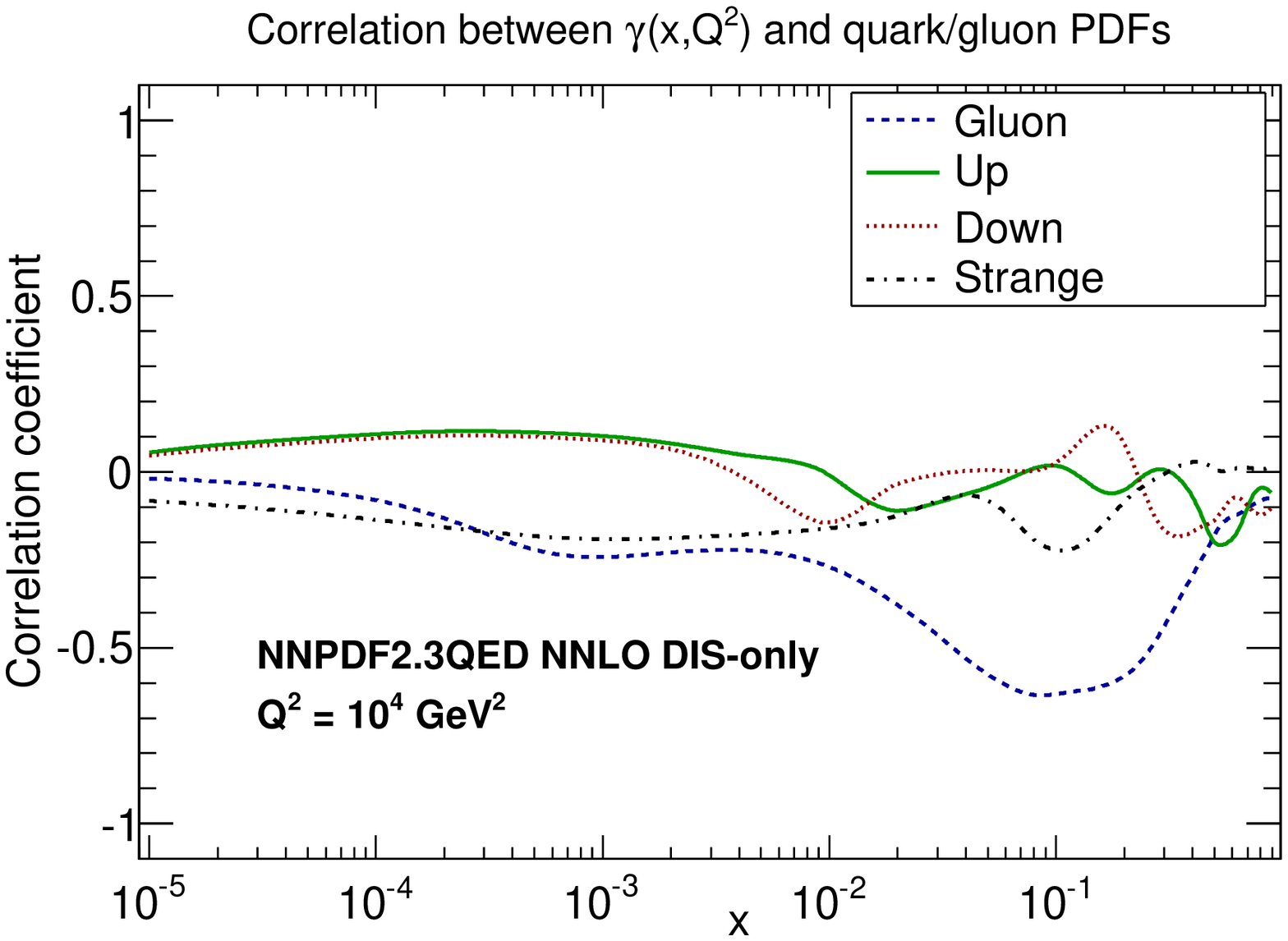}
    \par\end{centering}
  \caption{\label{fig:gammacorr} \small Correlation between the photon
    and other PDFs in the NNPDF2.3QED NLO DIS-only fit, shown as a
    function of $x$ at the input scale  $Q_0^2$=2
    GeV$^2$ (left) and at $Q^2=10^4$~GeV$^2$.}
\end{figure}
\begin{figure}[ht]
  \begin{centering}
    \includegraphics[scale=0.37]{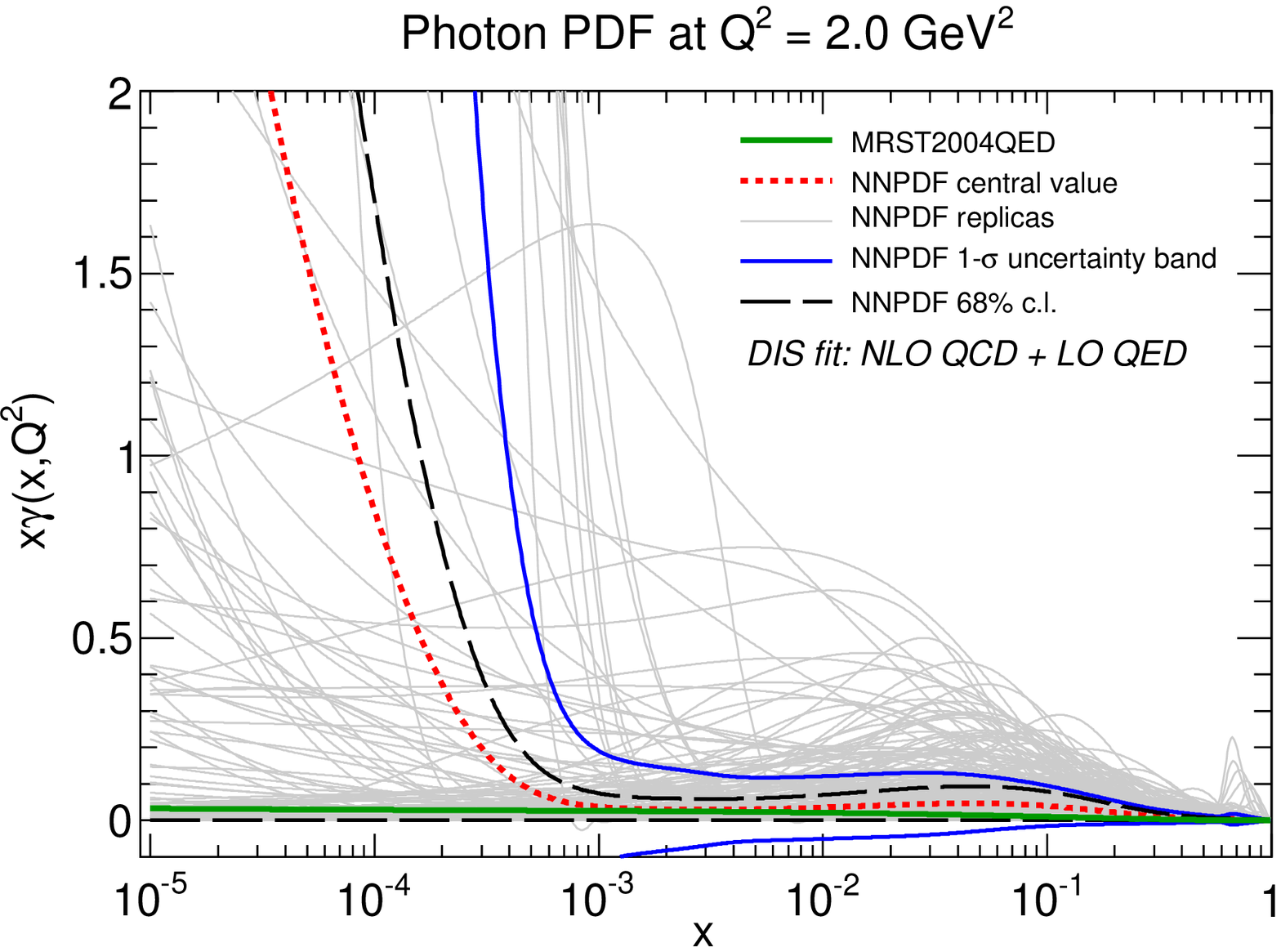}
    \includegraphics[scale=0.37]{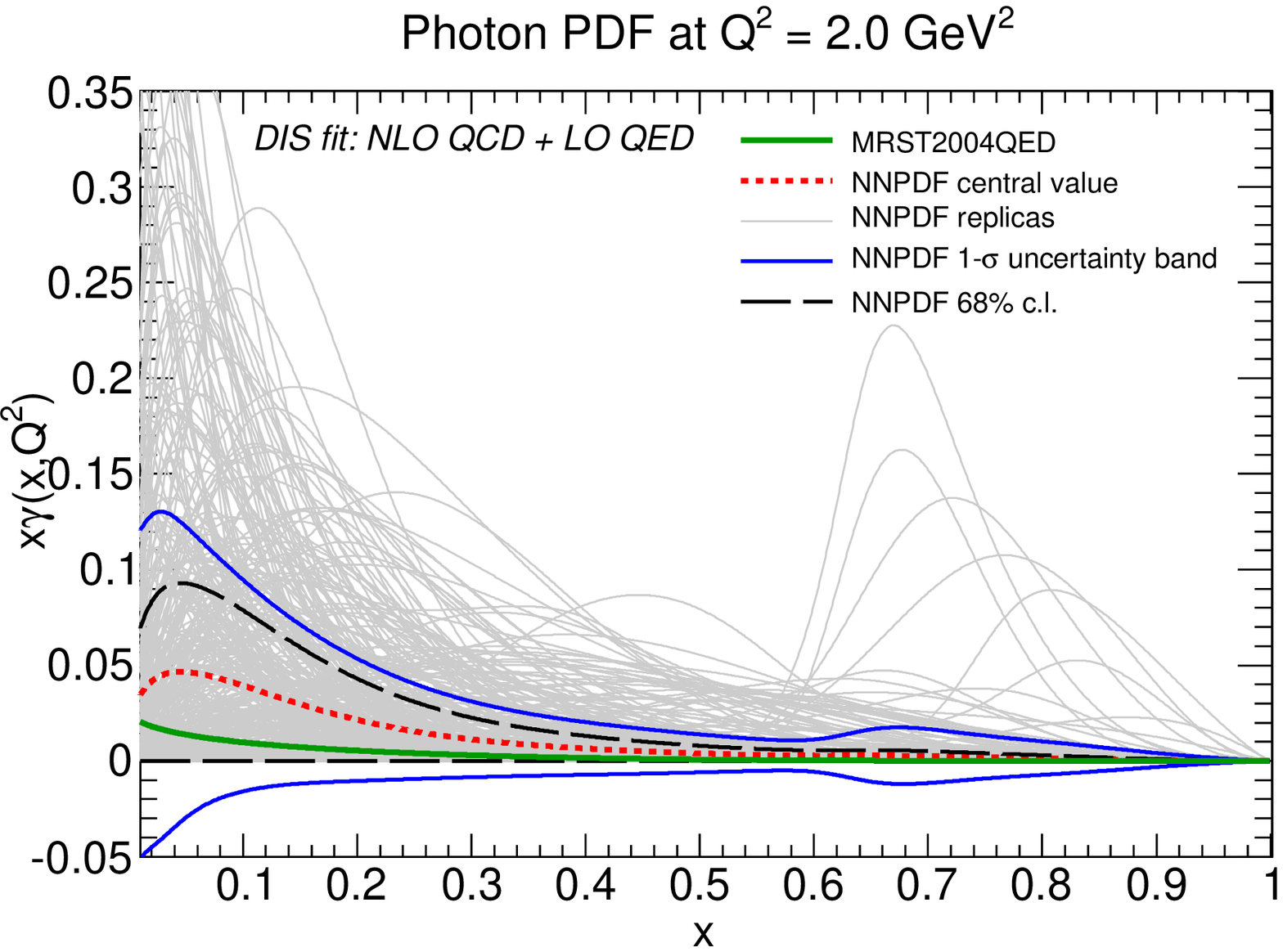}
    \par\end{centering}
  \caption{\label{fig:xpht-nlo} \small The photon PDF determined from
    the NNPDF2.3QED NLO DIS-only fit, in a linear (left plot) and
    logarithmic (right plot) scales, $N_{\text{rep}}=500$.  We show the
    central value (mean), the individual replicas and the
    PDF uncertainty band defined as a one $\sigma$ sigma interval and
    as a symmetric 68\%
    confidence level centered at the mean. The MRST2004QED photon PDF is
    also shown.  }
\end{figure}
In order to assess this difference quantitatively, 
in Fig.~\ref{fig:distances-nlo} we plot the distance between central
values and uncertainties of individual combinations of 
PDFs in the NLO QCD fit before
and after the inclusion of QED corrections. We refer to
Ref.~\cite{Ball:2010de} for a definition of the various combinations
of PDFs and of the distance.  
Recall that for a set of $N_{\rm rep}$ PDF replicas, $d\sim1$
corresponds to PDFs extracted from the same underlying distribution
(i.e. to statistically equivalent PDF sets), while $d\sim \sqrt{N_{\rm
    rep}}$ (so $d \sim 10$ in our case)
corresponds to PDFs extracted from distributions whose means or central
values differ by one $\sigma$.
The  distances are shown in Fig.~\ref{fig:distances-nlo} for the NLO
fit: it is clear that all PDFs but the gluon from the sets with and
without QED corrections are statistically equivalent, while the
gluon shows a change in the valence region of less than half $\sigma$. 
These results are unchanged  when QCD is treated at NNLO order.

The fact that the inclusion of a photon PDF has a negligible impact on
other PDFs can be also seen by determining the correlation between the
photon and other PDFs. Results are shown in Fig~\ref{fig:gammacorr}. The
correlation is negligible at the input  scale, meaning that the
particular shape of the photon in each replica has essentially no
effect on the other PDFs of that replica. In particular, this
correlation is much smaller than that which arises at a higher scale
(also  shown in Fig.~\ref{fig:gammacorr}), due to the mixing of PDFs
with the photon induced by PDF evolution.

Hence, at the initial scale $Q_0^2=2$~GeV$^2$ the sets with and without
QED corrections differ mainly because of the presence of a photon PDF
in the latter. The photon PDF  determined in the NLO fit is shown
in Fig.~\ref{fig:xpht-nlo} at $Q_0^2=2$~GeV$^2$: the individual
replicas, the mean value, the one-$\sigma$ range and the 68\%
confidence interval are all shown. The MRST2004QED photon PDF is also
shown.
It is clear that positivity imposes
a strong constraint on the photon PDF, which is only very loosely
constrained by DIS data. As a consequence, the probability
distribution of replicas is very asymmetric: some replicas may
have large positive values of $\gamma(x,Q^2)$, but positivity
always ensures that no replica goes below zero. It follows that the
usual gaussian assumptions cannot be made, and in particular there is
a certain latitude in how to define the uncertainty. 
Here and in the remainder
of this paper we will always define central values as the mean of the
distribution, and uncertainties as symmetric 68\% confidence levels centered at
the mean, namely, as the symmetric interval centered at the mean such
that 68\% of the replicas falls within it.  
All uncertainty bands will
be determined in this way, unless otherwise stated. Because of the accumulation
of replicas just above zero, the lower edge of the uncertainty band on
the photon PDF at the
initial scale turns out to be very close to zero. 
Again, results are essentially unchanged when the
fit is done using NNLO QCD theory.

\begin{figure}[ht]
\begin{centering}
\includegraphics[scale=0.37]{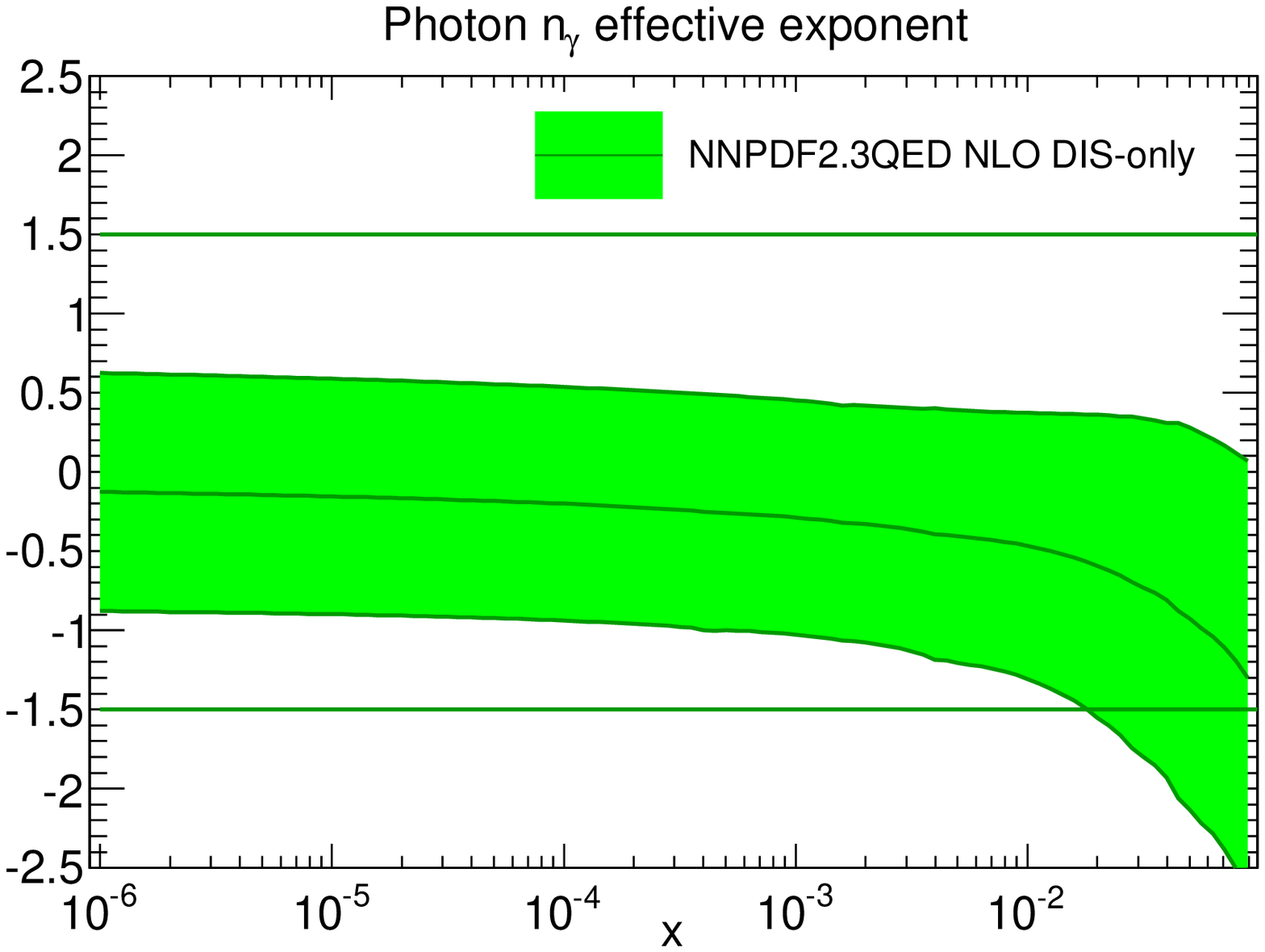}
\includegraphics[scale=0.37]{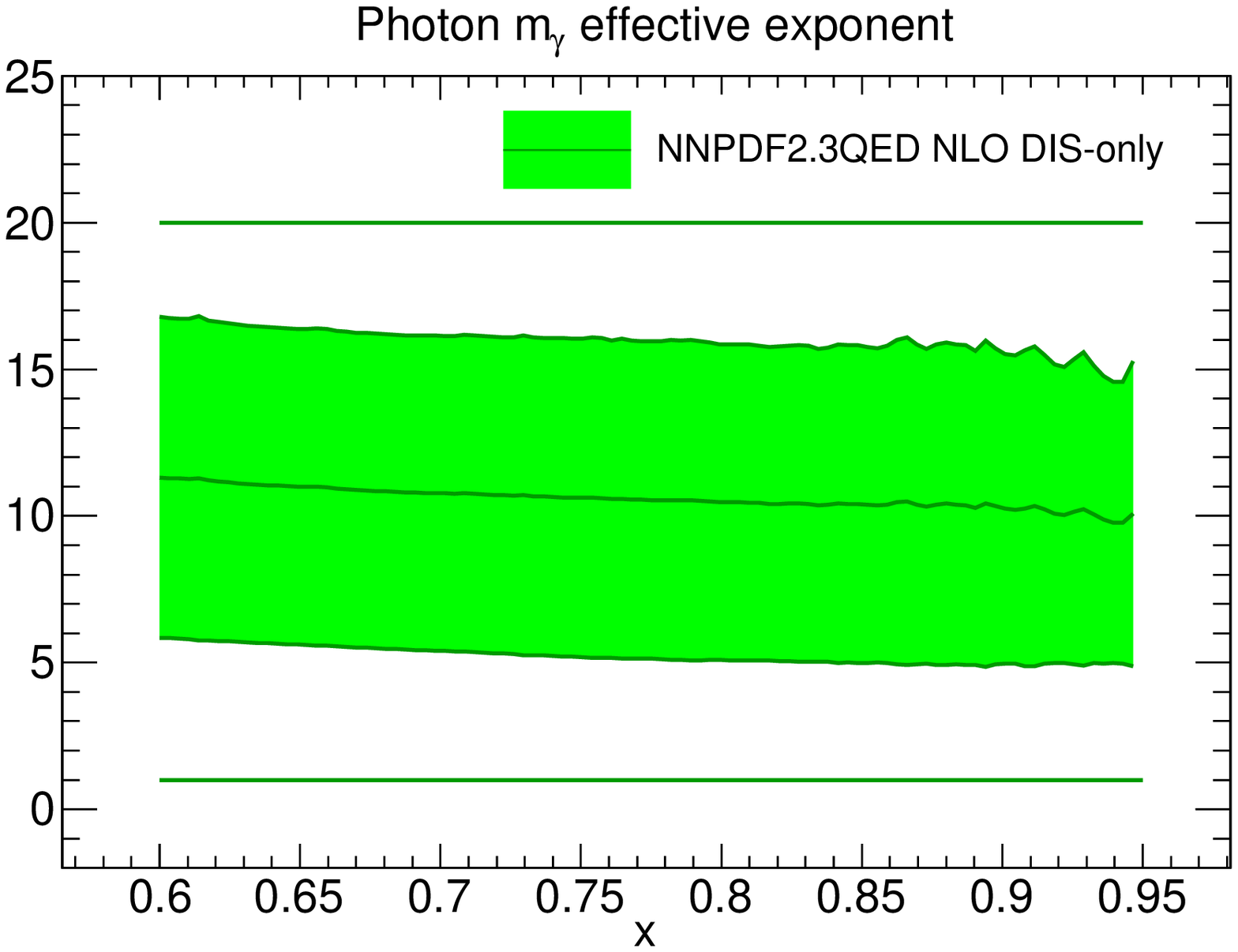}
\par\end{centering}
\caption{\label{fig:preproc} One-$\sigma$ range for the effective exponents
  Eq.~(\ref{eq:exp})  for the photon
  PDF, compared to the range of variation of the
  preprocessing exponents Eq.~(\ref{prerange}) (shown as horizontal lines).}
\end{figure}

As discussed in Sect.~\ref{sec:disqed}, we have determined the
effective exponents Eq.~(\ref{eq:exp}) for the photon
PDF, and compared them to the range of variation of the preprocessing
exponents Eq.~(\ref{prerange}). Given the very loose constraints that
the data impose on the photon PDF, it is especially important to make
sure that preprocessing imposes no bias. The comparison is shown in
Fig.~\ref{fig:preproc}: it is clear that the effective exponents are
well within the range chosen for the preprocessing exponents, so that
no bias is being introduced.

The photon PDF at the initial scale shown in Fig.~\ref{fig:xpht-nlo}
is essentially compatible with zero, and it
remains small even at the top of its uncertainty band; it is
consistent with the MRST2004QED photon PDF within its large uncertainty band.

\begin{figure}[ht]
  \begin{centering}
    \includegraphics[scale=0.43]{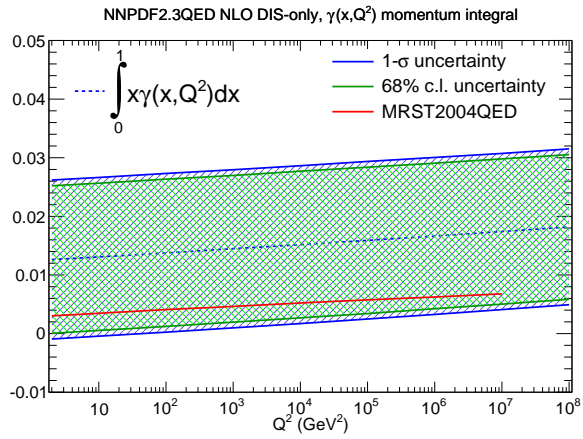}
    \par\end{centering}
  \caption{\small \label{fig:msr-gamma} The momentum fraction carried
   by the photon PDF in the NLO fit as a function of scale. The
   MRST2004QED result is also shown.}
\end{figure}

The momentum fraction carried by the photon  is accordingly small: it
is shown as a function of scale in  Fig.~\ref{fig:msr-gamma} for the
NLO fit; results at NNLO are very similar. 
At the input scale $Q_0^2=2$~GeV$^2$ we find
\be
\int_0^1 x \gamma \left(x,Q^{2}_0\right)= \lp 1.26 \pm 1.26 \rp\,\% \,  \, ,\label{eq:msr}
\ee
The symmetric 68\% confidence level uncertainty of Eq.~(\ref{eq:msr}) turns out to
be quite close to the standard deviation  
$\sigma=1.36\%$. Hence, even at the top of its uncertainty range
the photon momentum fraction hardly exceeds 2\%, and it is compatible
with zero to one $\sigma$. The momentum fraction carried by the the MRST2004QED 
photon (also shown in Fig.~\ref{fig:msr-gamma}) is well below 1\%,
and thus compatible with our own within uncertainties

\begin{figure}[ht]
\begin{centering}
\includegraphics[scale=0.80]{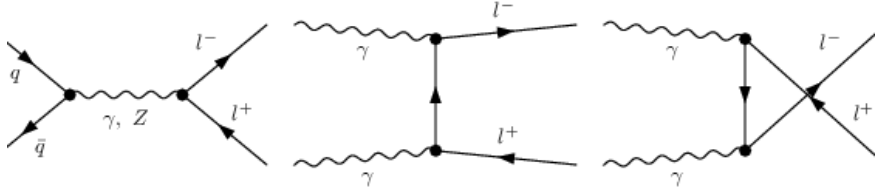}
\par\end{centering}
\caption{\label{fig:lhczborn} Feynman diagrams for the  Born-level partonic
  subprocesses which contribute to the production of dilepton pairs in
  hadronic collisions.
}
\end{figure}

\section{The photon PDF from $W$ and $Z$  production at the LHC}
\label{sec:lhcwz}

The photon PDF $\gamma(x,Q^2)$ 
determined in the previous section from a fit to DIS
data is affected by very large uncertainties. 
This suggests that its impact on predictions for hadron collider processes to which
the photon PDF contributes already at leading order could be
substantial, and thus, conversely, that data on such processes might provide 
further constraints. In this section we use the simplest of such processes,
namely, electroweak gauge boson production, to constrain the photon PDF.

At hadron colliders, the dilepton production process
receives contributions at Born level both from quark-initiated 
neutral current $Z/\gamma^*$ exchange and
from photon-initiated diagrams, see Fig.~\ref{fig:lhczborn}, and thus
the contributions from $\gamma(x,Q^2)$ must be
included even in a pure leading-order treatment of QED
effects. 
Photon-initiated  contributions  to
dilepton production at hadron colliders were
recently emphasized in Ref.~\cite{Dittmaier:2009cr}, where 
$O(\alpha)$
radiative corrections to this process~\cite{Baur:1998kt,Dittmaier:2001ay,Baur:2001ze,Baur:2004ig,Arbuzov:2007db,Arbuzov:2005dd,Brensing:2007qm,Balossini:2009sa,CarloniCalame:2007cd,Dittmaier:2009cr} were reassessed, and also kinematic cuts to
enhance the sensitivity to $\gamma(x,Q^2)$ were suggested.

\begin{figure}
\begin{centering}
\includegraphics[trim=0cm 0.5cm 0cm 0cm, clip=true, scale=0.80]{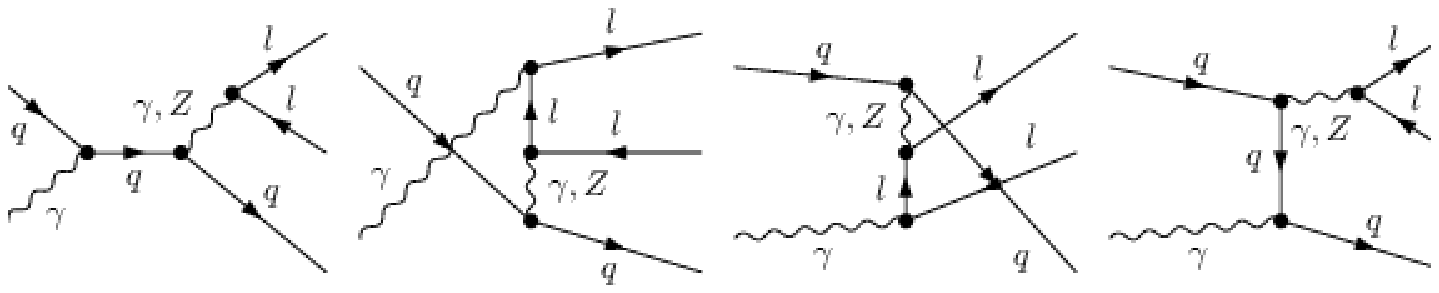}\\
\includegraphics[scale=0.60]{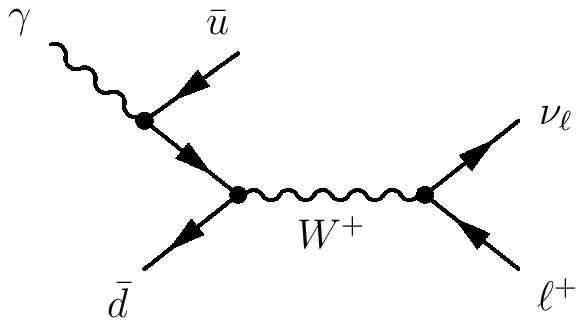}\qquad
\includegraphics[scale=0.60]{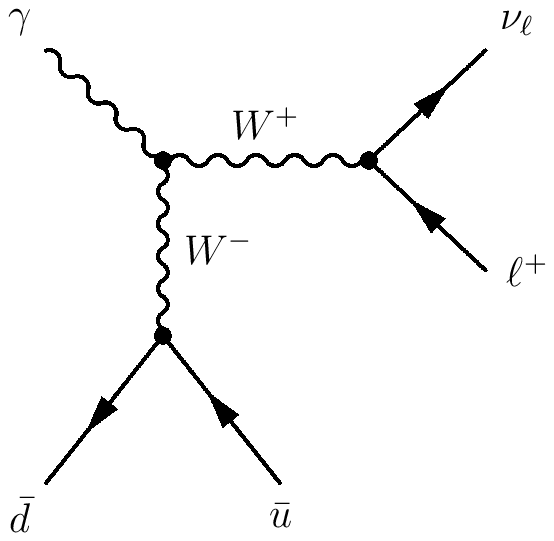}
\par\end{centering}
\caption{\label{fig:lhcwz2} Some  Feynman diagrams for $O(\alpha)$
photon-initiated partonic subprocesses which contribute to neutral
current (top row) and charged current (bottom row) dilepton pair
production in hadronic collisions.}
\end{figure}

Beyond the Born approximation, radiative corrections to the neutral-current
 process, as well as the charged-current process, which starts 
at  $O\lp \alpha\rp$ (see Fig.~\ref{fig:lhcwz2} for some
representative Feynman diagrams)
may be comparable in size to the Born
level contribution, because the suppression due to the extra power of
$\alpha$ might be compensated by the enhancement
arising from the larger size of the quark-photon parton
luminosity in comparison to the photon-photon luminosity. However, a full
inclusion of $O(\alpha)$ corrections would require solving evolution
equations to NLO in the QED and mixed QED+QCD terms, so it is beyond
the scope of this work; we will nevertheless discuss an approximate
inclusion of such corrections which, while not allowing us to claim
more than LO accuracy in QED, should ensure that NLO QED corrections
are not unnaturally large.

We use neutral and charged-current 
Drell-Yan production data from the LHC to further
constrain the photon PDF, thereby arriving at our final NNPDF2.3QED
PDF sets.
As discussed in the introduction, we do this by  combining the
photon PDF from NNPDF2.3 DIS-only set discussed in the previous section
with the standard NNPDF2.3 PDF set, and then using gauge boson
production data to reweight the result. We discuss first this two-step
fitting procedure, and then the ensuing NNPDF2.3QED PDF set and its features.

\subsection{The prior NNPDF2.3QED and its reweighting}
\label{sec:combination}

As a first step towards the determination of a PDF set with inclusion
of QED corrections, we would like to use the photon PDF determined in
the previous section from a fit to DIS data in conjunction with PDFs
which retain all the information provided by the full NNPDF2.3 data
set, which, on top of DIS, includes Drell-Yan and jet production data
from the Tevatron and the LHC.

We have seen in the previous section that all PDFs determined
including QED corrections are statistically equivalent to their
standard counterparts determined when QED corrections are not
included, with the only exception of the gluon, which undergoes a
change by less than half $\sigma$ in a limited kinematic
region. Furthermore, the photon in each PDF replica is essentially
uncorrelated to the shape of other PDFs which are input to perturbative
evolution, the only significant  correlation being due to the mixing
induced by the evolution itself.
We
can therefore simply combine the photon PDF obtained from the DIS fit
of the previous section with the standard NNPDF2.3 PDFs at the
starting scale $Q_0^2$.
 This
procedure entails a certain loss of accuracy, which in particular
appears as a violation of the momentum sum rule of the order of the
momentum fraction carried by the photon at the initial scale
Eq.~(\ref{eq:msr}), 
namely of order 1\%. This is the accuracy to
which the momentum sum rule would be verified if it were not imposed
as a constraint in the fit~\cite{Ball:2011uy}.

The information contained in LHC Drell-Yan production data is included
in the fit through the Bayesian reweighting method presented in Ref.~\cite{Ball:2011gg,Ball:2010gb}. 
This method allows for the inclusion of new data without having to
perform a full refit, by using Bayes' theorem to modify the prior
probability distribution of PDF replicas in order to account for the
information contained in the new data. The ensuing replica set
contains an amount of information, and thus allows for the computation
of observables with an accuracy, that corresponds to an effective
number of replicas $N_{\rm eff}$, which may be determined from the
Shannon entropy of the reweighted set. 

\begin{table}
\centering
\begin{tabular}{c|c|c|c|c|c}
\hline
Dataset  &  Observable & Ref.  &  $N_{\rm dat} $ &$\lc \eta_{\rm min}, \eta_{\rm max}\rc$  &  
$\lc M^{\rm min}_{\rm ll}, M^{\rm max}_{\rm ll}\rc$\\
\hline
\hline
LHCb  $\gamma^*/Z$ Low Mass &  $d\sigma(Z)/dM_{ll}$   &  ~\cite{LHCb-CONF-2012-013}  & 9  & [2,4.5]  & 
[5,120] GeV
  \\
ATLAS $W,Z$  &  $d\sigma(W^{\pm},Z)/d\eta$  &   ~\cite{Aad:2011dm} & 30  & [-2.5,2.5]  & 
[60,120] GeV
  \\
 ATLAS $\gamma^*/Z$ High Mass  &  $d\sigma(Z)/dM_{ll}$  &   ~\cite{Aad:2013iua} & 13  & [-2.5,2.5]  & 
[116,1500] GeV
  \\
 \hline
\end{tabular}
\caption{\small Kinematical coverage of the three LHC datasets used to determinethe photon PDF.  \label{tab:expdata}
}
\end{table}

In our case, the new data only
constrain significantly the photon PDF,  hence we need to guarantee that
good accuracy is obtained by  starting with a large number of
photon replicas. The initial
prior set is thus obtained combining 500 photon PDF replicas with a
standard set of 100 NNPDF2.3 replicas. In practice, this is done by
simply producing five copies of the NNPDF2.3 100 replica set, and
combining each of them at random with one of the 500 photon PDF
replicas obtained from the QED fit to DIS data discussed in the
previous section. The procedure is performed at NLO and NNLO,  in each
case combining the photon PDF from the combined QED+QCD fit to DIS data
with the other PDFs from the corresponding standard NNPDF2.3 set. 
Furthermore, the
procedure is repeated for three different values of
$\alpha_s=0.117,\>0.118,\>0.119$. We find no dependence of the photon
PDF on the value of $\alpha_s$, though there are minor differences
between the photon determined using NLO or NNLO QCD theory in the DIS
fit.
\begin{figure}[ht]
\begin{centering}
\includegraphics[scale=0.36]{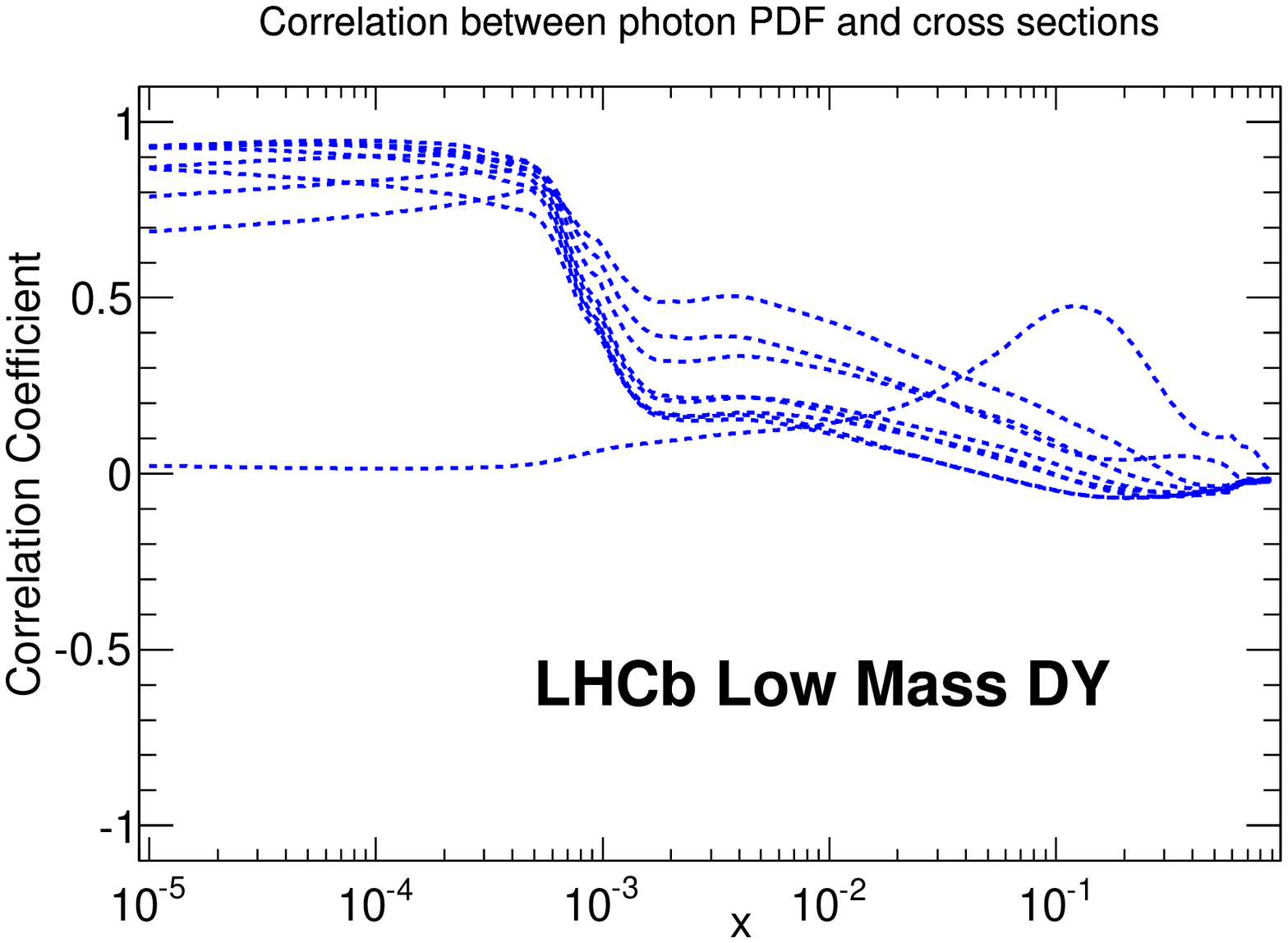}
\includegraphics[scale=0.36]{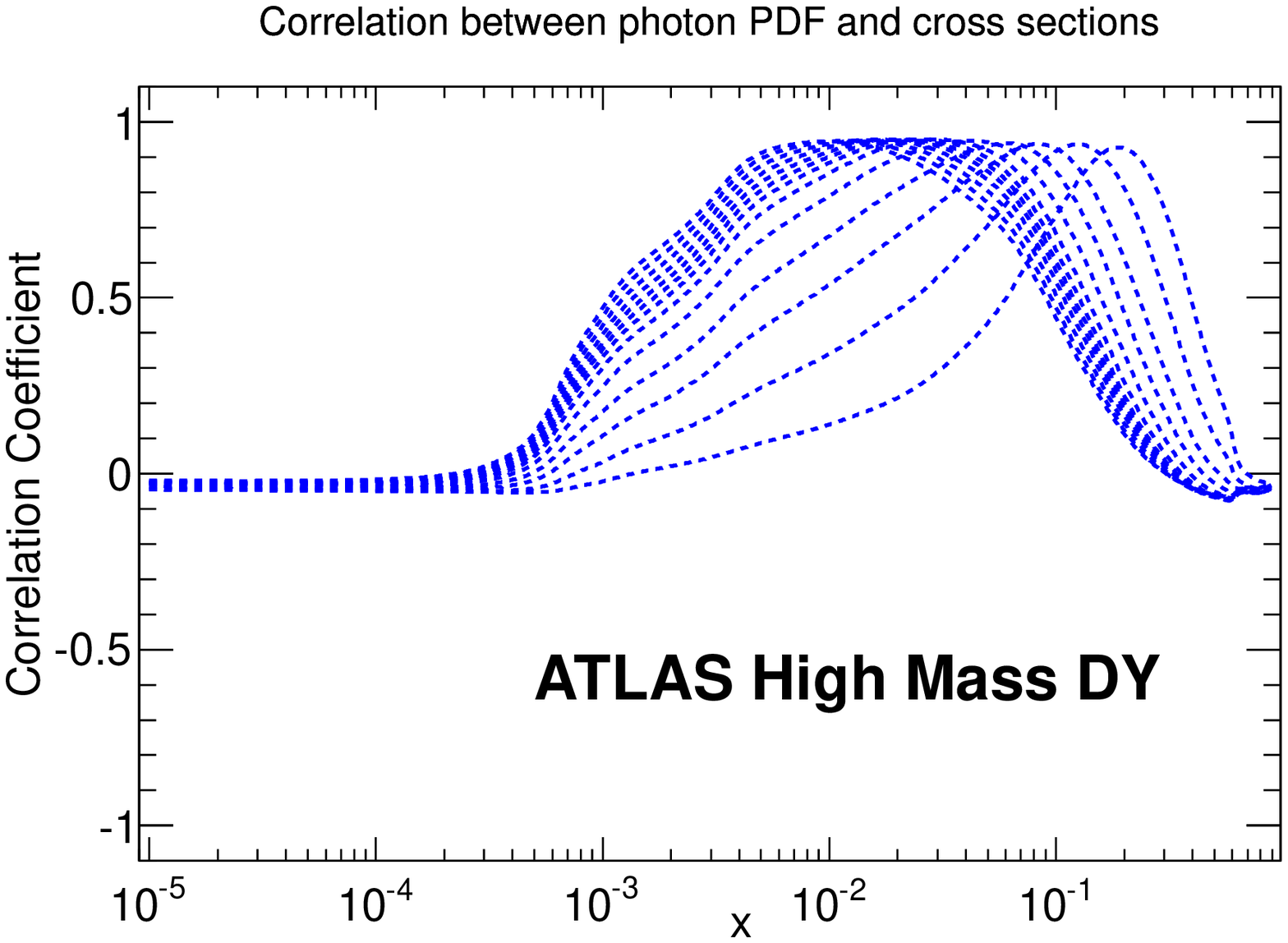} \\
\includegraphics[scale=0.36]{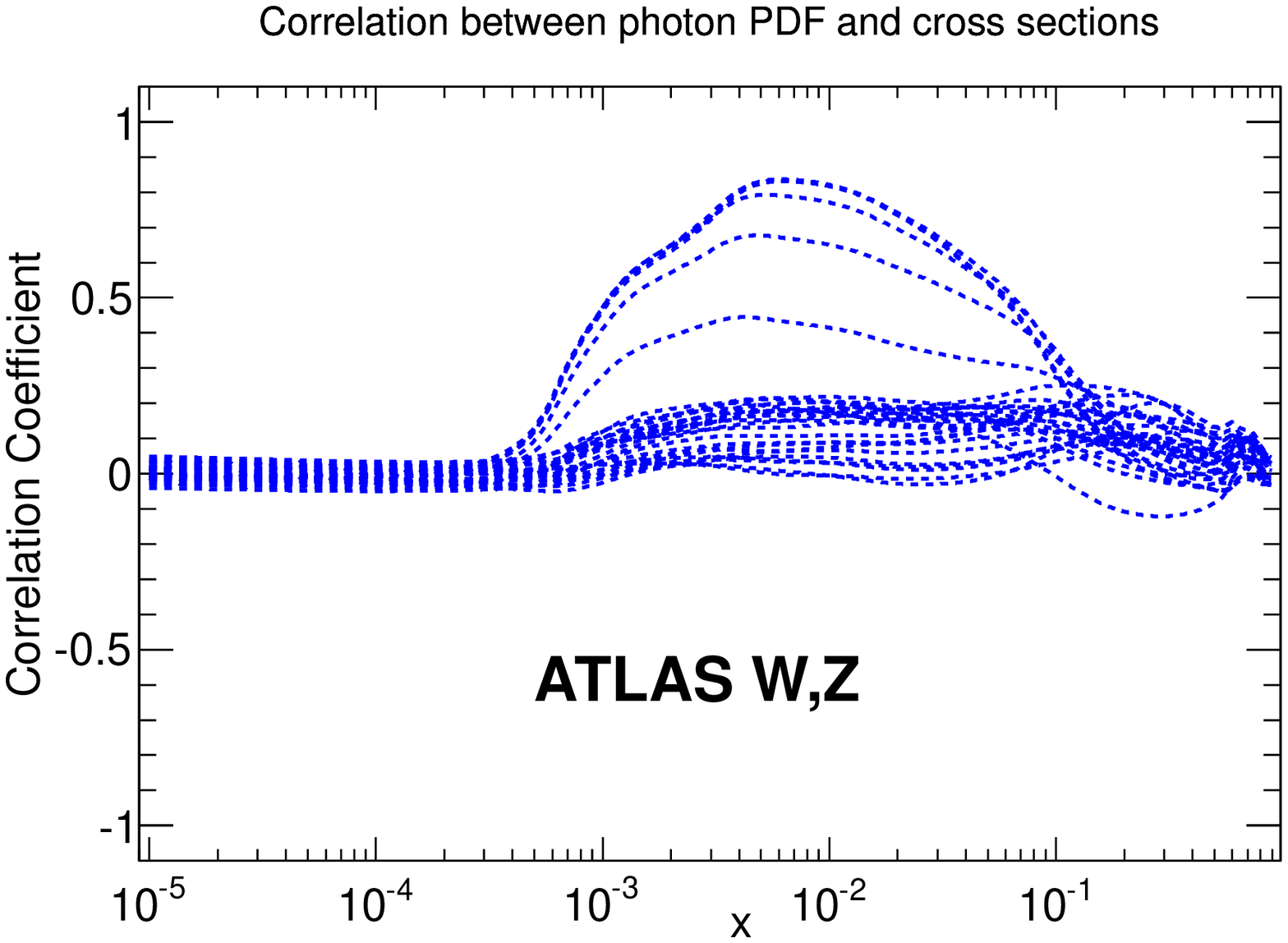}
\par\end{centering}
\caption{\label{fig:correlations} Correlation between the photon PDF
  and the LHC data of Tab.~\ref{tab:expdata}, shown as function of $x$
  for $Q^2=10^4$~GeV$^2$. Each curve corresponds
  to an individual data bin.}
\end{figure}

In each case, the set of $N_{\rm rep}=500$ replicas is then evolved to
all scales using combined QED+QCD evolution. Note that this in
particular implies that no further violation of the momentum sum rule
is introduced on top of that which was present at the initial scale,
up to approximations introduced when solving the evolution
equations.

Reweighting is performed using the following LHC datasets:
\begin{itemize}
\item LHCb low-mass $Z/\gamma^*$ Drell-Yan production from the 2010 run~\cite{LHCb-CONF-2012-013}
\item ATLAS inclusive  $W$ and $Z$ production data from the 2010 run~\cite{Aad:2011dm}
\item ATLAS high-mass $Z/\gamma^*$ Drell-Yan production from the 2011 run~\cite{Aad:2013iua},
\end{itemize}
whose kinematic coverage is summarized  in Table~\ref{tab:expdata}.
Using data with three different mass ranges for the dilepton pairs,
below, at, and above the $W$ and $Z$ mass, 
guarantees that both the low $x$ (from low mass) and high $x$ (from
high mass) regions are covered.

 \begin{figure}[t]
\begin{centering}
\includegraphics[scale=0.37]{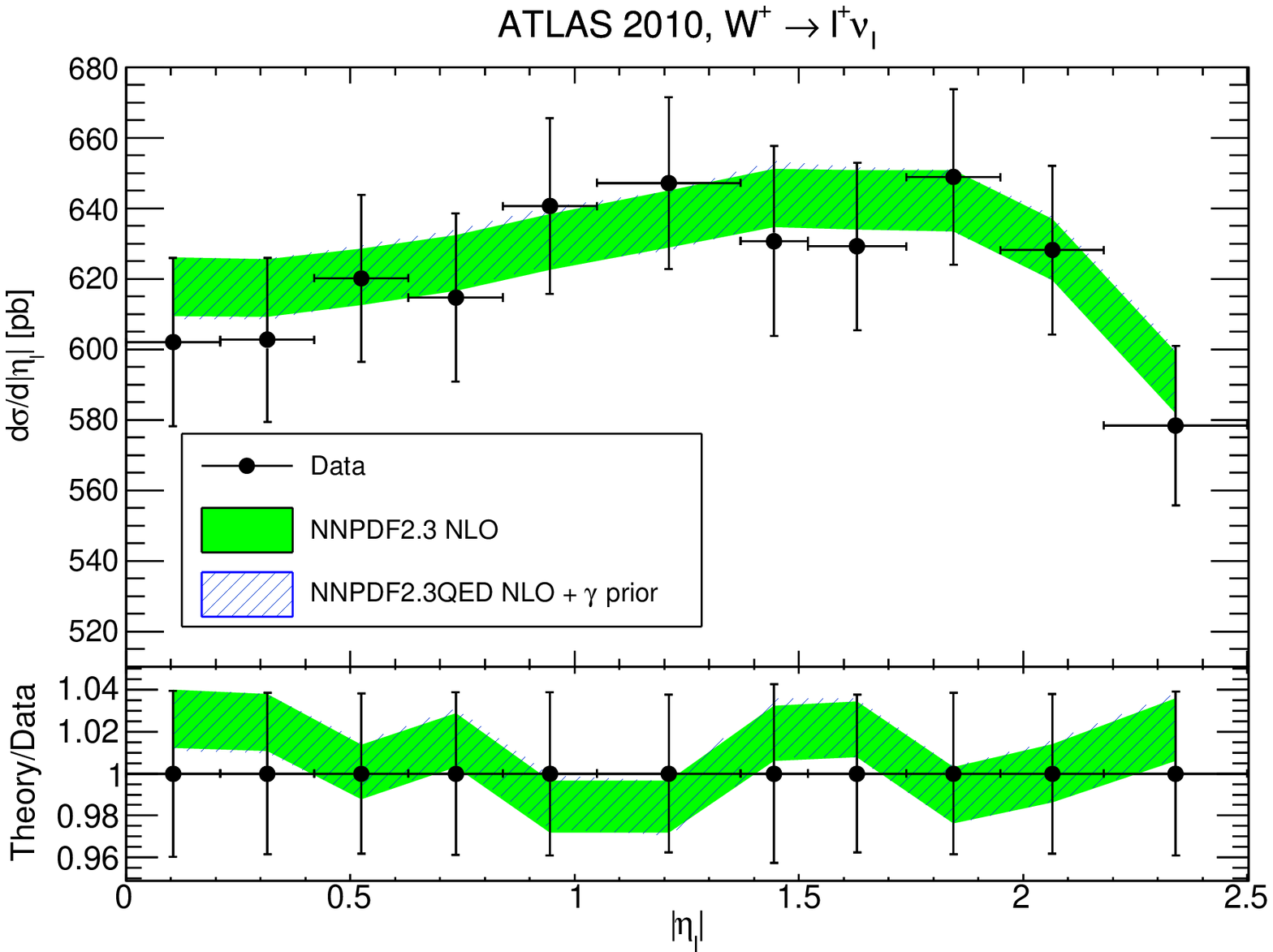}
\includegraphics[scale=0.37]{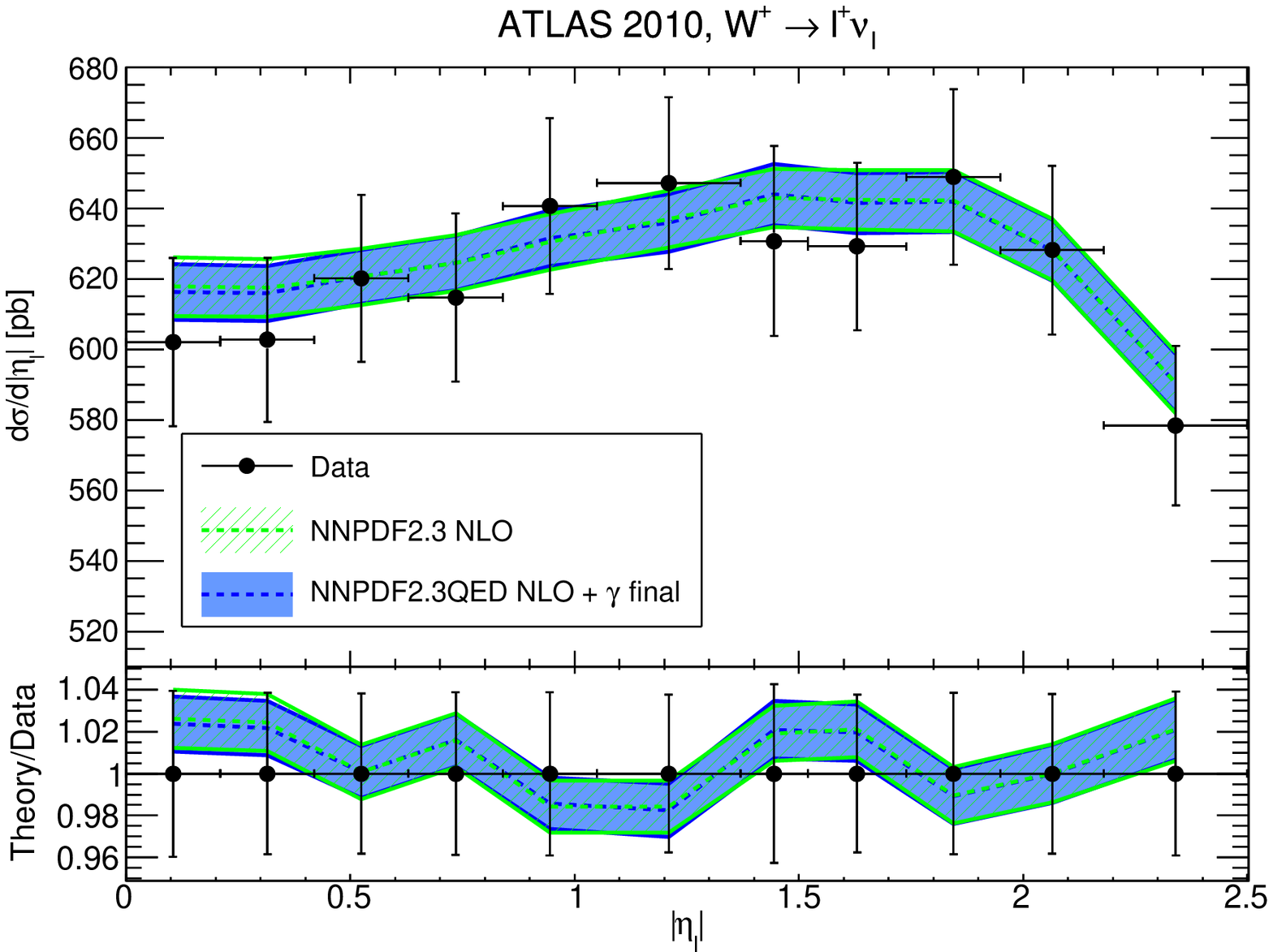}
\includegraphics[scale=0.37]{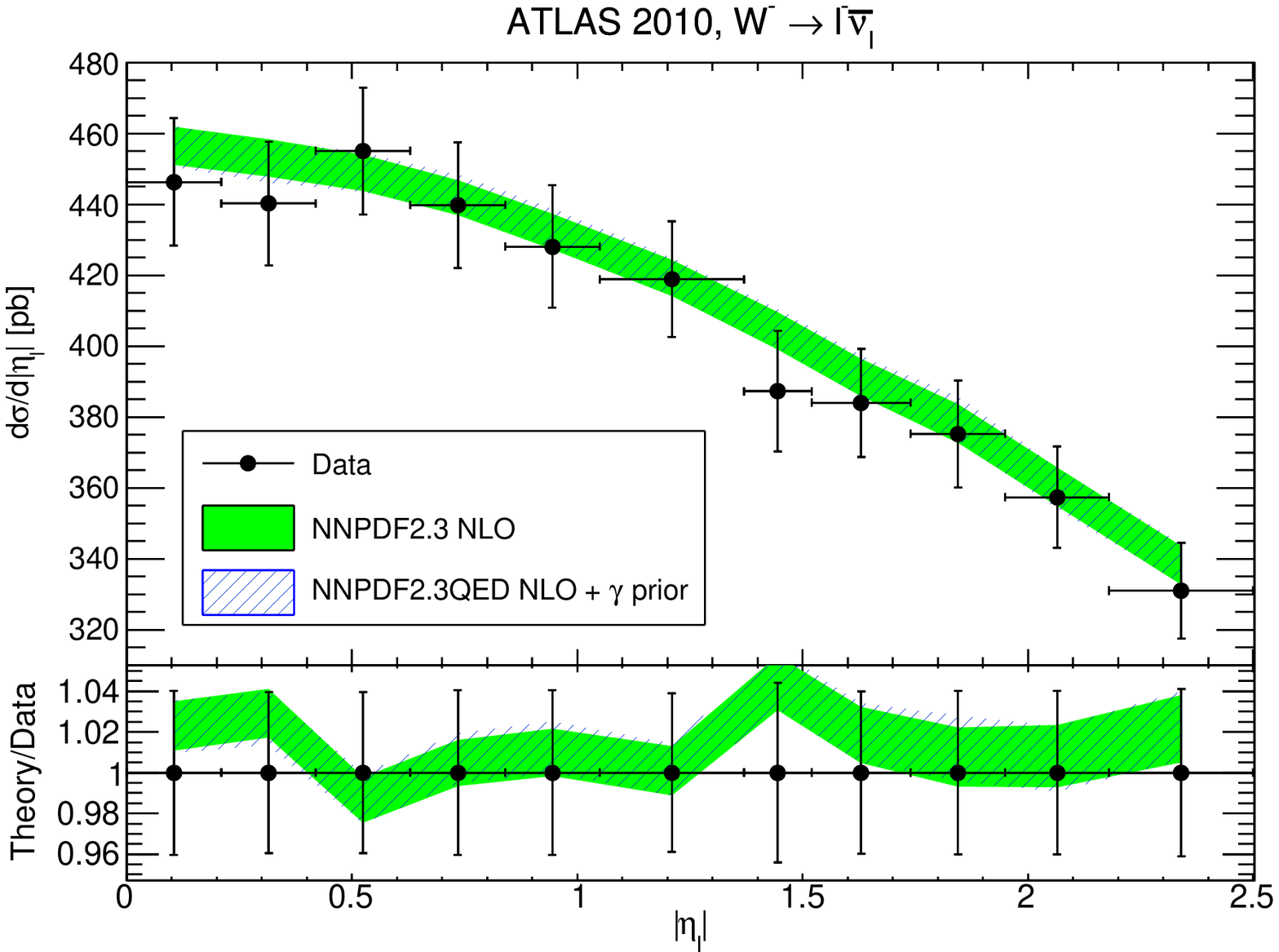}
\includegraphics[scale=0.37]{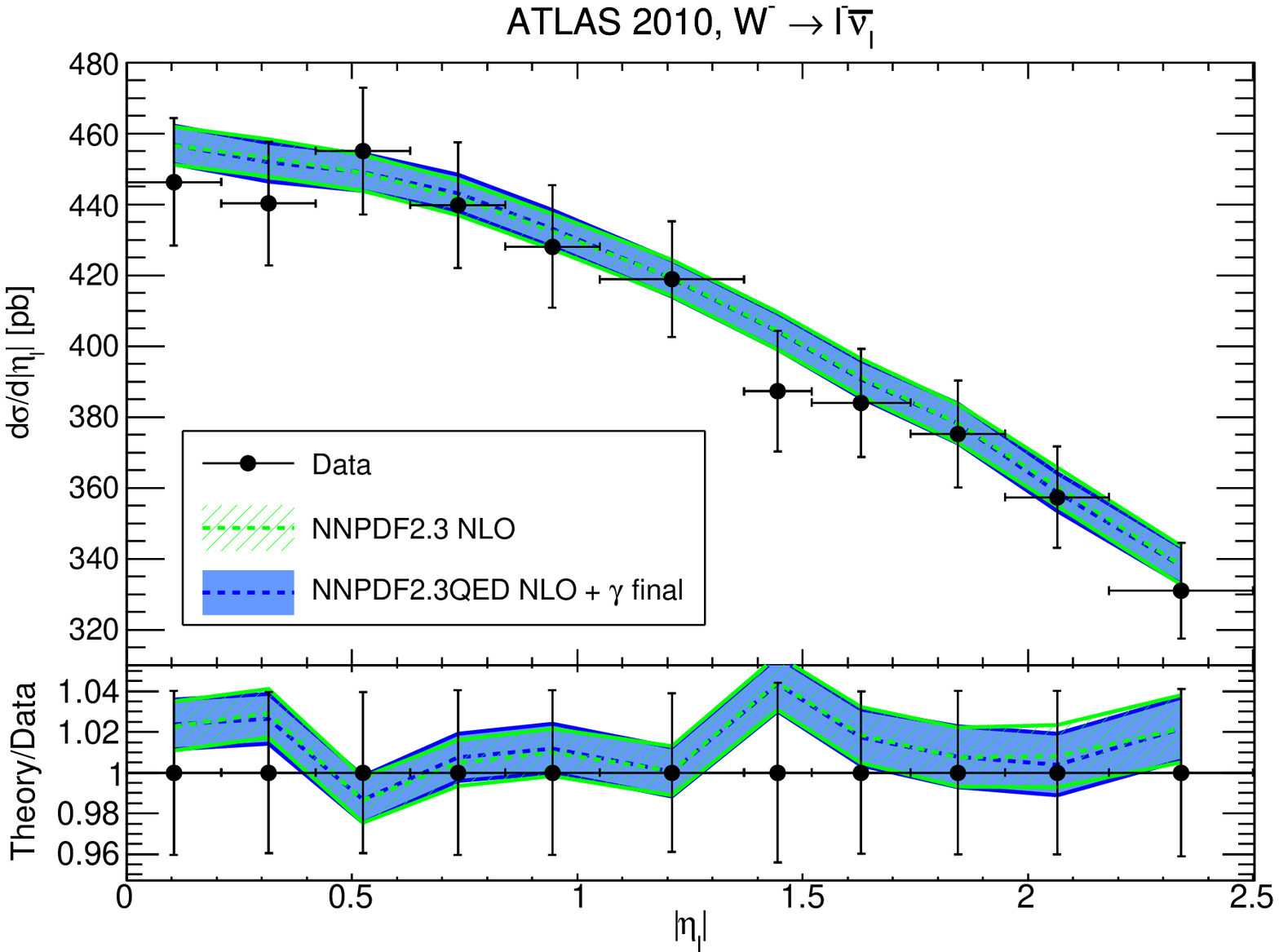}
\par\end{centering}
\caption{\label{fig:rwlhc} \small Comparison of the ATLAS $W$ production
data with NLO theoretical predictions obtained using
PDFs before (left) and after (right) reweighting with the data of
Tab.~\ref{tab:expdata}. In all plots we also show for comparison results obtained
using the default NNPDF2.3 PDF set, with all QED corrections switched
off. From top to bottom: $W^+$ and $W^-$. Error bands on the
theoretical prediction correspond to one $\sigma$
uncertainties. Experimental error bars give the total combined
statistical and systematic uncertainty.
}
\end{figure}

For all the ATLAS data the experimental covariance matrix is
available, hence the $\chi^2$ may be computed fully accounting for
correlated systematics. This is not the case for LHCb: hence, the
low-mass data are treated adding statistical and systematic errors in
quadrature, and only including normalization errors in the covariance
matrix. 
We have checked that if reweighting is performed using the
diagonal covariance matrix, statistically indistinguishable results
are obtained. This means that within the large uncertainty of the
photon PDF, and due to the small impact of QED corrections on the
quark and gluon PDFs, the lack of information on correlations for the
LHCb experiment is immaterial. 
However, this implies that $\chi^2$ values quoted for LHCb should only be
taken as indicative.
Unfortunately, the CMS off-peak Drell-Yan data~\cite{CMSdy} is not
yet publicly available, and thus could not be used in the present analysis.

 \begin{figure}[ht]
\begin{centering}
\includegraphics[scale=0.37]{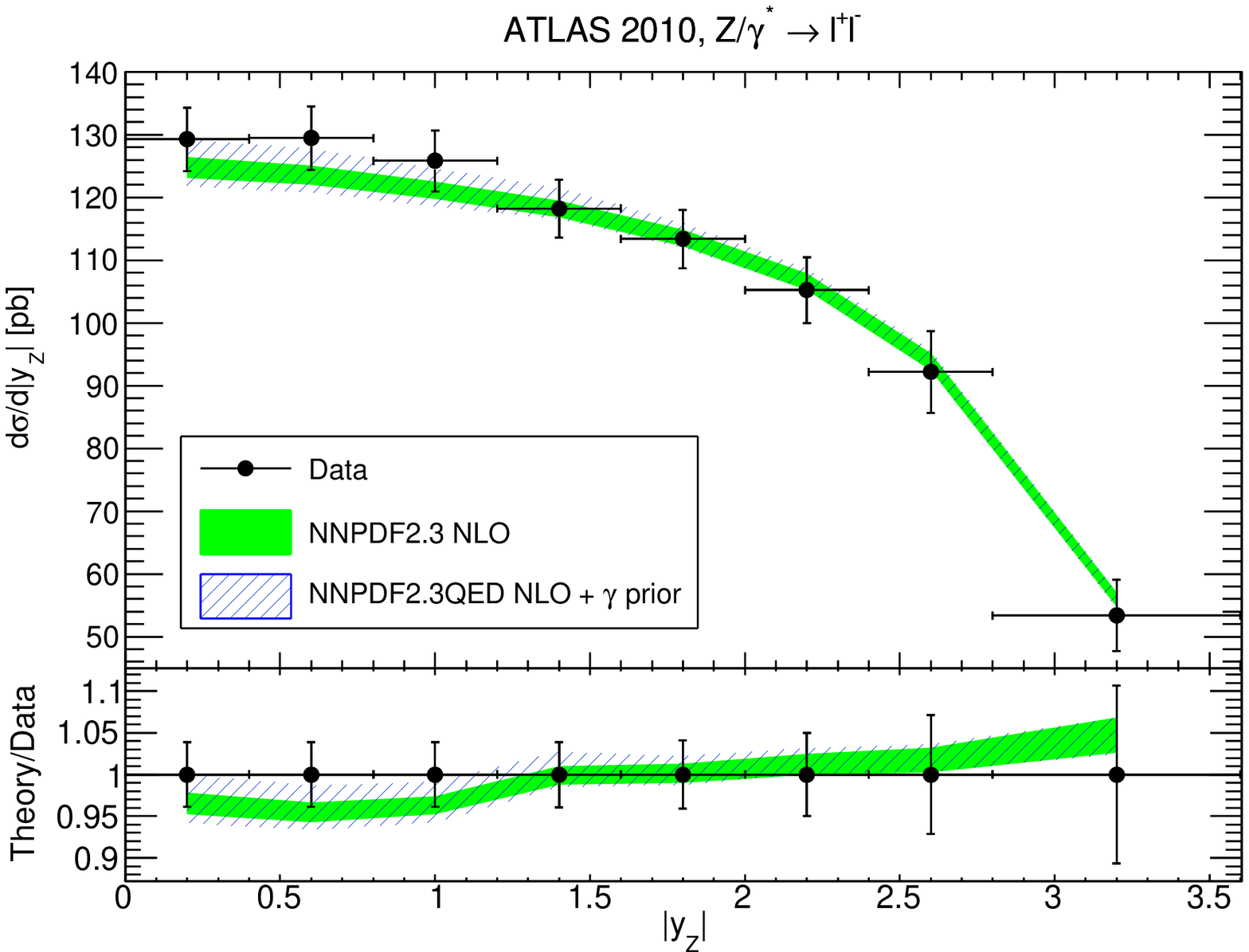}
\includegraphics[scale=0.37]{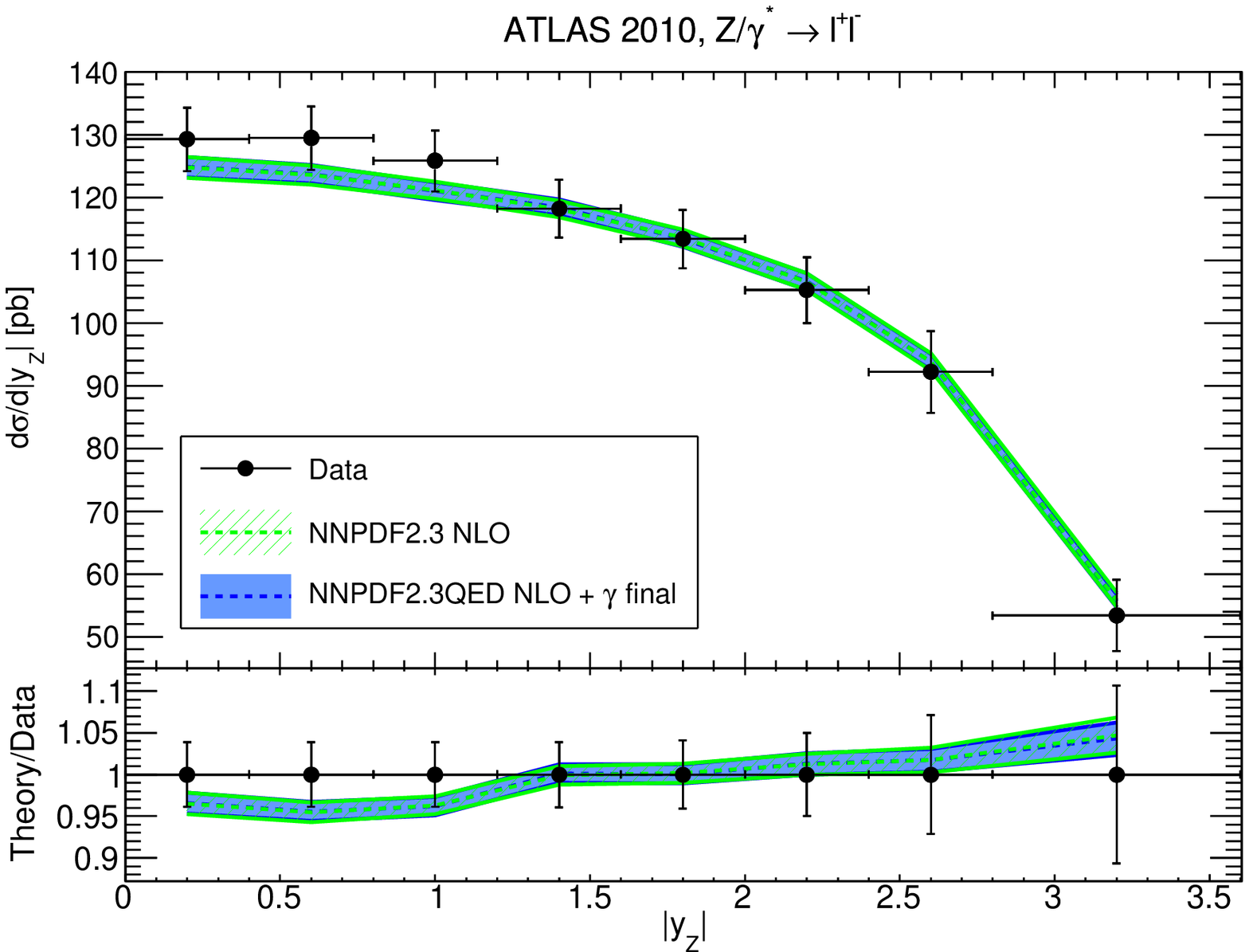}
\par\end{centering}
\caption{\label{fig:rwlhca} \small Same as Fig.~\ref{fig:rwlhc}, but
  for the neutral current data. 
}
\end{figure}

The range of $x$ for the photon PDF which is affected by each of the
datasets of Table~\ref{tab:expdata} can be determined quantitatively
by computing the correlation coefficient (see~\cite{Demartin:2010er} and Sect. 4.2 of
Ref.~\cite{Alekhin:2011sk}) between a given observable and the PDFs. The
correlation coefficients computed using the NNPDF2.3QED NLO prior set
 are shown in Fig.~\ref{fig:correlations} for  each 
 bin in the
experiments in Table~\ref{tab:expdata}. It is clear that the LHC
data guarantee a good kinematic coverage for all $10^{-5}\lsim x\lsim
0.1$. 
The correlation is weaker for real $W$ and $Z$ production data,
where the $s$-channel quark contribution dominates as the propagator
goes on shell. 
The high-mass (low-mass) Drell-Yan data is thus essential to pin
down $\gamma(x,Q^2)$ at large (small) Bjorken-$x$, where uncertainties
are the largest. Indeed, a preliminary determination of the photon
distribution~\cite{Carrazza:2013bra}, which did not use the LHCb data,
had significantly larger uncertainties at small $x$, consistently with
the expectations based on the correlation plot of
Fig.~\ref{fig:correlations}.

 \begin{figure}[ht]
\begin{centering}
\includegraphics[scale=0.37]{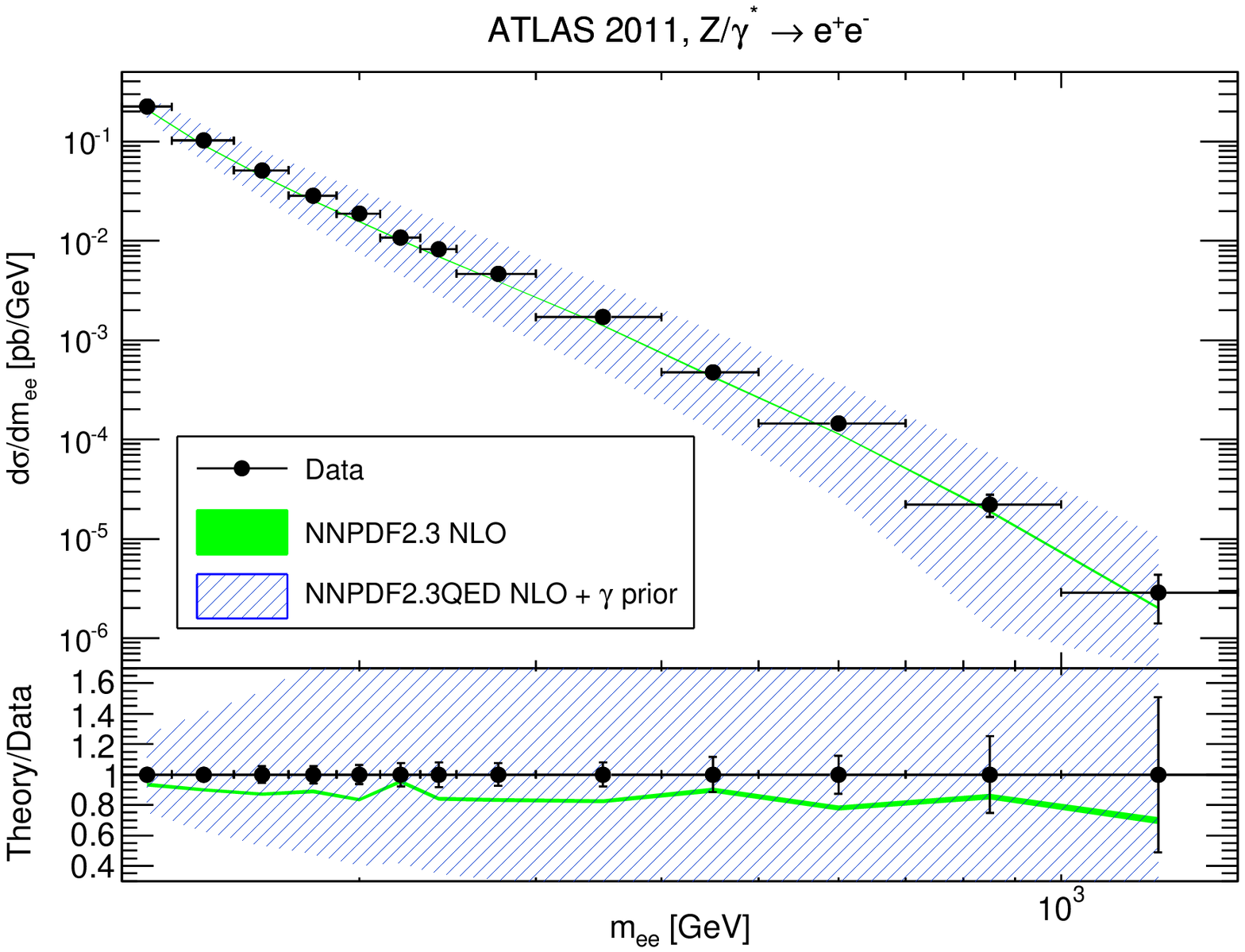}
\includegraphics[scale=0.37]{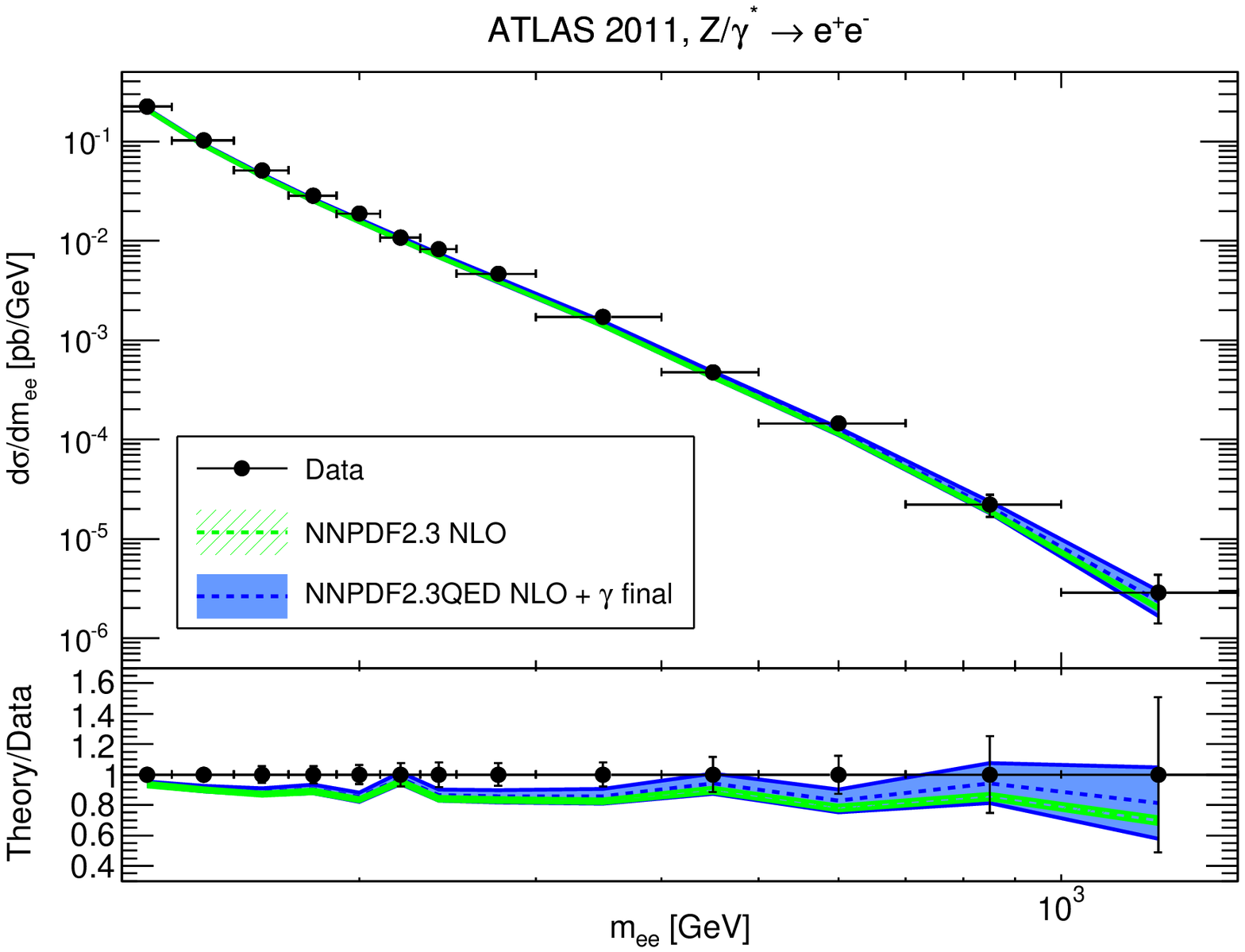}
\par\end{centering}
\caption{\label{fig:rwlhc1} \small  Same as Fig.~\ref{fig:rwlhc}, but
  for the ATLAS high-mass neutral-current data.}
\end{figure}

Theoretical predictions  for the datasets in Table~\ref{tab:expdata} have been
computed at NLO and NNLO in QCD using  {\tt DYNNLO}~\cite{Catani:2010en},
supplemented with Born-level and $O\lp \alpha \rp$ QED corrections
using {\tt HORACE}~\cite{Balossini:2009sa,CarloniCalame:2007cd}. 
Results from  {\tt DYNNLO} and  {\tt HORACE} have been combined additively,
avoiding double counting, in order to obtain
a consistent combined QCD+QED theory prediction. The additive
combination of QED and QCD corrections avoids introducing
$O(\alpha\alpha_s)$ terms, which are beyond the accuracy of our
calculation.
In the {\tt DYNNLO}
calculation, the renormalization and
factorization scale have been set to
the invariant mass of the dilepton pair in each bin. The {\tt HORACE}
default settings, with the renormalization and
factorization set to the mass of the gauge boson, have been
used for the ATLAS high-mass data, but we have also  checked that for
this data the
choice is immaterial, in that the LO results obtained using {\tt
  DYNNLO} and  {\tt HORACE} with the respective scale settings agree
with each other. 

For the LHCb low-mass data we have used a
modified version of  {\tt HORACE} in which the scale choice is the same as in 
{\tt DYNNLO}, since for these low scale data the choice of
renormalization and factorization scale does make a significant
difference. Note that the smallest mass values reached by these data
correspond to momentum fractions $x\sim 10^{-3}$ in the central
rapidity regions, for which, at the scale of the data,
 fixed order (unresummed) results are
expected to be adequate (see Ref.~\cite{Ball:2007ra}, in particular
Fig.~1). Indeed we shall see that our results are perturbatively stable
in that the photon PDF at NLO and NNLO is very similar for all $x$
(see Figs.\ref{fig:photonrw}-\ref{fig:photonrwnnlo} below).

The same selection and  kinematical cuts as in the corresponding experimental analysis has been adopted:
in particular, the same requirements concerning lepton-photon final state recombination
and the treatment of final state QED radiation have been implemented in the 
{\tt HORACE} computations.

It should be noticed that, whereas the LHCb and ATLAS high-mass data
are only being included now in the fit, the $W$ and $Z$ production
data were already included in the original NNPDF2.3 PDF determination
(where they turned out to have a moderate impact). Therefore, in
principle a modified version of NNPDF2.3 in which these data are
removed from the fit should have been used as a prior.  In practice,
however, this would make very little difference. We
have verified that the inclusion of QED evolution affects minimally
the prediction for this data, where differences are at the same level
of the Monte Carlo integration uncertainty, as can also be seen from
Fig.~\ref{fig:qedfig1}, recalling (see Fig.~\ref{fig:correlations})
that the main impact of this data is in the $x\sim0.01$ region. This
means that the contributions to this process in the
reweighting and in the original NNPDF2.3 fit in practice only differ because of
the inclusion of the photon contribution. Furthermore, we have
explicitly verified that if the ATLAS $W$ and $Z$ production data are
excluded from the fit, the photon is systematically modified by a
small but non-negligible amount
(less then half $\sigma$ at most) in the region
$x\sim 10^{-3}$ where these data are expected to carry information (see
Fig.~\ref{fig:correlations}), while all other PDFs are essentially unaffected. 

Whereas our computation is only accurate to leading order in QED, we
did include $O(\alpha)$ corrections to the electroweak gauge boson
production process through {\tt HORACE}, with the aim of avoiding 
unnaturally large NLO QED corrections. This
raises several issues which we now discuss in turn.
 \begin{figure}[t]
\begin{centering}
  \includegraphics[scale=0.37]{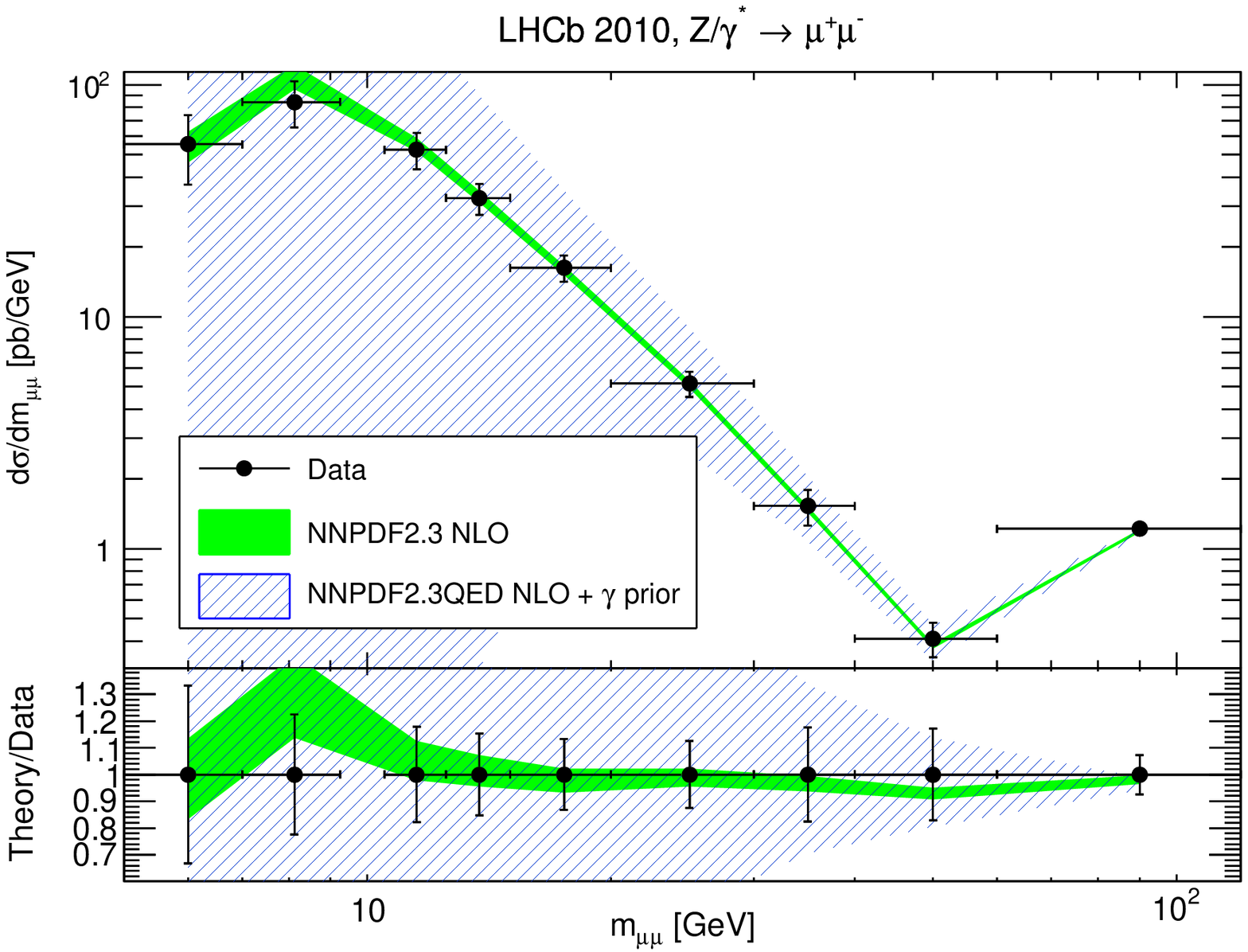}
  \includegraphics[scale=0.37]{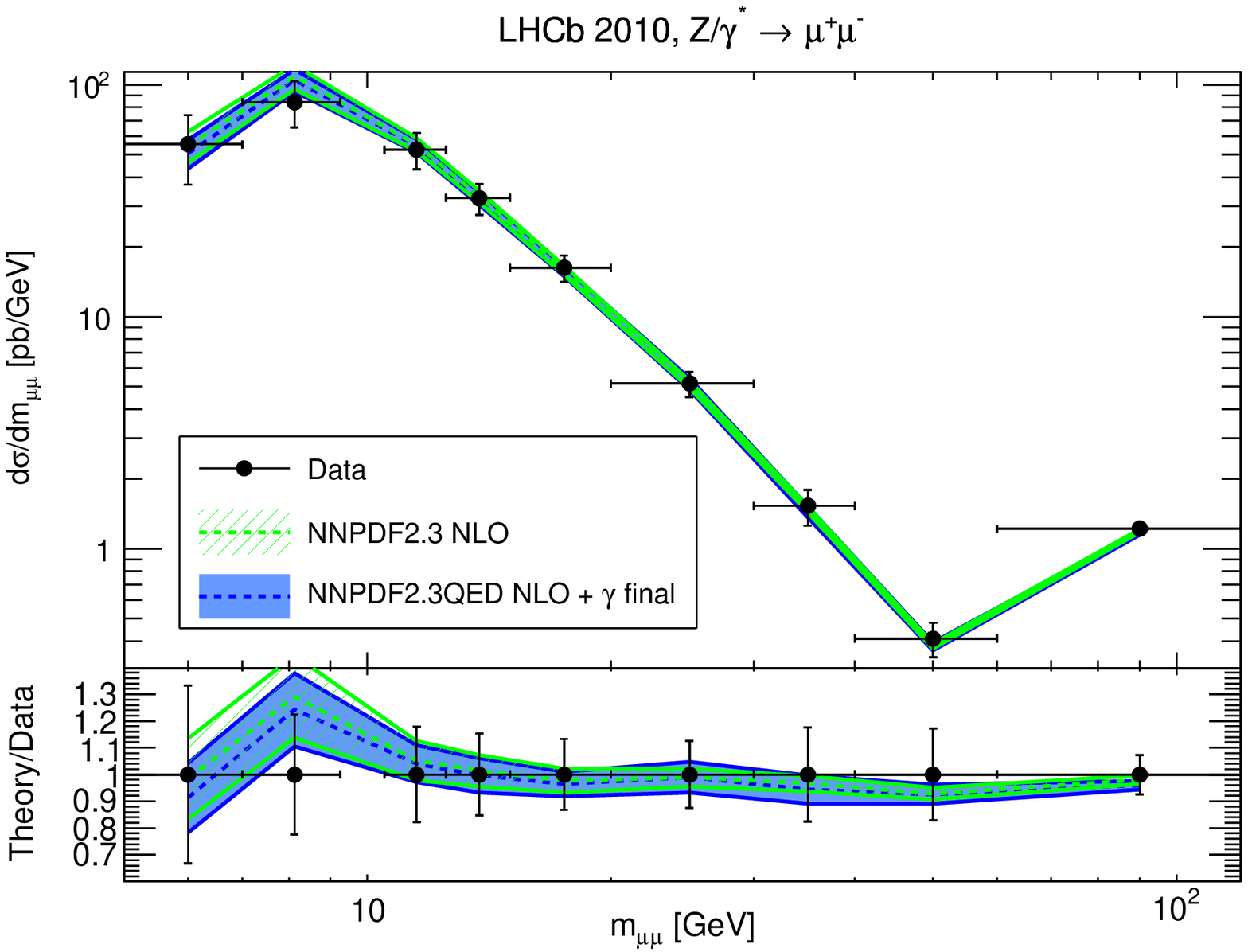}
\par\end{centering}
\caption{\label{fig:rwlhc2} \small Same as Fig.~\ref{fig:rwlhc}, but
  for the LHCb low-mass neutral current data.}
\end{figure}

As pointed out in Refs.~\cite{Diener:2005me,Dittmaier:2009cr}, usage
of the leading-order expressions in QED for the DIS coefficient
functions can be viewed as the choice of the DIS factorization scheme,
in which deep-inelastic coefficient functions are taken to coincide to
all orders with their leading-order expression, with higher order
corrections factorized into the PDFs. Therefore, use of the DIS scheme
for the QED corrections to the Drell-Yan process ensures that
predictions for Drell-Yan obtained with PDFs determined using DIS data
and  LO QED are actually accurate up to NLO, modulo any NLO corrections from 
QED evolution. Therefore, we have used the
DIS-scheme expressions for NLO corrections to Drell-Yan  as
implemented in  {\tt HORACE}. 
Of course, in practice, there will be NLO QED evolution effects, 
even though there is a certain overlap between the
kinematic region of the HERA DIS data and that of the LHC Drell-Yan
data, so we cannot claim NLO QED accuracy. However we expect this 
procedure to lead to greater stability of our results upon the 
inclusion of NLO QED corrections.

Radiative corrections related to final-state QED radiation have
already been subtracted from the ATLAS data, but not from the LHCb
data. Therefore, for ATLAS we have only included photon-induced
processes in the {\tt HORACE} runs, while for LHCb we have also included
explicit $O(\alpha)$ contributions from final-state QED
radiation. Electroweak corrections, which are not subtracted from any
of the
data and which are not included in our calculation,
could be potentially relevant in the high-mass
region~\cite{Dittmaier:2009cr}. However, in practice they are always
much smaller than the statistical uncertainty on the ATLAS data.

Finally, to NLO in QED the scheme used in defining electroweak
couplings should be specified. The {\tt DYNNLO} code uses the
so-called $G_\mu$ scheme for the electroweak couplings, while {\tt
  HORACE} also uses the $G_\mu$ scheme for charged-current production,
but the improved Born approximation for neutral-current production. We
have verified the differences in predictions between the two scheme
are negligible in comparison
to the statistical uncertainties of  the Monte Carlo integrations. 
 \begin{figure}[ht]
\begin{centering}
\includegraphics[scale=0.37]{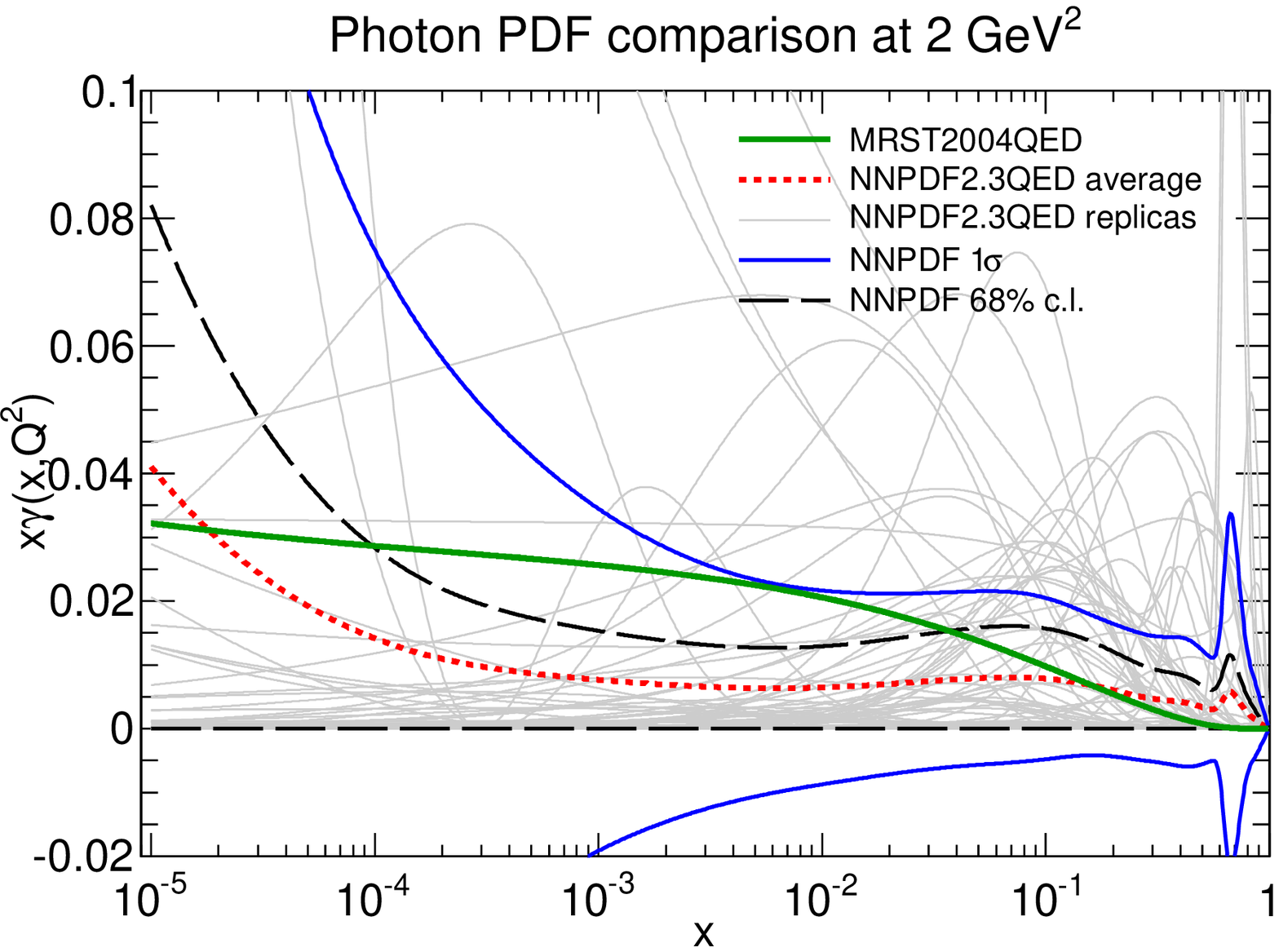}
\includegraphics[scale=0.37]{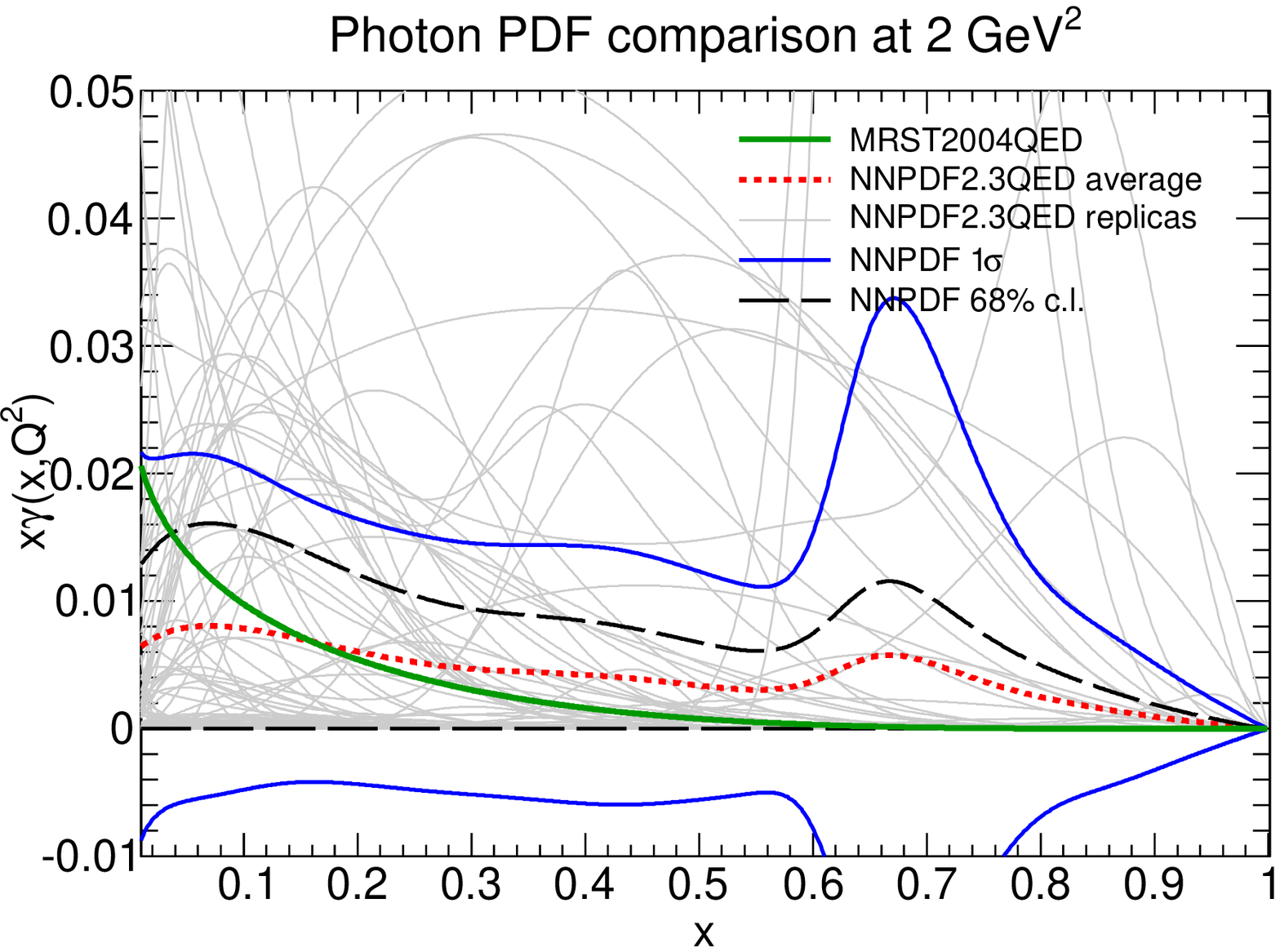}
\includegraphics[scale=0.37]{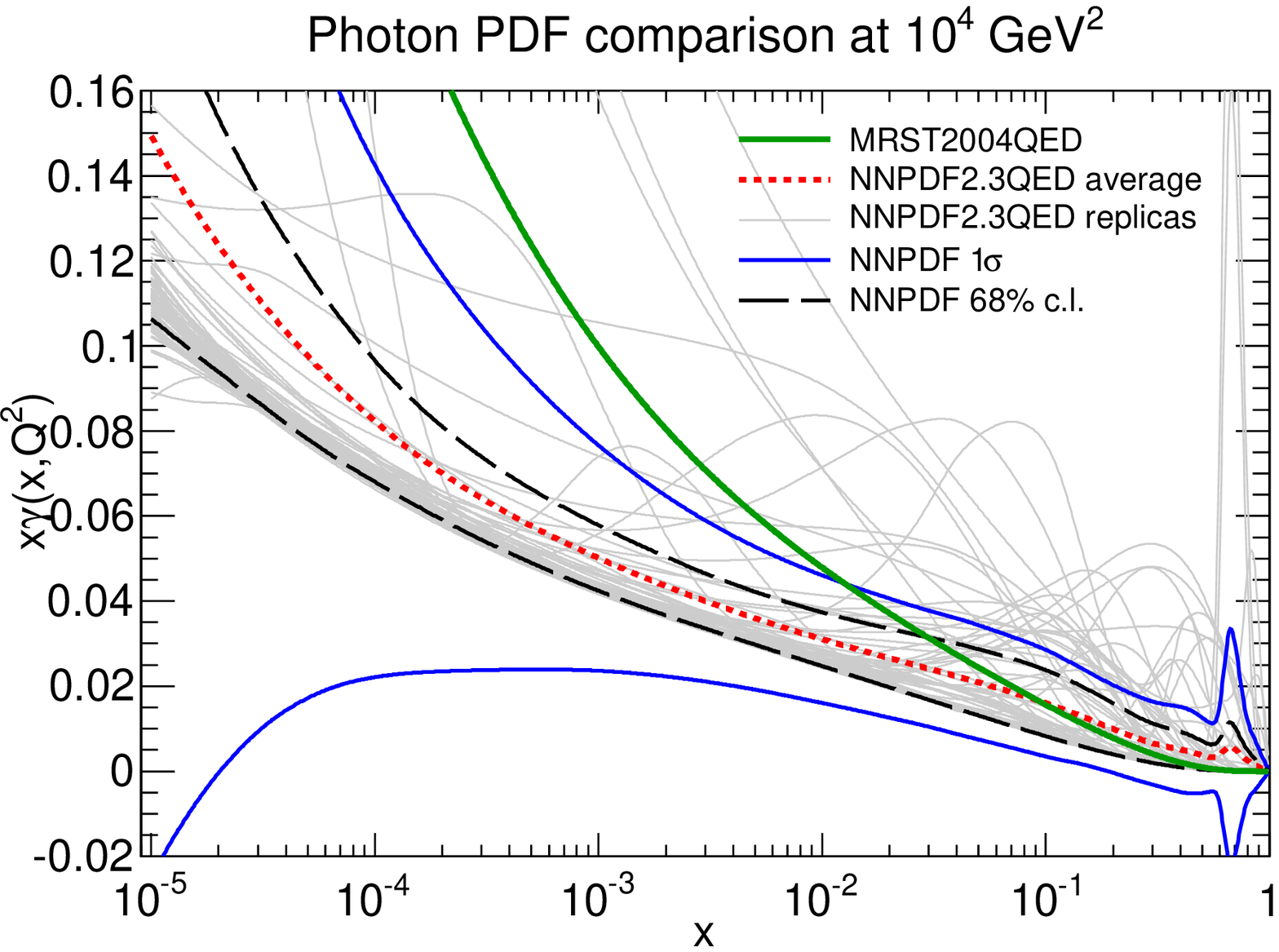}
\includegraphics[scale=0.37]{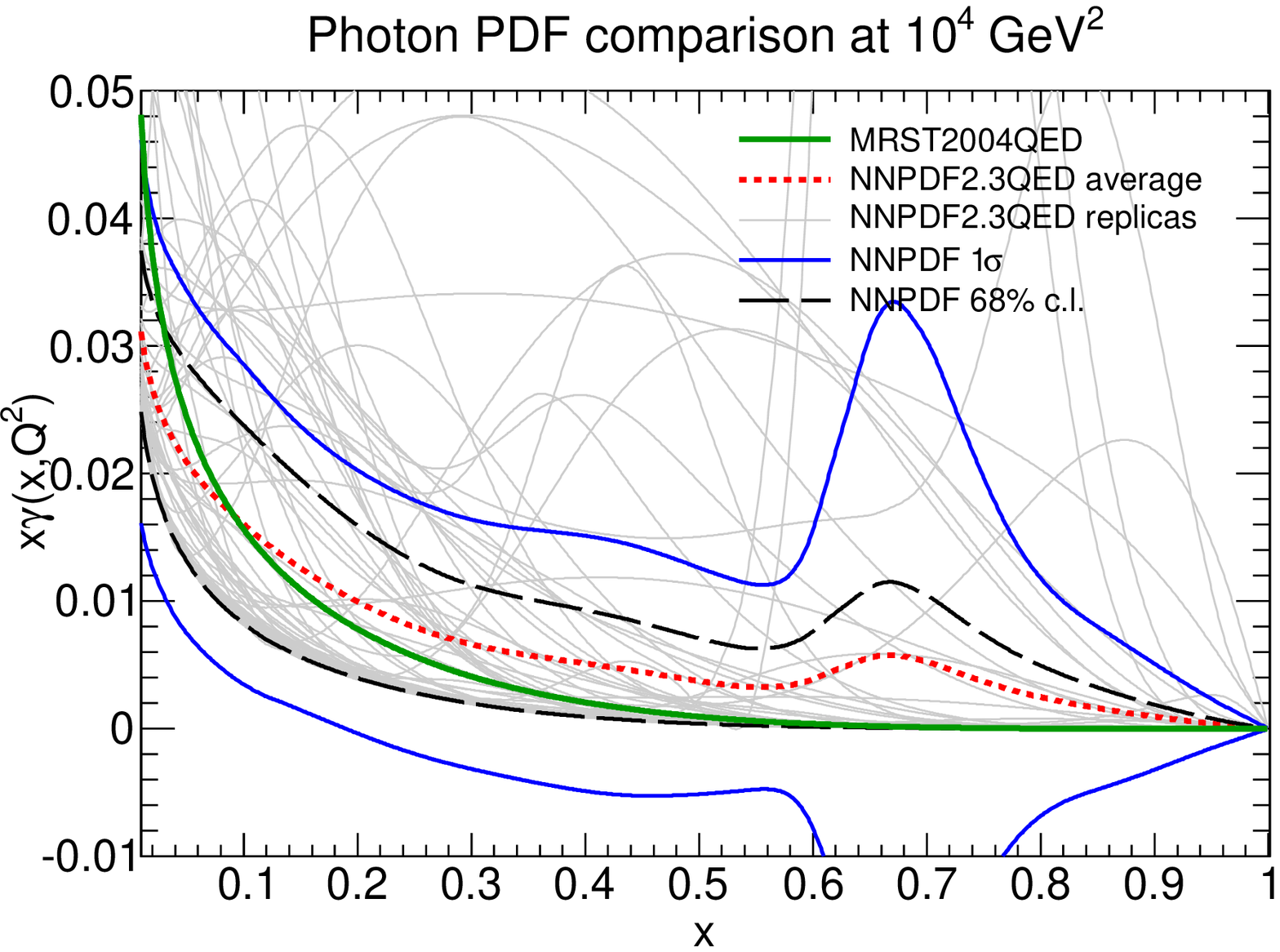}
\par\end{centering}
\caption{\label{fig:photonrw} \small  The NNPDF2.3QED  NLO photon PDF 
at $Q^2=2$ GeV$^2$ and $Q^{2}=10^{4}$ GeV$^{2}$ plotted vs. $x$ on a
log (left) or linear (right) scale. The 100 replicas are shown, along
with the mean, the one-$\sigma$, and the 68\% confidence level
ranges. The MRST2004QED photon PDF is also shown for comparison.
}
\end{figure}

 \begin{figure}[ht]
\begin{centering}
\includegraphics[scale=0.37]{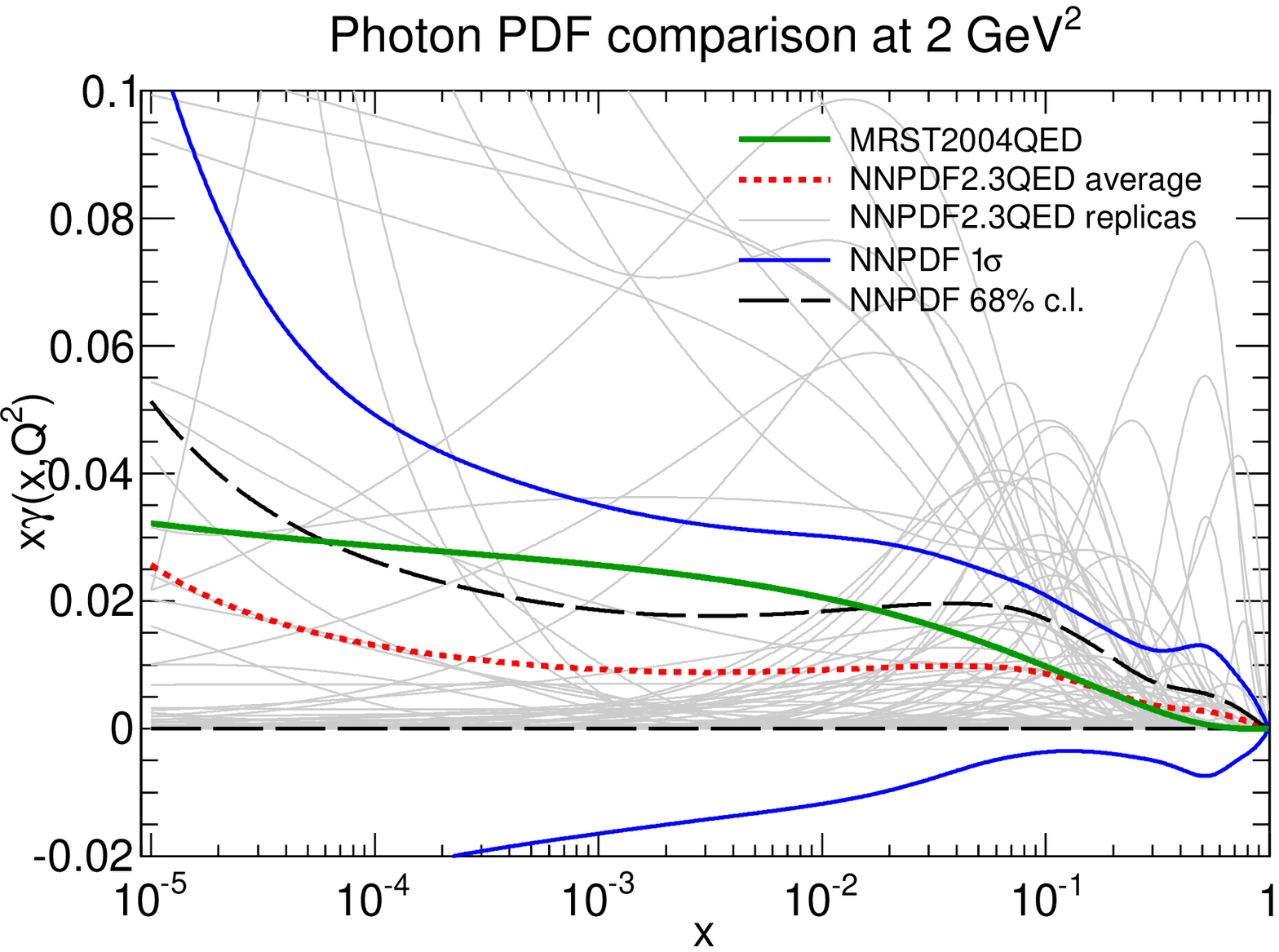}
\includegraphics[scale=0.37]{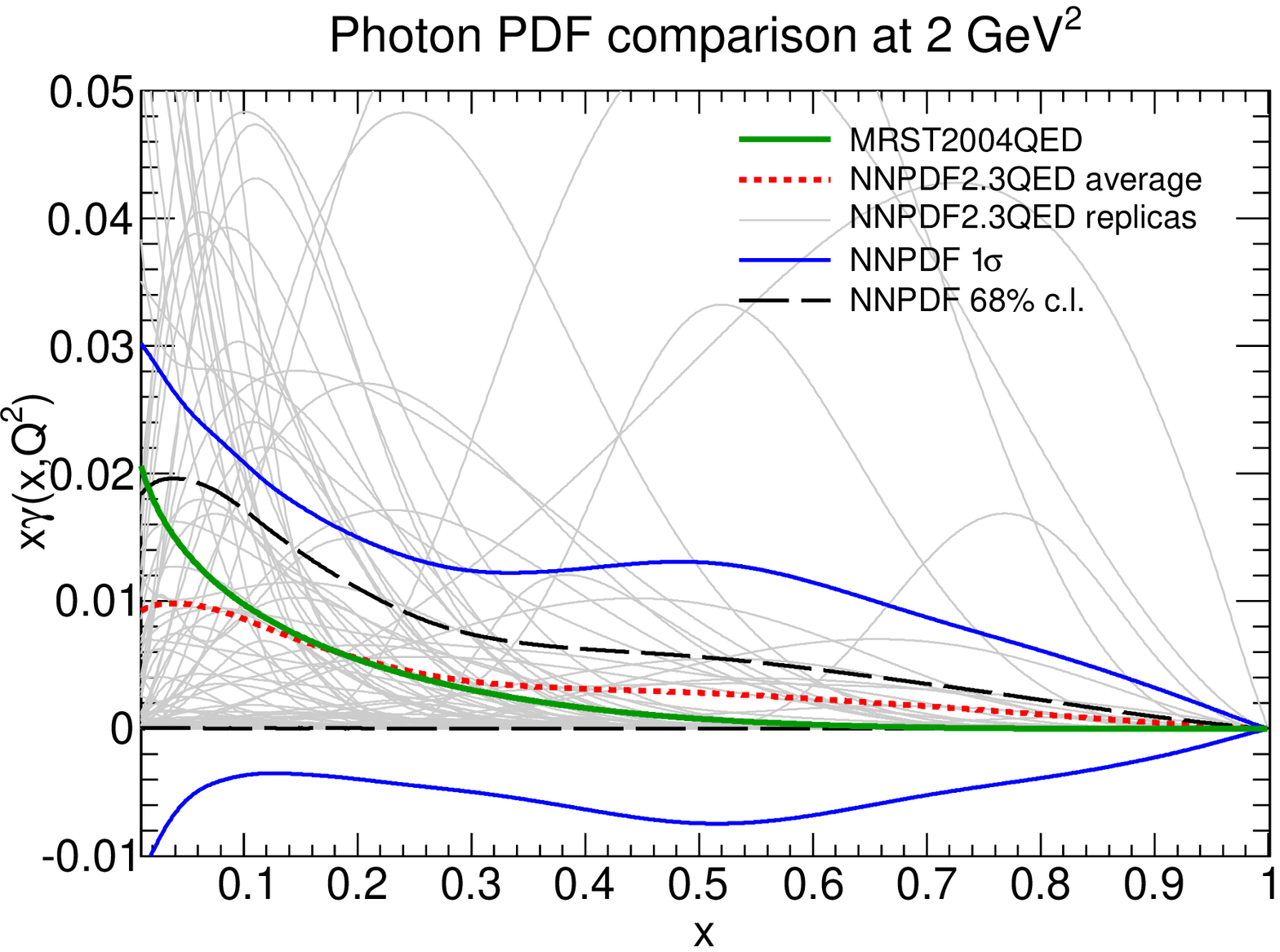}
\includegraphics[scale=0.37]{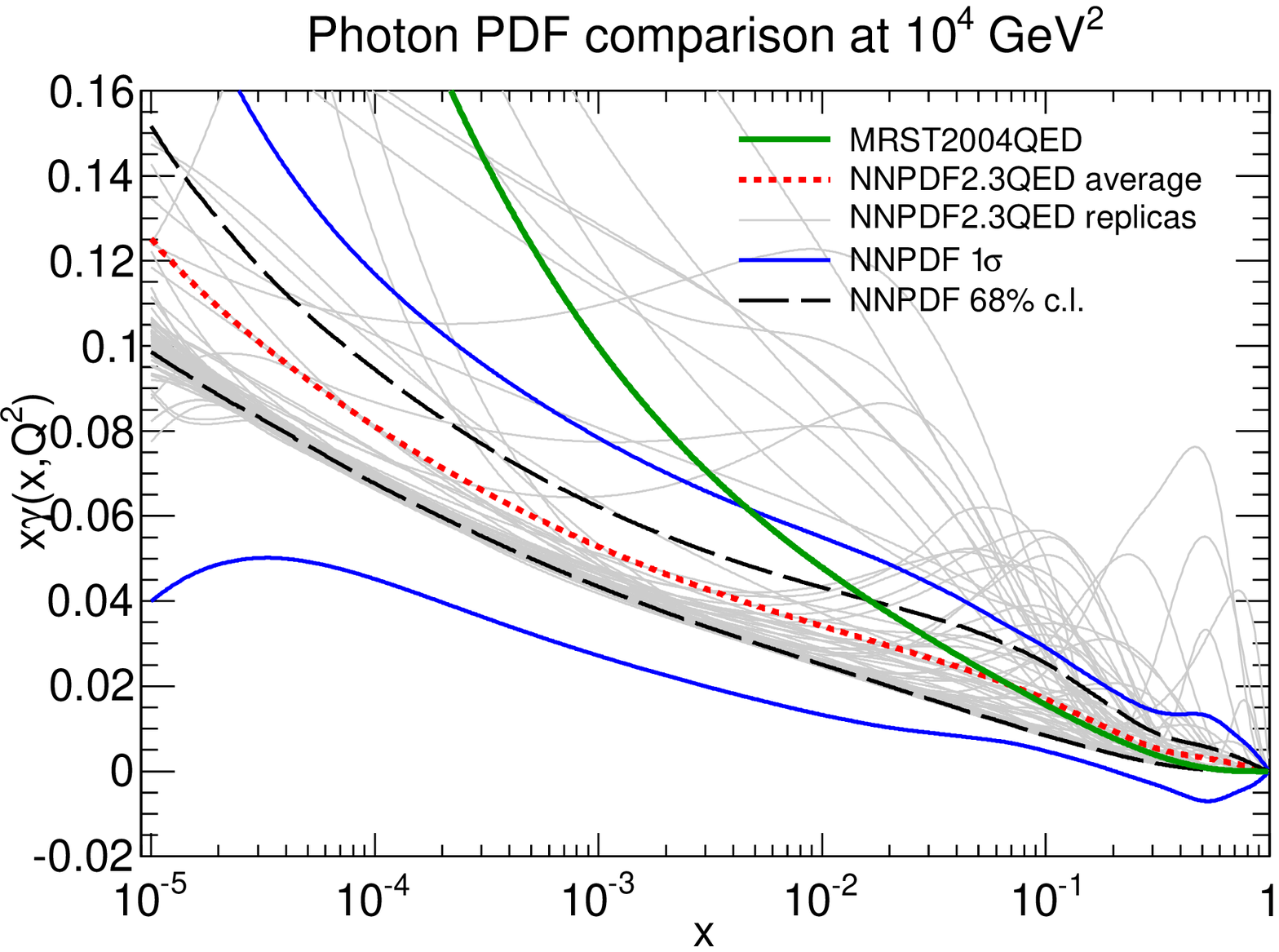}
\includegraphics[scale=0.37]{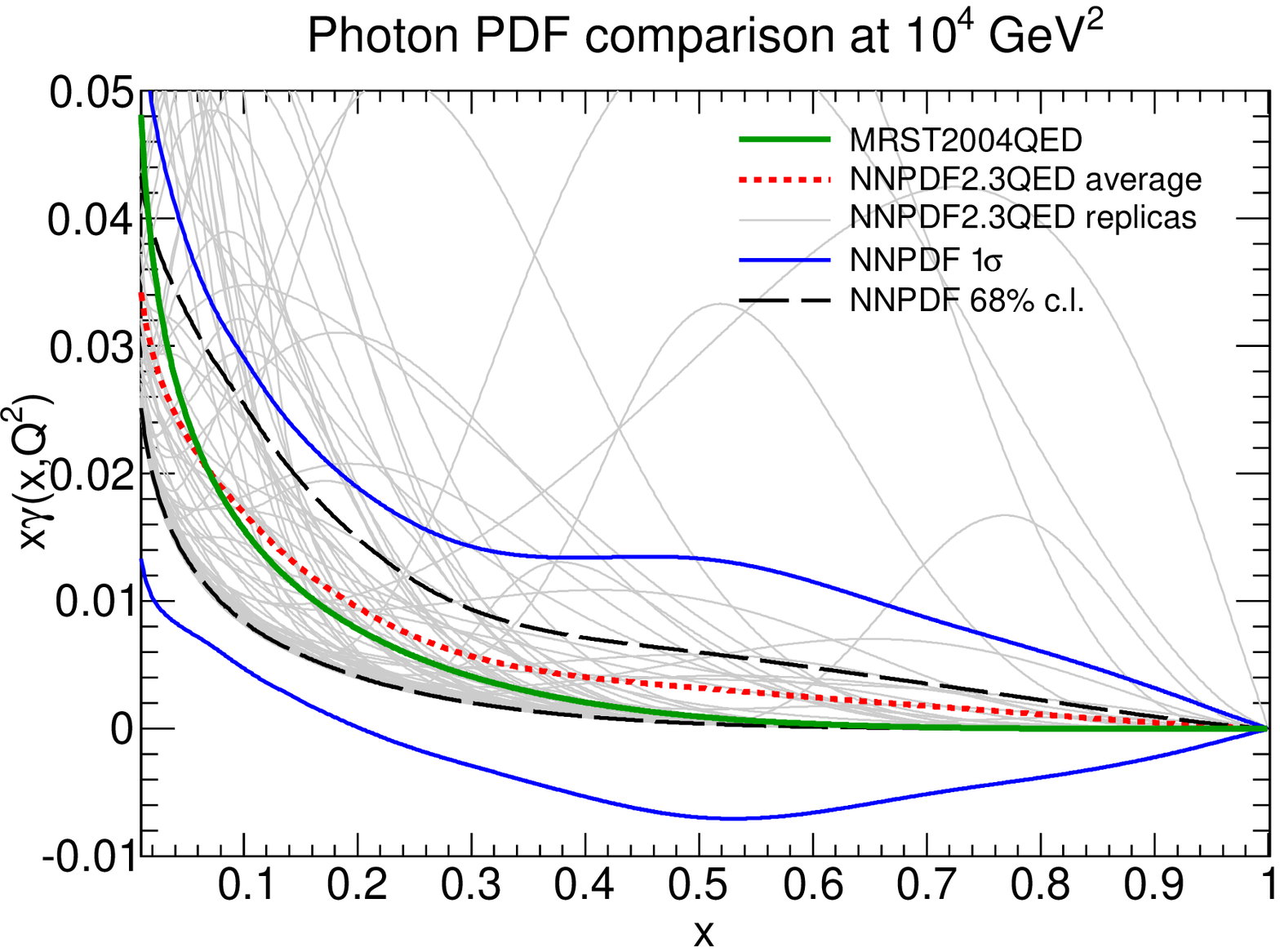}
\par\end{centering}
\caption{\label{fig:photonrwnnlo} \small Same as~\ref{fig:photonrw}
  for the  NNPDF2.3QED NNLO PDF set.
}
\end{figure}

\begin{table}
\centering
\begin{tabular}{c||c||c|c|c}
\hline
\multicolumn{5}{c}{NLO} \\
\hline
&  LHCtot  & ATLAS $W,Z$  & ATLAS high mass DY  & LHCb low-mass DY \\ 
\hline
\hline
$\chi^2_{\rm in}$  & 2.02 & 1.20 & 3.78 & 2.20 \\
$\chi^2_{\rm rw}$  &  1.00 &  1.15 & 1.01 & 0.29\\
\hline
$N_{\rm eff}$  & 287 & 364 & 326 & 267 \\
$\la \alpha \ra$  & 1.41 & 1.24 & 1.53 & 0.89\\
\hline
\end{tabular}
\\ \vspace{0.5cm}
\begin{tabular}{c||c||c|c|c}
\hline
\multicolumn{5}{c}{NNLO} \\
\hline
&  LHCtot  & ATLAS $W,Z$  & ATLAS high mass DY  & LHCb low-mass DY \\ 
\hline
\hline
$\chi^2_{\rm in}$  &  2.01 & 1.37  &  3.44 & 2.06 \\
$\chi^2_{\rm rw}$  &  1.08  &  1.21 & 1.00 & 0.66 \\
\hline
$N_{\rm eff}$  & 197 & 297 & 330 &  363 \\
$\la \alpha \ra$  & 1.48 & 1.33 & 1.52 & 1.20 \\
\hline
\end{tabular}
\caption{ \small Reweighting parameters in the construction of the
  final NNPDF2.3 sets. All $\chi^2$ values are defined as in Tab.~\ref{tab:chi2}.\label{tab:rwnlo}}
\end{table}

\subsection{The NNPDF2.3QED set}
\label{sec:nnpdf23qed}

The NNPDF2.3QED PDF set is obtained by performing a reweighting of the
prior $N_{\rm rep}=500$ replica set with the data of
Table~\ref{tab:expdata}. The procedure is performed at NLO and NNLO
in QCD, with three different values of $\alpha_s$ in each case. The
theoretical prediction used for reweighting is computed as discussed
in the previous section, and the $\chi^2$ used for reweighting is
then determined from its comparison to the data, using the 
fully correlated systematics for the two
ATLAS experiments, for which the covariance matrix is available, but
adding statistical and systematic errors in quadrature for LHCb, for
which information on correlations is not available.
The ensuing weighted set of replicas is
then unweighted~\cite{Ball:2011gg} to obtain a standard set of
$N_{\rm rep}=100$ replicas.

The parameters of the reweighting are collected in
Table~\ref{tab:rwnlo}: we show the $\chi^2$  (divided by the number of
data points) for the data of
Table~\ref{tab:expdata} before and after reweighting, the effective
number of replicas after reweighting, and the mean value of $\alpha$,
the parameter which measures the consistency of the data which are
used for reweighting with those included in the prior set, by
providing the factor by which the uncertainty on the new data must be
rescaled in order of the two sets to be consistent (so $\alpha\sim 1$
means consistent data). Values are given for reweighting performed
using each individual dataset, and the three datasets combined. All
$\chi^2$ values are computed using the experimental definition of the
covariance matrix as in Table~\ref{tab:chi2}; 
the same form of the covariance matrix has also
been used for reweighting for simplicity, as this choice is immaterial
as discussed above.

\begin{figure}[ht]
\begin{centering}
\includegraphics[scale=0.74]{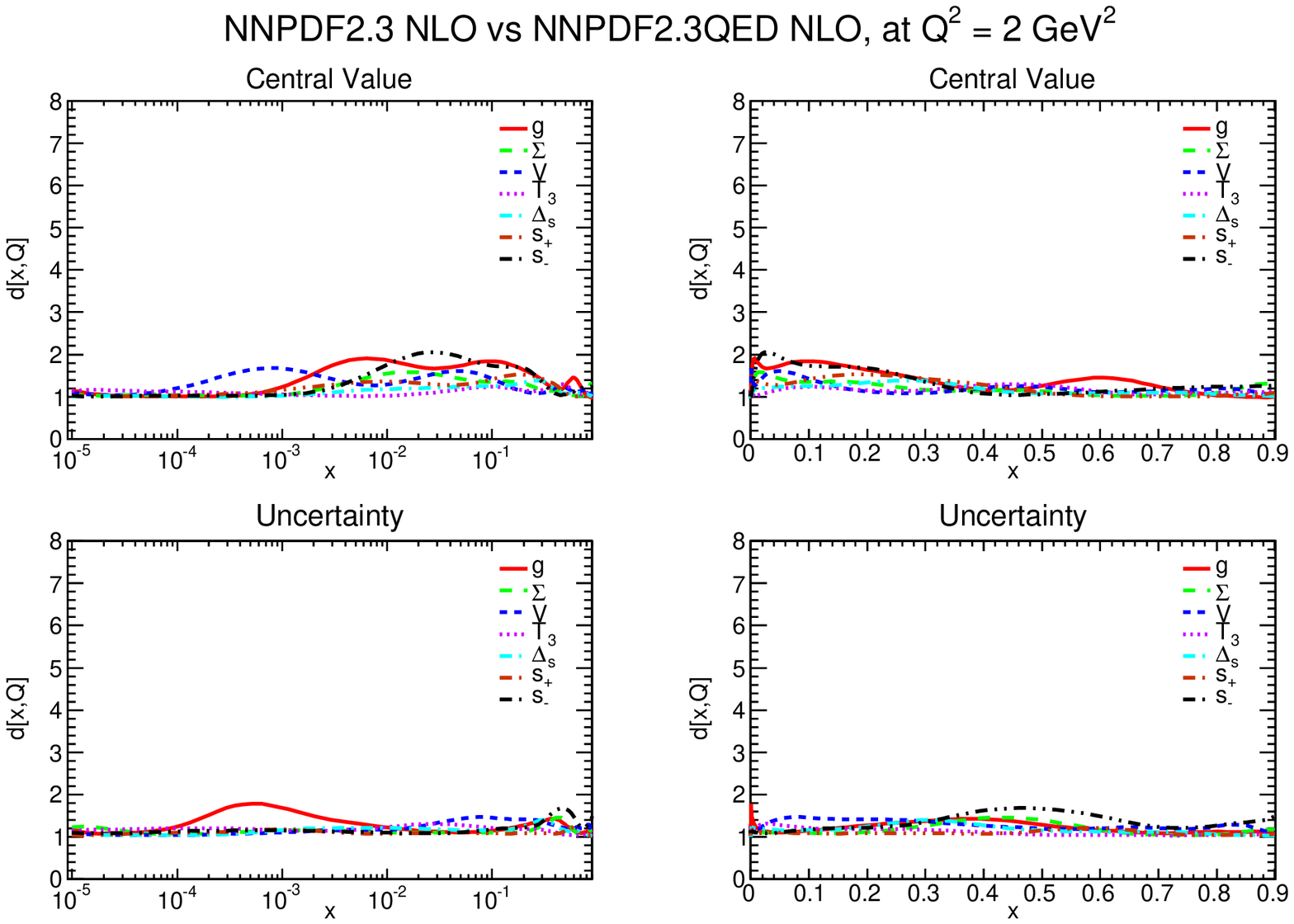}
\par\end{centering}
\caption{\label{fig:distances2} 
Distances between  PDFs in the NNPDF2.3  and the NNPDF2.3QED NLO sets,
at the input scale of $Q_0^2$=2~GeV$^2$. Distances between central values (top) and uncertainties
    (bottom) are shown, on a logarithmic (left) and linear (right)
    scale in $x$.
}
\end{figure}

\begin{figure}[ht]
\begin{centering}
\includegraphics[scale=0.74]{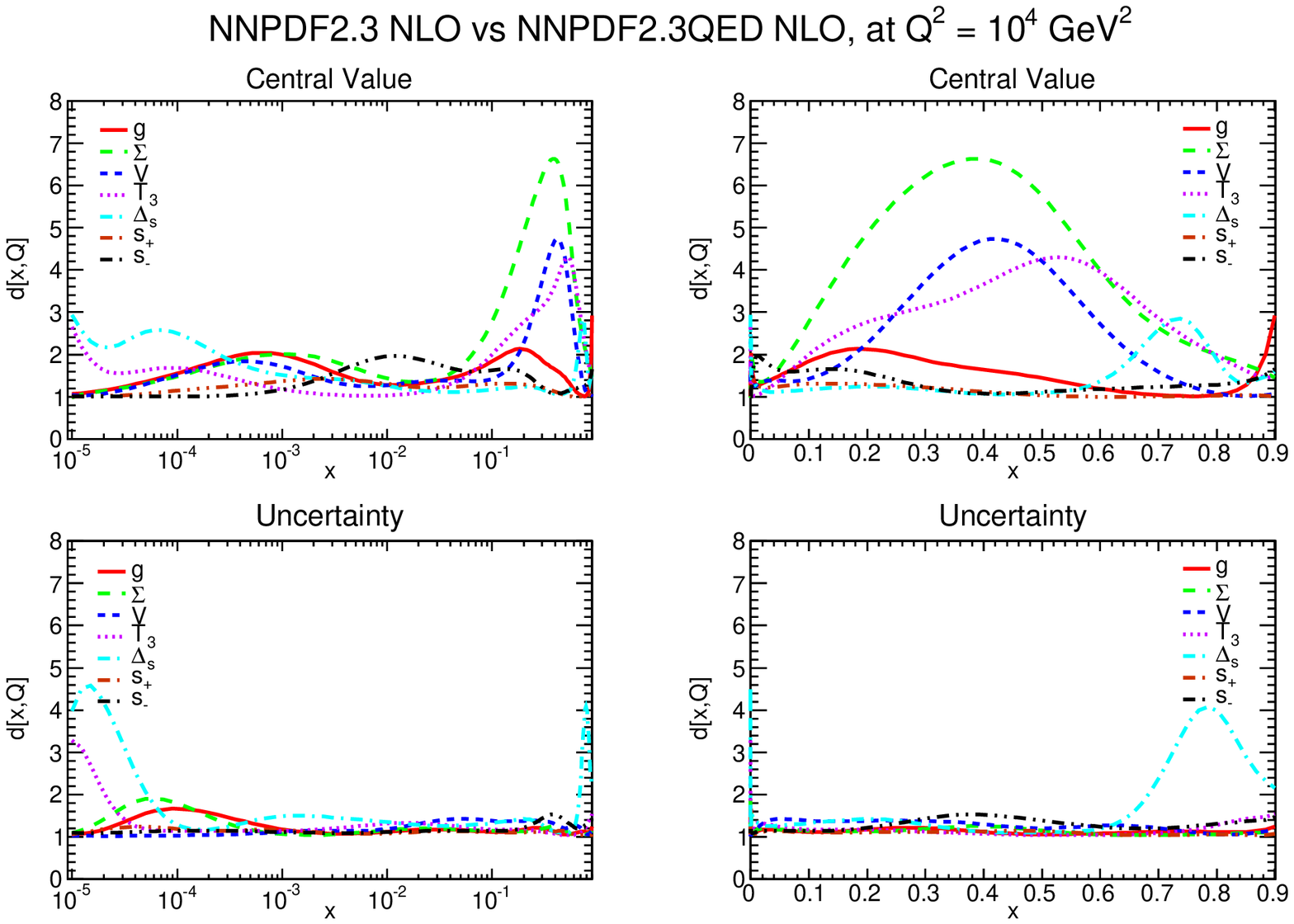}
\par\end{centering}
\caption{\label{fig:distances2bis} Same as Fig.~\ref{fig:distances2} but now
computed at
$Q^2=10^{4}$~GeV$^2$.
}
\end{figure}

In all cases the final effective number replicas turns out to be
$N_{\rm eff}>100$, thereby guaranteeing the accuracy of the
final unweighted set. All sets show good compatibility with the
prior datasets. 
The final $\chi^2$
values show that the reweighted set provides an essentially perfect fit
to the data; the low values for LHCb are a consequence of the fact
that for this experiment the correlated systematics is not available
so statistical and systematic errors are added in quadrature. 
Before reweighting the $\chi^2$ of
individual replicas shows wide fluctuations: indeed, its average and variance
over the starting replica sample are given by
$\langle \chi^2\rangle=25.6 \pm 164.4$. After reweighting the value
becomes $\langle \chi^2\rangle= 1.117\pm 0.098$, thus showing that the
$\chi^2$ of indvidual replicas has become on average almost as good as
that of the central reweighted prediction.

\begin{table}[ht]
\begin{center}
{\small
\centering
\begin{tabular}{|c|c|c|c|}
\hline
     & NNPDF2.3QED NLO & NNPDF2.3QED NNLO & MRST2004QED \\ 
\hline
\hline $\gamma$; $Q^2=2$~GeV$^2$  &
 $\lp 0.42 \pm 0.42\rp$\% & $\lp 0.34 \pm 0.34 \rp$\%  & $0.30$\% \\  [2ex] 
\hline $\gamma$; $Q^2=10^4$~GeV$^2$  & $\lp 0.68 \pm 0.42\rp$\% &
$\lp 0.61 \pm 0.34\rp$\% & $0.52$\% \\   [2ex] 
\hline\hline
 total; $Q^2=2$~GeV$^2$
  & $\lp 100.43 \pm 0.44\rp$\%  &  $\lp 100.32 \pm 0.34 \rp$\% & $99.95$\%\\   [2ex] 
\hline  total; $Q^2=10^4$~GeV$^2$ & $\lp 100.38 \pm 0.43\rp $\% & 
$\lp  100.29 \pm 0.36 \rp \%$ & $99.92$\% \\   [2ex] 
\hline
\end{tabular}}
\caption{\small \label{tab:msr} Momentum fractions (in percentage)
  carried by the photon PDF (upper two rows) and by the sum of all
  partons in the proton (lower two rows) in the NNPDF2.3QED 
  NLO, NNLO and  MRST2004QED PDF sets at two different scales}
\end{center}
\end{table}

\begin{figure}[ht]
\begin{centering}
\includegraphics[scale=0.37]{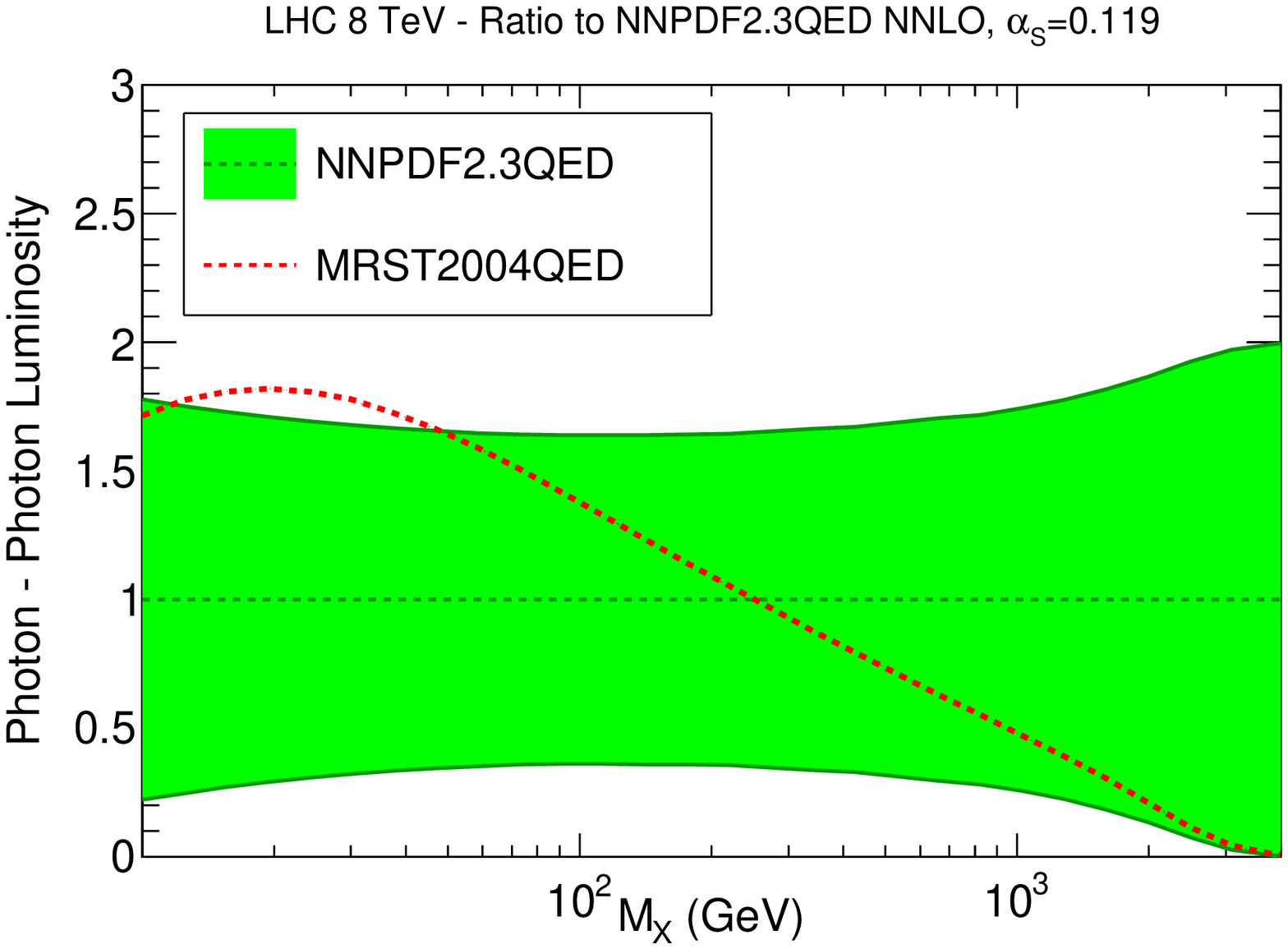}
\includegraphics[scale=0.37]{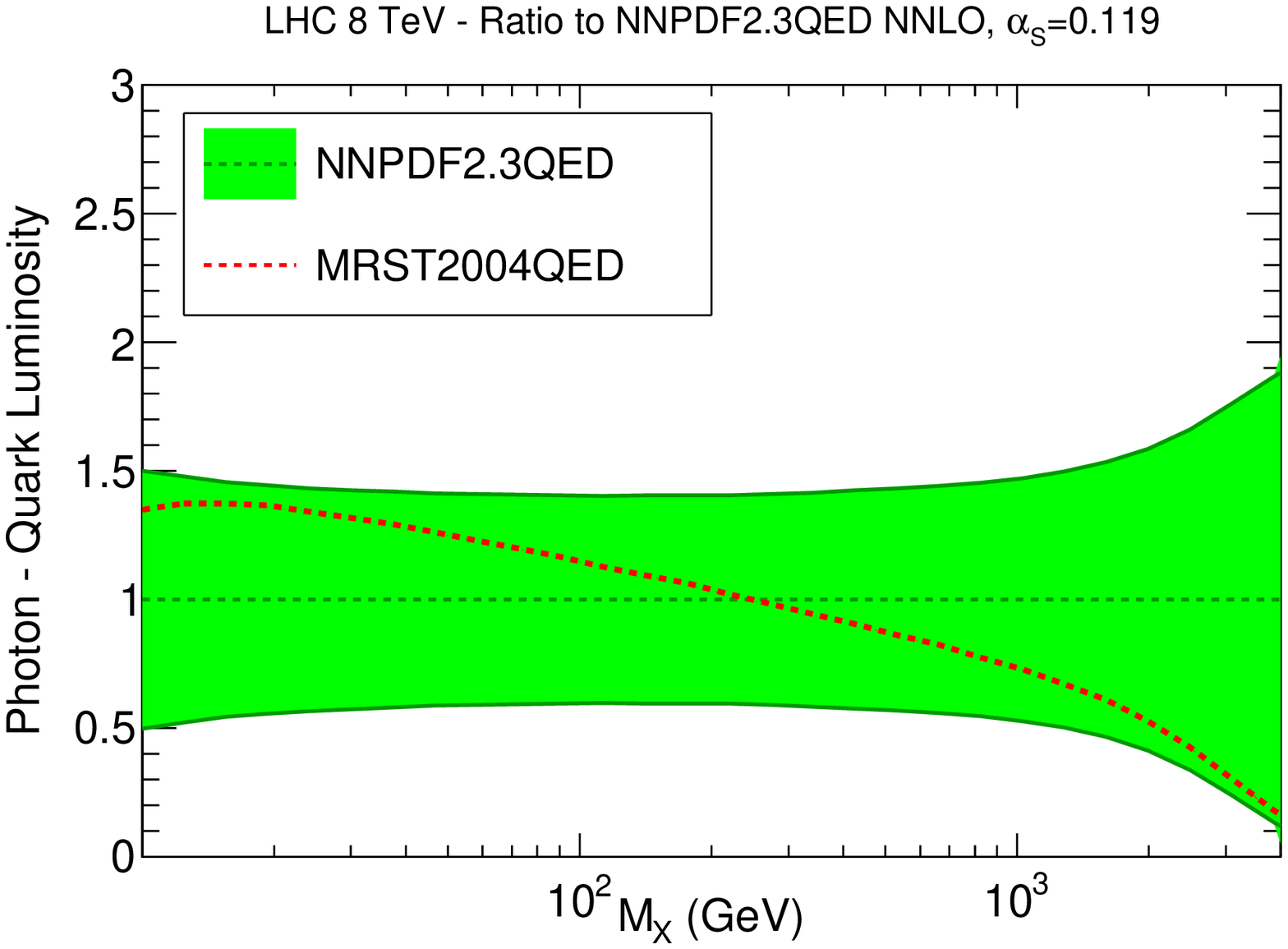}
\par\end{centering}
\caption{\label{fig:lumimrst} \small The photon-photon $\gamma \gamma$ (left) and 
photon-quark $\gamma q$ (right) parton luminosities at the LHC~8~TeV computed using
  MRST2004QED PDFs, shown as a ratio to the NNPDF2.3QED result. The
  68\% confidence level on the latter is also shown.}
\end{figure}

A first assessment of the impact of the photon-induced corrections and
their effect on  the  photon PDF can be obtained by comparing the data
to the theoretical prediction obtained using pure QCD theory and the
default NNPDF2.3 set, QCD+QED with the prior photon PDF, and QED+QCD
with the  final NNPDF2.3QED set. The comparison is shown in
Figs.~\ref{fig:rwlhc}-\ref{fig:rwlhc2} for the NLO sets (the NNLO
results are very similar): in the left plots we show the QED+QCD
prediction obtained using the prior PDF set, and in the right plots
the prediction obtained using the final reweighted sets, compared in
both cases to the pure QCD prediction obtained using {\tt DYNNLO} and the
NNPDF2.3 set. At the $W,Z$ peak, the impact of QED
corrections is quite small, though, in the case of neutral current 
production, to
which the photon-photon process contributes at Born level, when the
prior photon PDF is used one can see
the widening of the uncertainty band due to the large uncertainty of
the photon PDF of Fig.~\ref{fig:xpht-nlo}. At low or  high mass, as
one moves away from 
the peak, the
large uncertainty on the prior photon PDF 
induces an increasingly large uncertainty  on the theoretical
prediction, 
substantially larger
than the data uncertainty. This means that these data do constrain the
photon PDF and  indeed after reweighting the uncertainty is
substantially reduced. 

 \begin{figure}[ht]
\begin{centering}
\includegraphics[scale=0.37]{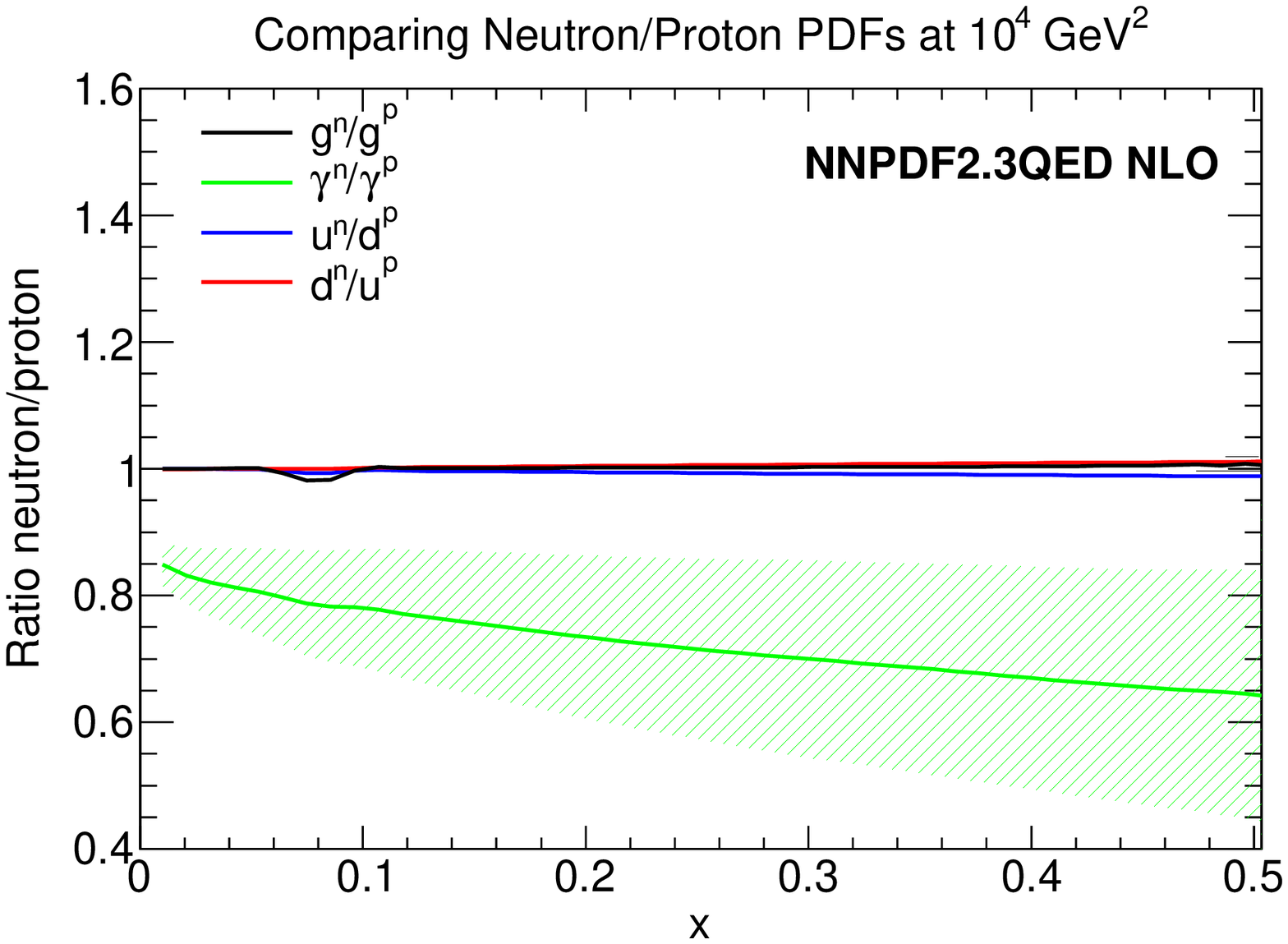}
\includegraphics[scale=0.37]{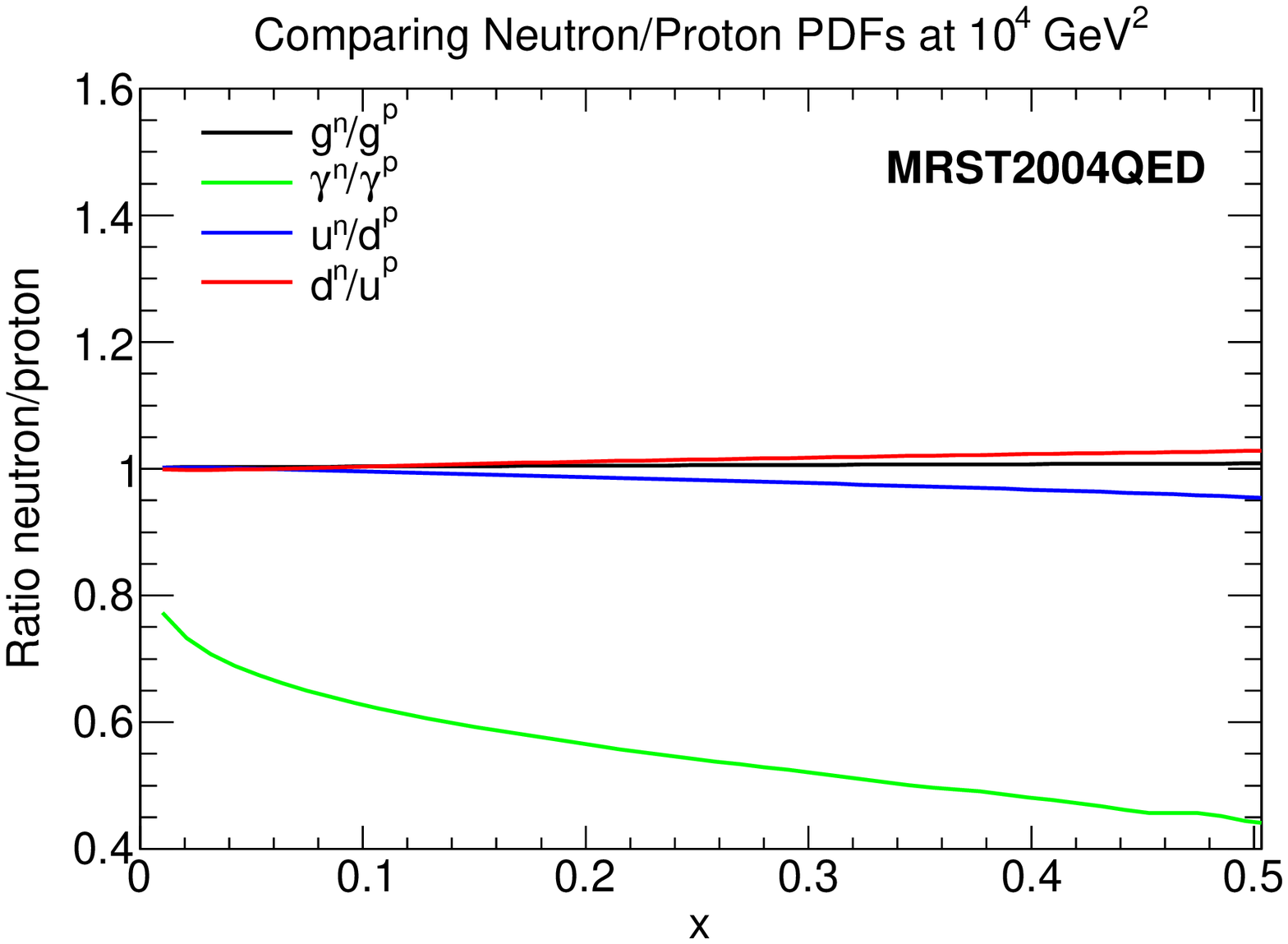}
\par\end{centering}
\caption{\small \label{fig:isospinLHC} The ratio of the neutron to the
  proton PDFs in the NNPDF2.3QED NLO set at $Q^2=10^4$ GeV$^2$ (left)
  and MRST2004QED set (right).  Results for the photon, gluon, up and
  down quark are shown.  Error bands correspond to one-$\sigma$
  uncertainties.
}
\end{figure}

The final
NNPDF2.3QED photon PDF obtained in the NLO and NNLO fits is
respectively shown at $Q^2_0=2$ GeV$^2$ in Fig.~\ref{fig:photonrw} 
and Fig.~\ref{fig:photonrwnnlo}. 
We display individual replicas,
the central (mean) photon, and the one-$\sigma$ and 68\% confidence
level ranges, as well as the MRST2004QED result. 
The improvement in accuracy in comparison to the prior
PDF of Fig.~\ref{fig:xpht-nlo} is apparent, especially at small and at
large $x$.
Note also that, especially at large $x$, where the experimental information
remains scarce (recall Fig.~\ref{fig:correlations}), the positivity
bound still plays an important role in constraining the photon PDF. 
Indeed, at the starting scale $Q_0$  the lower edge of the uncertainty
band (determined as discussed in Sect.~\ref{sec:disfit}) is again very close to
the positivity constraint, and consequently, even after having used
the LHC data, the probability distribution of the
photon PDF is significantly asymmetric, departing substantially
from Gaussian.  This
should be kept in mind in phenomenological applications, in particular
when computing uncertainties.

In Table~\ref{tab:msr} we show the momentum fraction carried by the photon PDF
in NNPDF2.3QED at NLO and NNLO, both at a low and high scale: it is about half of a
percent, compatible with zero within uncertainties, and mildly
dependent on scale. The MRST2004QED values, also shown, are consistent
within uncertainties. Note that the standard deviation would be almost
twice the 68\% confidence level interval given in the table.
We also give the total momentum, which deviates from unity because of
the slightly inconsistent procedure that we have followed in
constructing the prior set, by combining the photon from a fit to DIS
data with the other PDFs from the global NNPDF2.3 fit
as discussed in Sect.~\ref{sec:combination} above. We also see that
the total momentum fraction is not quite scale independent, because of
the approximation introduced when neglecting terms of
$O(\alpha\alpha_s)$ in the solution of the combined QED+QCD evolution
equations. Both effects are well below the 1\% level. 

All other PDFs at the initial scale $Q_0$ are left unaffected by the
reweighting. 
This can be seen by computing the distances between PDFs
in the starting NNPDF2.3 set and in the final NNPDF2.3QED set; they
are displayed in Fig.~\ref{fig:distances2}, at 
the scale $Q^2_0=2$~GeV$^2$ at which PDFs are parametrized: it is
apparent that the distances are compatible with statistically
equivalent PDFs. 
It is interesting to repeat the same comparison at 
$Q^2=10^4$~GeV$^2$ (Fig.~\ref{fig:distances2bis}): in this case,
statistically significant differences start appearing, as a
consequence of the fact that the statistically equivalent starting
PDFs in the two sets are then
evolved respectively with and without QED corrections. However,
the differences are below the one-$\sigma$ level (and concentrated
at large $x$), consistent with the
conclusion that the new data are compatible with those used for the
determination of the NNPDF2.3 PDF set.

In Figs.~\ref{fig:photonrw}-\ref{fig:photonrwnnlo} the photon PDF from
the MRST2004QED set is also shown for comparison. The MRST2004QED photon
PDF is based on a model; an alternative (not publicly available) 
version of it, in which 
consitituent rather than current quark masses are used as model
parameters, has been  used~\cite{Aad:2011dm} to estimate the
model uncertainty, though consitituent masses are considered to be
less appropriate by the authors of Ref.~\cite{Martin:2004dh}.
The MRST2004QED
photon  turns
out to be in good agreement with the central NNPDF2.3QED prediction at
medium and large $x$, but at small $x\lsim0.03$ it grows more quickly,
and for $x\le 10^{-2}$ it is larger and well outside the NNPDF2.3QED
uncertainty band.

It is  also interesting to compare the PDFs
from the NNPDF2.3QED and MRST2004QED sets at the level of the
parton luminosities which enter the computation of hadronic
processes. 
This comparison is shown in Fig.~\ref{fig:lumimrst}. The two
luminosities are in good agreement for invariant masses of
the final state $M_X\sim100$~GeV,  but the
agreement is less good for higher or lower final-state masses, with
the MRST2004QED rather smaller at high mass and larger at  low mass,
where, for  $M_X\sim20$~GeV it is outside the NNPDF2.3QED
uncertainty band.
As we will see in the next section, these differences translate into
differences in the predictions for electroweak processes at the LHC.

So far, we have shown results for the PDFs of the proton. Note, however that,
as discussed in Sect.~\ref{sec:disfit},  even though we assume that
isospin holds at the scale at which PDFs are parametrized,
QED corrections to
perturbative evolution 
introduce a violation of the isospin symmetry at all other scales.
Therefore, we provide independent NNPDF2.3QED  PDF sets for proton and
neutron. The size of isospin violation is expected to be comparable to
the QED corrections themselves, so very small for quark and gluon
distributions but more significant for the photon PDF. The expectation
is borne out by  Fig.~\ref{fig:isospinLHC} 
where  the ratio of the neutron to the proton
PDF at $Q^2=10^4$ GeV$^2$  in 
NNPDF2.3QED NLO is compared to that in MRST2004QED set. The comparison shows
that while the amount of isospin violation in the MRST2004QED photon PDF, 
which had a built-in model of
non-perturbative isospin violation, is somewhat larger than our own, especially
at large $x$, the difference is within  the PDF uncertainty,
as anticipated in Sect.~\ref{sec:disqed}. The amount of isospin
violation on quark and gluon PDFs is extremely  small, on the scale of
PDF uncertainties, both for MRST2004QED and NNPDF2.3QED. The same
conclusions hold if the NNLO set is used.

\section{Implications for HERA and LHC phenomenology}

\label{sec:searches}

As first examples of the use of the NNPDF2.3QED PDF set, we consider now
several processes which are sensitive to photon-initiated corrections.
In particular, we will discuss direct photon production at HERA, 
backgrounds for searches for new massive electroweak gauge bosons and $W$
pair production  at small $p_T$ and large
invariant mass. 

\subsection{Direct photon production at HERA}
\label{sec:directphotons}

 \begin{figure}[t]
\begin{centering}
\includegraphics[scale=0.37]{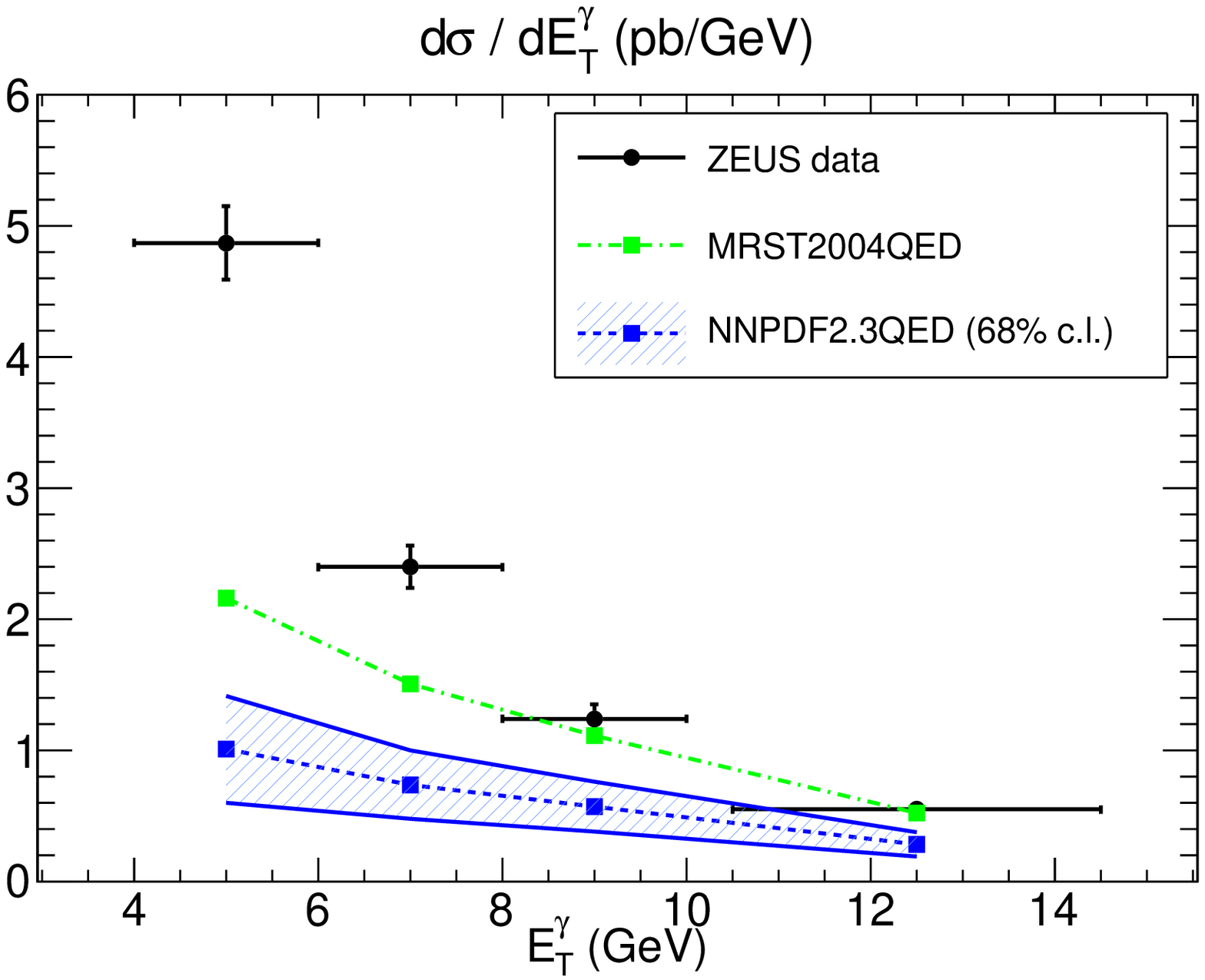}
\includegraphics[scale=0.37]{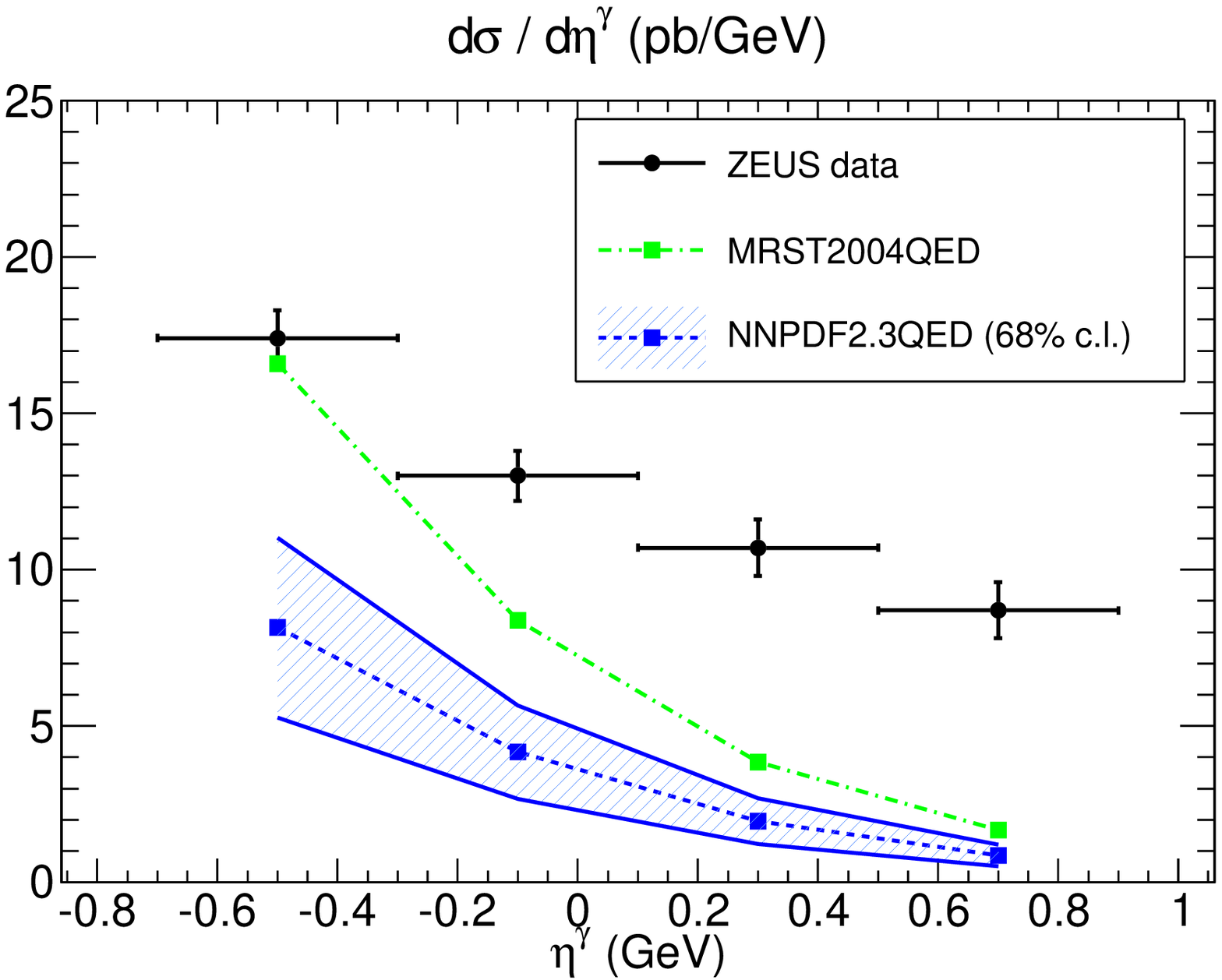}
\par\end{centering}
\caption{\label{fig:photondis} Comparison between the ZEUS
  data~\cite{Chekanov:2009dq} for the photon transverse energy (left)
  and rapidity (right) distributions in deep-inelastic isolated photon
  production and the leading log theoretical prediction obtained
  using NNPDF2.3QED and MRST2004QED PDFs.}
\end{figure}

Deep-inelastic isolated photon production 
provides a direct handle on the photon parton distribution of the
proton, through Compton scattering of the incoming electron off the
photon component of the proton (see Ref.~\cite{DeRujula:1998yq} and references therein). 
At the leading log level, this $O(\alpha^2)$ partonic subprocess is the
only contribution. In practice, however, the
$O(\alpha^3)$ quark-induced contributions~\cite{GehrmannDeRidder:2006wz} may be comparable (as for
the Drell-Yan
process  discussed in Sect.~\ref{sec:lhcwz}) because of the larger
size of the quark distribution. 
In Ref.~\cite{Martin:2004dh}, the total cross-section for this process
computed at the leading log level using MRST2004QED PDFs was shown to be
in reasonable agreement with  HERA integrated cross sections for
prompt photon production data~\cite{Chekanov:2004wr}. 

However, more recent HERA data~\cite{Chekanov:2009dq} for the rapidity
and transverse energy distribution of the photon do not agree well
with either the fixed order~\cite{GehrmannDeRidder:2006wz} or the leading
log~\cite{DeRujula:1998yq,Martin:2004dh} results for all values of
the kinematics, suggesting that a calculation matching the leading-log
resummation to the fixed order result would be necessary  in order to obtain
good agreement. In the absence of such a calculation, we did not use
these data for the determination of the photon PDF.

Theoretical predictions obtained using the leading log
calculation~\cite{Martin:2004dh} and the NNPDF2.3QED or MRST2004QED PDF sets
are compared in Fig.~\ref{fig:photondis}
to the ZEUS data of Ref.~\cite{Chekanov:2009dq}. 
These predictions have been obtained using the code  of Ref.~\cite{Martin:2004dh}. 
The selection cuts are the same as in~\cite{Chekanov:2009dq}, namely \be 10 \le Q^2 \le 300~{\rm GeV}^2 \, , \quad
4 \le  E_T^{\gamma} \le 15~{\rm GeV} \, , \quad -0.7 \le \eta^{\gamma} \le 0.9 \, .
\ee
The fact that the prediction  is in better agreement with
the data at large $E_T$ is consistent with the expectation that the leading
log approximation which is being used is more reliable in this
region. However, as already mentioned, a fully matched calculation
would be needed in order to consistently combine the leading log and
fixed order results.

 \begin{figure}[ht]
\begin{centering}
\includegraphics[scale=0.37]{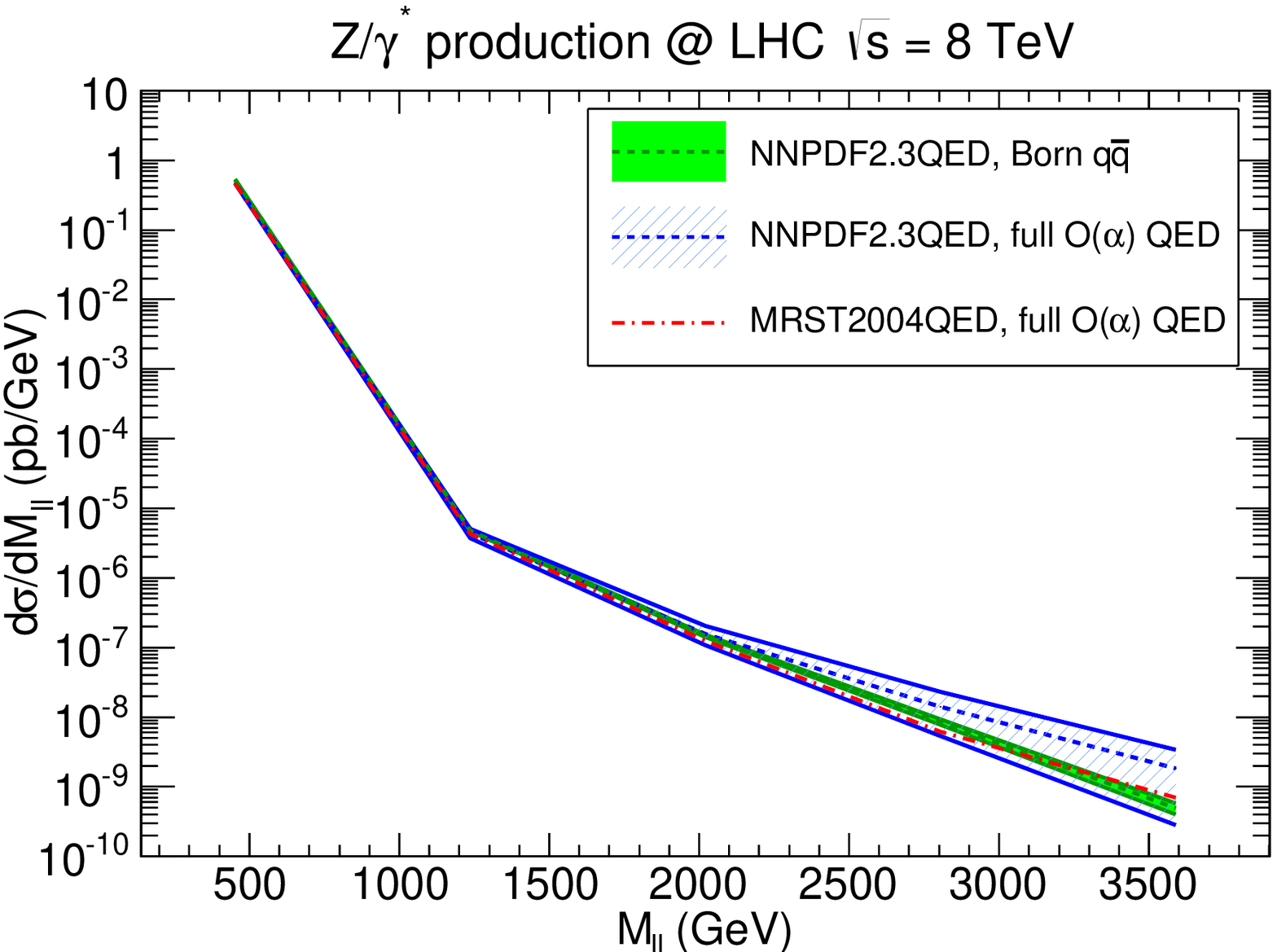}
\includegraphics[scale=0.37]{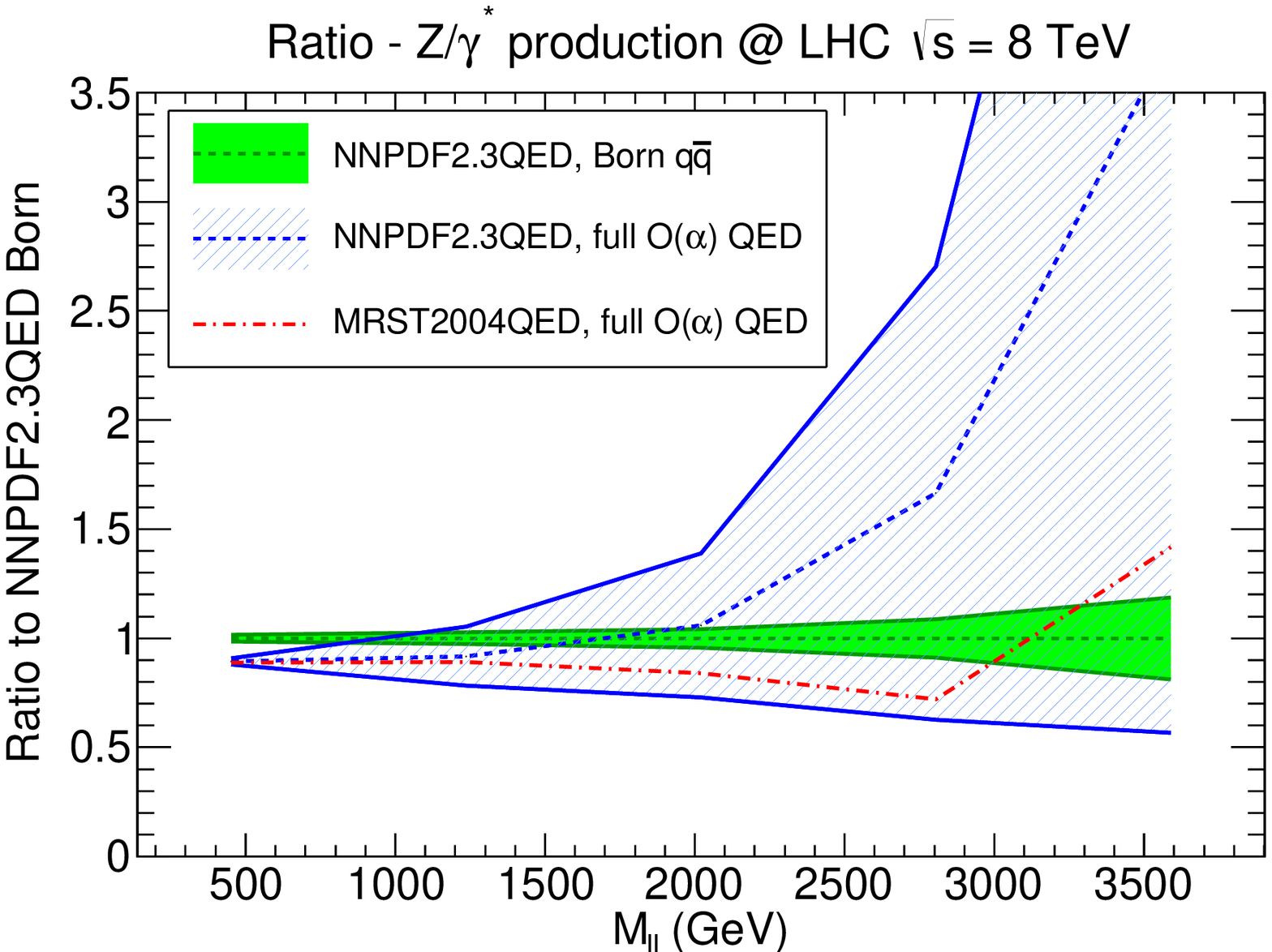}
\includegraphics[scale=0.37]{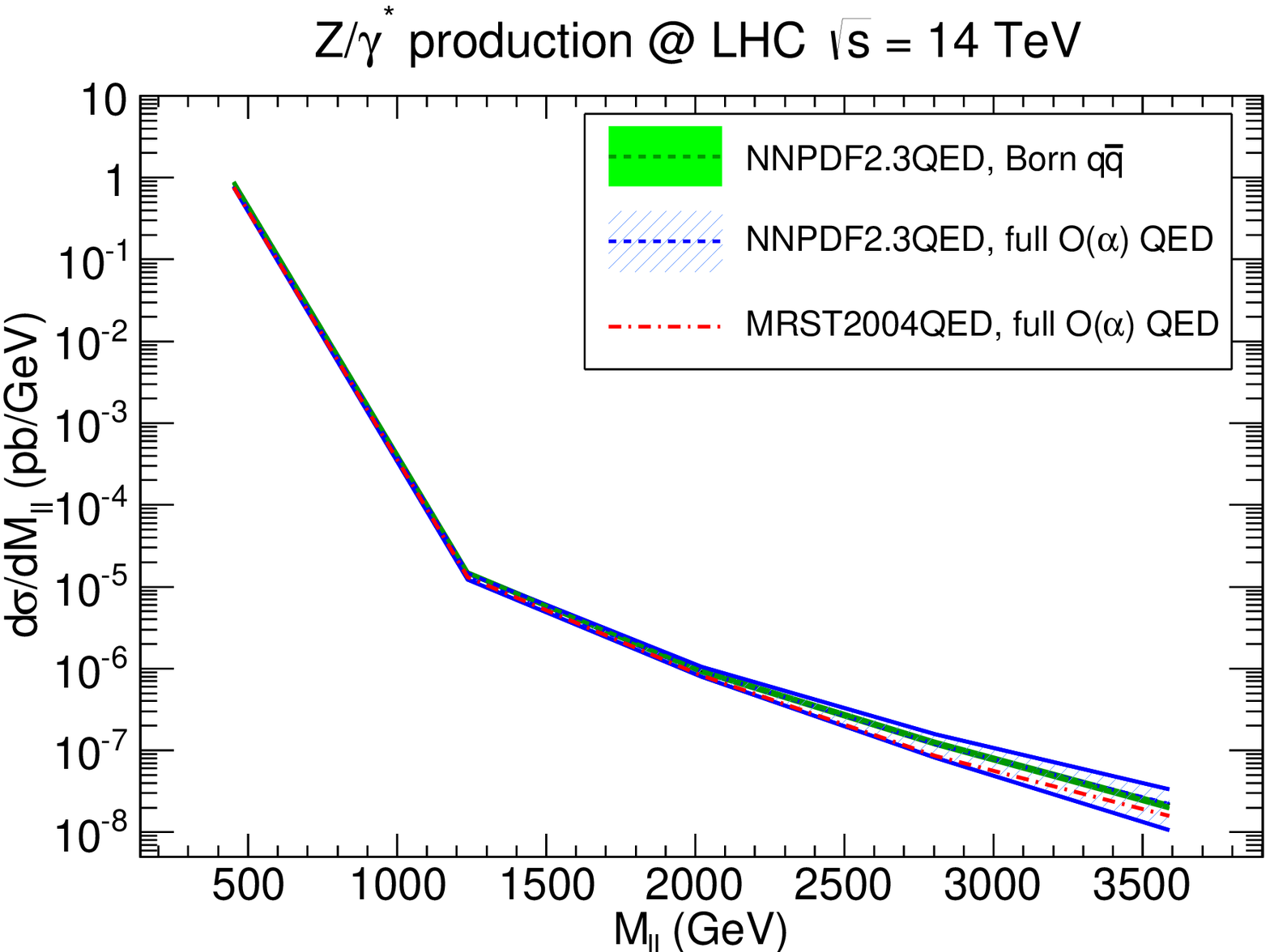}
\includegraphics[scale=0.37]{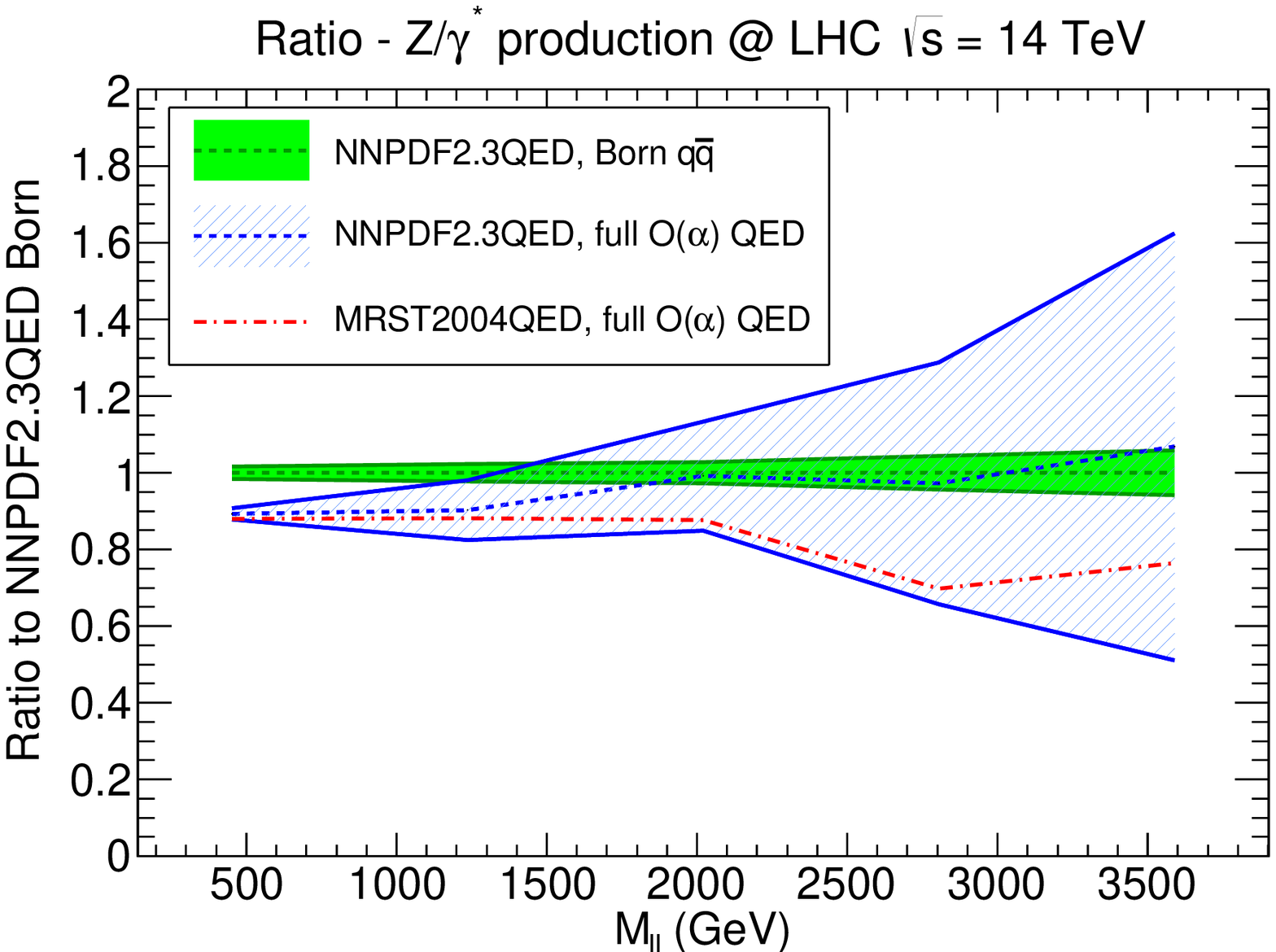}
\par\end{centering}
\caption{\label{fig:Zproduction} Neutral current Drell-Yan
  production at the LHC as a
  function of the invariant mass of the dilepton pair using
  NNPDF2.3QED and MRST2004QED PDFs.
Theoretical predictions for the Born $q\bar{q}$  
and the full $O\lp \alpha\rp$ process (including
photon-induced contributions) at the LHC 8 TeV (top) and LHC 14 TeV (bottom),
are shown both on an absolute scale (left) or as a ratio
to the central value of the Born $q\bar{q}$  cross section
from NNPDF2.3QED.}
\end{figure}

\subsection{Searches for new massive electroweak gauge bosons}
\label{sec:wzprimesearch}

Heavy electroweak gauge bosons, denoted generically by $W'$ and $Z'$,
have been actively searched at the 
LHC (see
e.g.~\cite{Chatrchyan:2012meb,Chatrchyan:2011dx,Chatrchyan:2011wq,Collaboration:2011dca}),
with current
limits for $M_{V'}$ between 1 and 2 TeV depending on the model
assumptions.
The main background for such searches is the off-resonance production
of $W$ and $Z$ bosons respectively.
At such large invariant masses of the dilepton pair, photon-induced contributions, of the type
shown in Figs.~\ref{fig:lhczborn}--\ref{fig:lhcwz2}, are potentially large.

\begin{figure}[ht]
\begin{centering}
\includegraphics[scale=0.37]{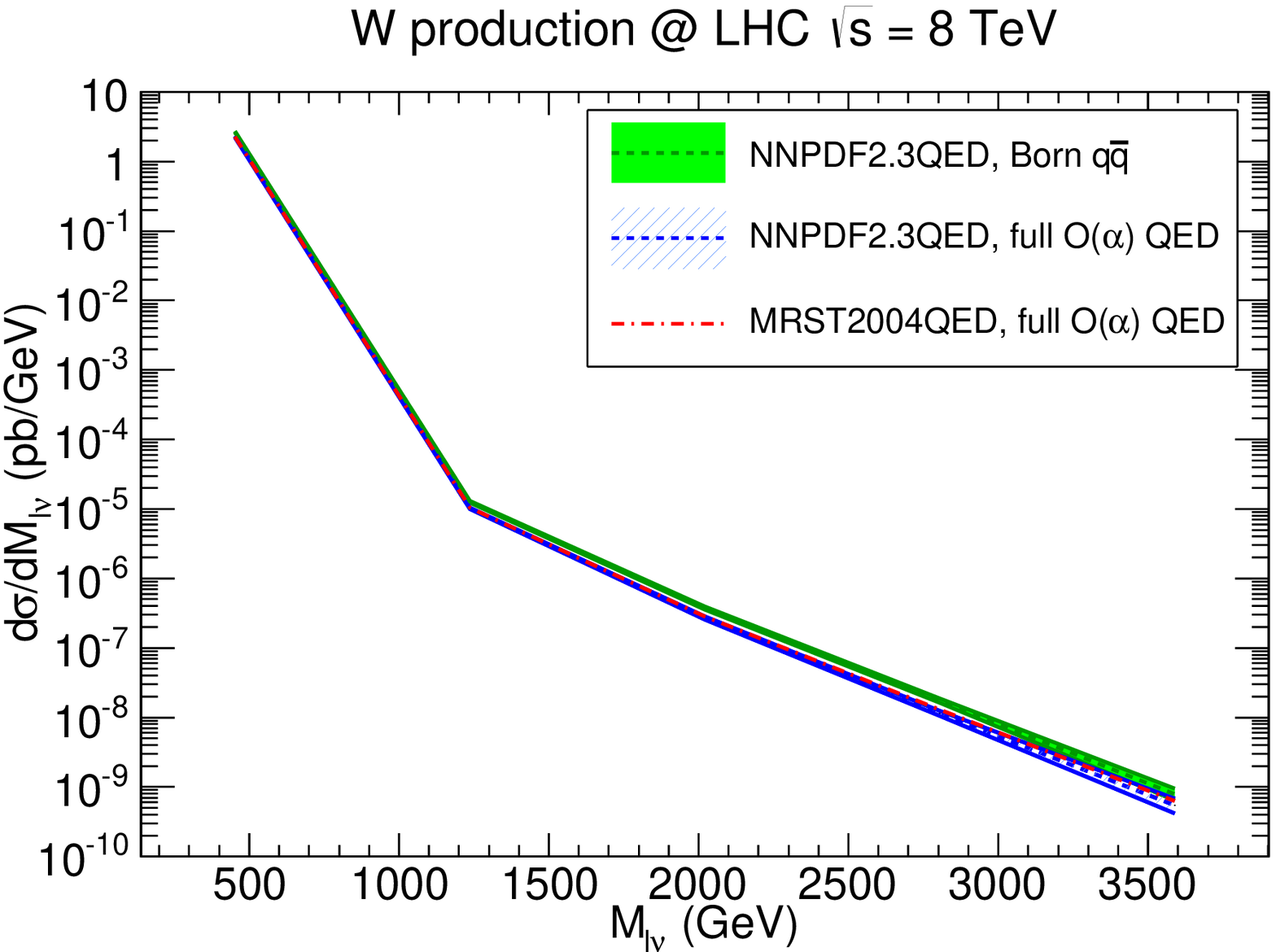}
\includegraphics[scale=0.37]{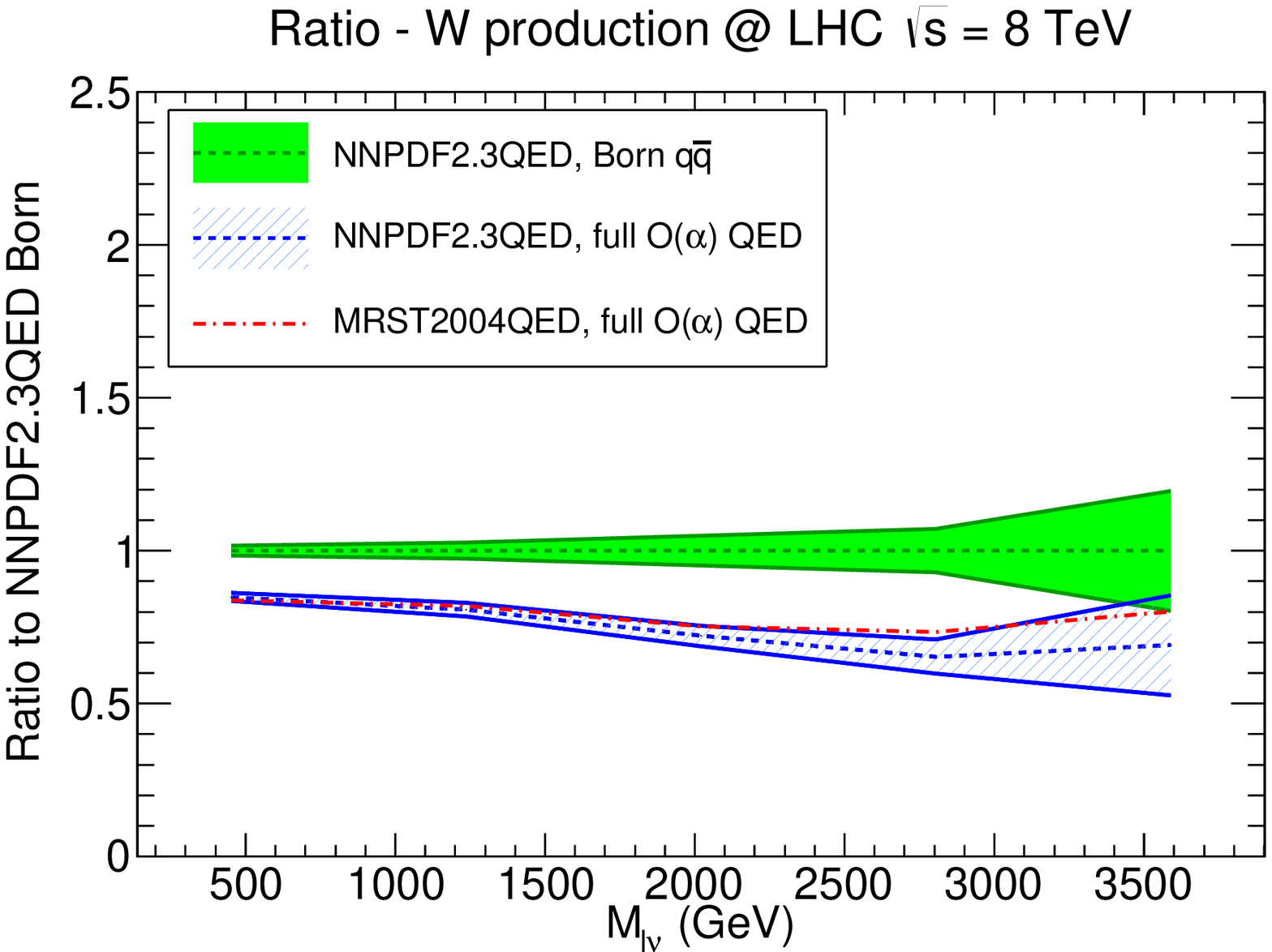}
\includegraphics[scale=0.37]{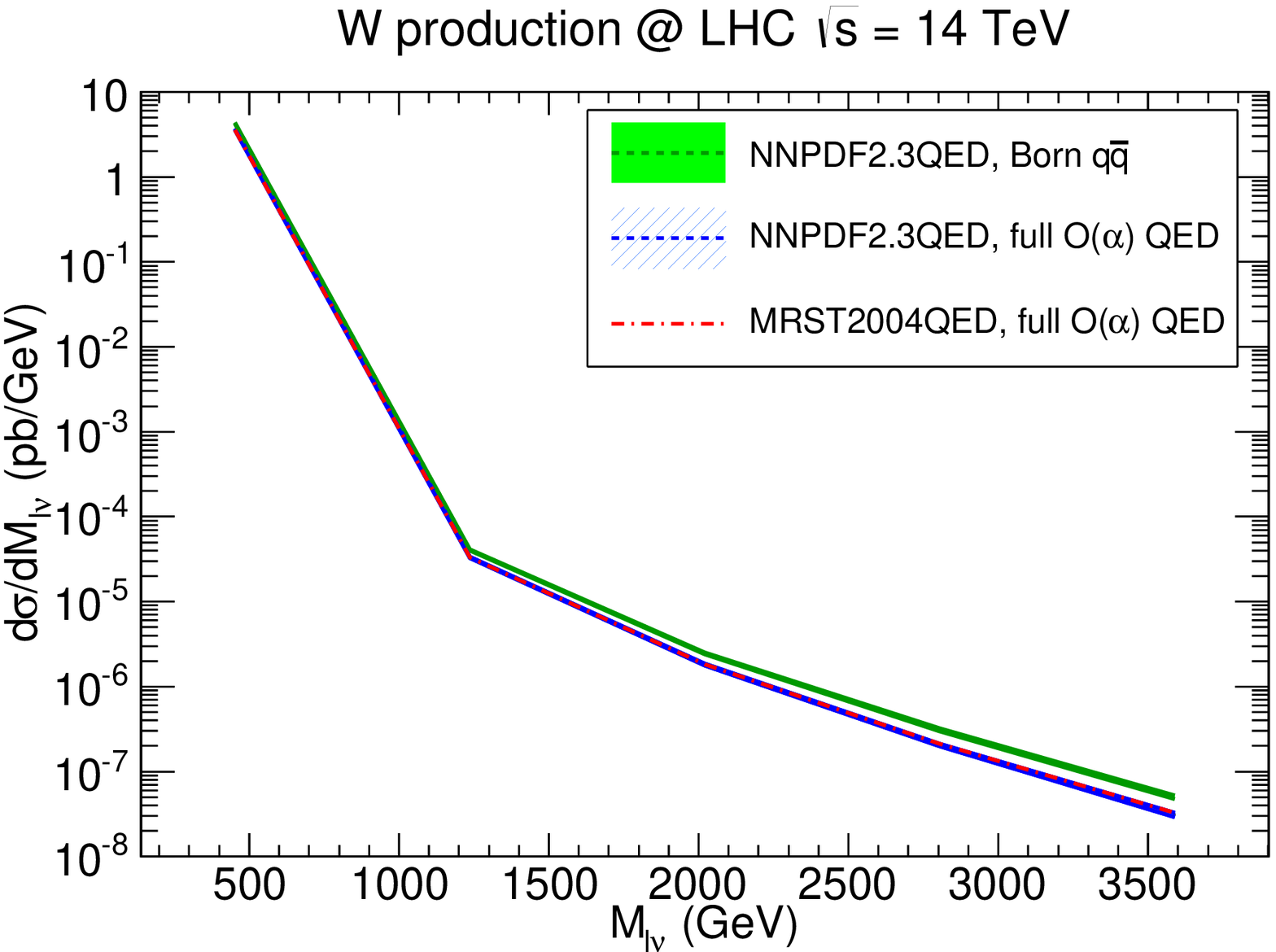}
\includegraphics[scale=0.37]{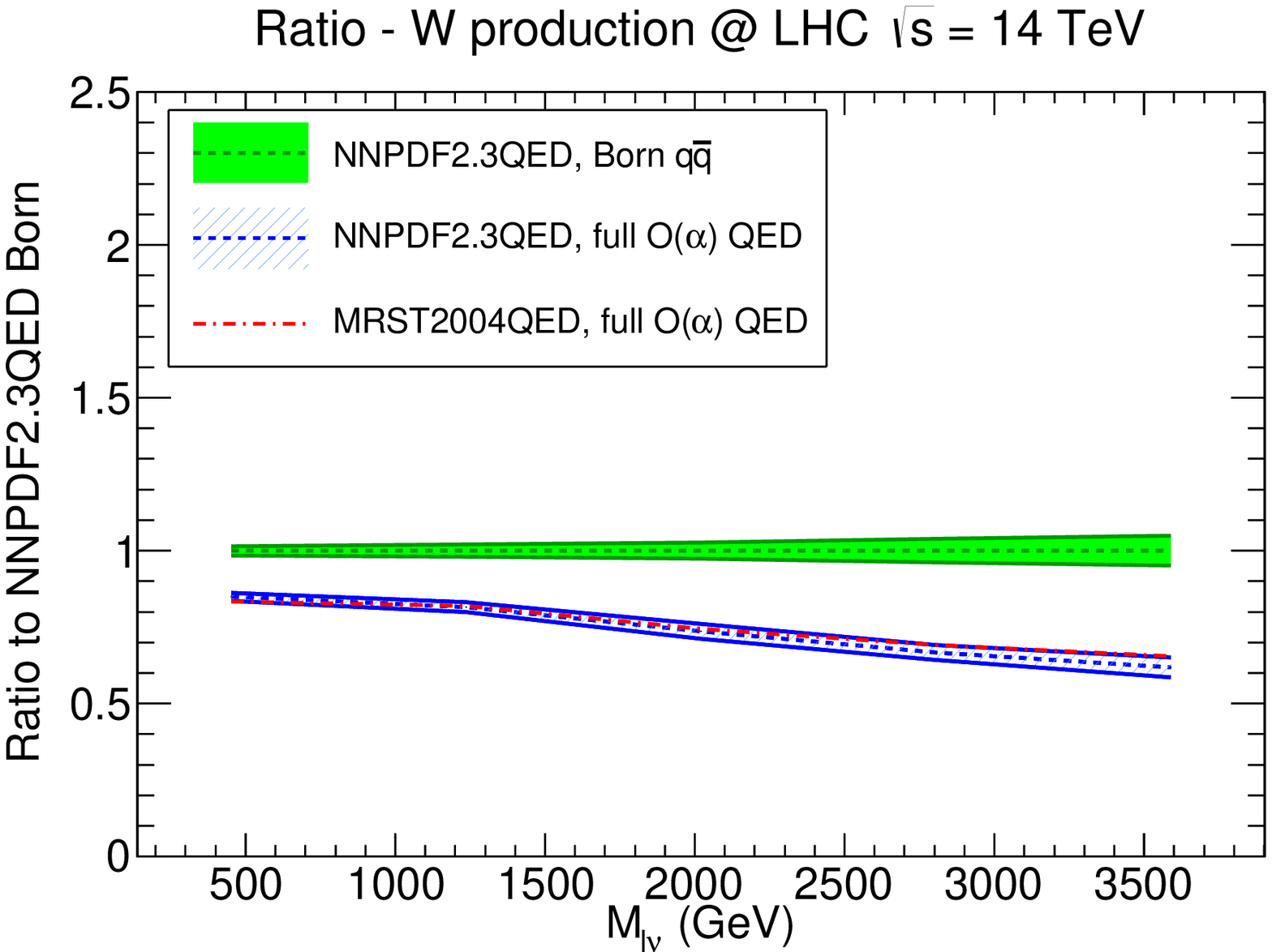}
\par\end{centering}
\caption{\label{fig:Wproduction} Same as Fig.~\ref{fig:Zproduction}
  but for high-mass charged-current production.}
\end{figure}

We have thus computed  the theoretical predictions for high mass off-shell
$W$ and $Z$ production using NNPDF2.3QED.
We have calculated separately the $q\bar{q}$ initiated Born contributions, 
the Born term supplemented by photon-initiated processes, and the full set of
  $O\lp \alpha \rp$
QED corrections, all determined with {\tt HORACE} (hence using LO QCD
theory)
and the various electroweak scheme
choices discussed in Sect.~\ref{sec:nnpdf23qed}.
We have used the following kinematical cuts, roughly corresponding
to those used in the ATLAS and CMS searches
\be
 p_t^l \ge 25~{\rm GeV} \, , \quad
| \eta^{\gamma} |\le 2.4 \, ,
\ee
and we have generated enough statistics to properly populate the highest mass bins and
reduce the impact of statistical fluctuations.
Results are displayed in Fig.~\ref{fig:Zproduction}, for
the neutral-current and in Fig.~\ref{fig:Wproduction}
for charged-current dilepton production respectively. %
They are provided for LHC~8 TeV  and LHC~14 TeV,
shown both in an absolute scale  and as a ratio
to the central value of the Born $q\bar{q}$  cross section
from NNPDF2.3QED, using the NLO set.

The contribution  from the photon-induced diagrams is generally not
negligible. Especially in the neutral current case, in which the
photon-induced contribution starts at  Born level, the uncertainty
induced by the QED corrections in the large invariant mass region is
substantial, because the LHC data we used to constrain the photon PDF
(recall in particular Tab.~\ref{tab:expdata} and
Fig.~\ref{fig:correlations}) have little effect there: the uncertainty
is of order
20\% for $M_{ll} \sim$ 1 TeV at LHC 8 TeV, and it reaches the
50\%  level for $M_{ll} \sim$ 2 TeV.
Of course, for a given value of $M_{ll}$, the photon-induced uncertainties
decrease when going to 14 TeV, since smaller values of $x$ are probed,
closer to the region of the data used for the current PDF determination.

Currently, the uncertainty on QED corrections is typically estimated
by varying the photon PDF  between its MRST2004QED value and zero. Our
results suggest that this might underestimate the size of the
photon-induced contribution; it certainly does underestimate the
uncertainty related to our current knowledge of it. This 
follows directly from the behavior of the luminosities of Fig.~\ref{fig:lumimrst}. 
In order to obtain more reliable exclusion
limits for $Z'$ and $W'$ at the LHC,
 a more accurate determination of
the photon PDF at large $x$ might be necessary. This could come from
the inclusion in the global PDF fit of new observables that are
particularly sensitive to the photon PDF,
such as $W$ pair production, as we now discuss.

\begin{figure}[ht]
\begin{center}
\input{diaguuWWphot.tex}
\end{center}
\vspace{-0.8cm}
\caption{\label{WWphotons} 
\small Tree-level diagrams for the LO processes $\gamma\gamma \to \mathrm{W}^-\mathrm{W}^+$, from
Ref.~\cite{Bierweiler:2012kw}.}
\end{figure}

\subsection{$W$ pair production at the LHC}

The production of pairs of electroweak gauge bosons is important,
specifically for the determination of
triple and quartic gauge boson
couplings~\cite{Chatrchyan:2013oev,Chatrchyan:2011tz,ATLAS:2012mec},
and it is a significant background to searches~\cite{Chatrchyan:2012ypy,Chatrchyan:2012rva,Chatrchyan:2012kk,Aad:2013wxa,Aad:2012nev} since several
extensions to the Standard Model including warped extra dimensions~\cite{Randall:1999vf}
and dynamical electroweak symmetry-breaking models~\cite{Andersen:2011yj,Eichten:2007sx} predict the
existence of heavy resonances decaying to pairs of electroweak gauge
bosons.

We consider now specifically the production of $W$ boson pairs for large
values of the invariant mass $M_{WW}$ and moderate values of the
transverse momentum $p_{T,W}$.
Photon-induced contributions to this process start at Born level (see
Fig.~\ref{WWphotons}), and their contribution can be substantial, in particular
at large values of $M_{WW}$. 
NLO QCD corrections,  as well as the formally  NNLO 
but numerically significant gluon-gluon initiated
contributions, are known, 
and available in public codes such as {\tt MCFM}~\cite{Campbell:2011bn}.
Fixed-order electroweak corrections to $W$ pair production are also known~\cite{Bierweiler:2012kw},
as well as the resummation of large Sudakov electroweak logarithms
at NNLL accuracy~\cite{Kuhn:2011mh}; a recent review of theoretical
calculations is in  Ref.~\cite{Baglio:2013toa}.
%
 \begin{figure}[ht]
\begin{centering}
\includegraphics[scale=0.37]{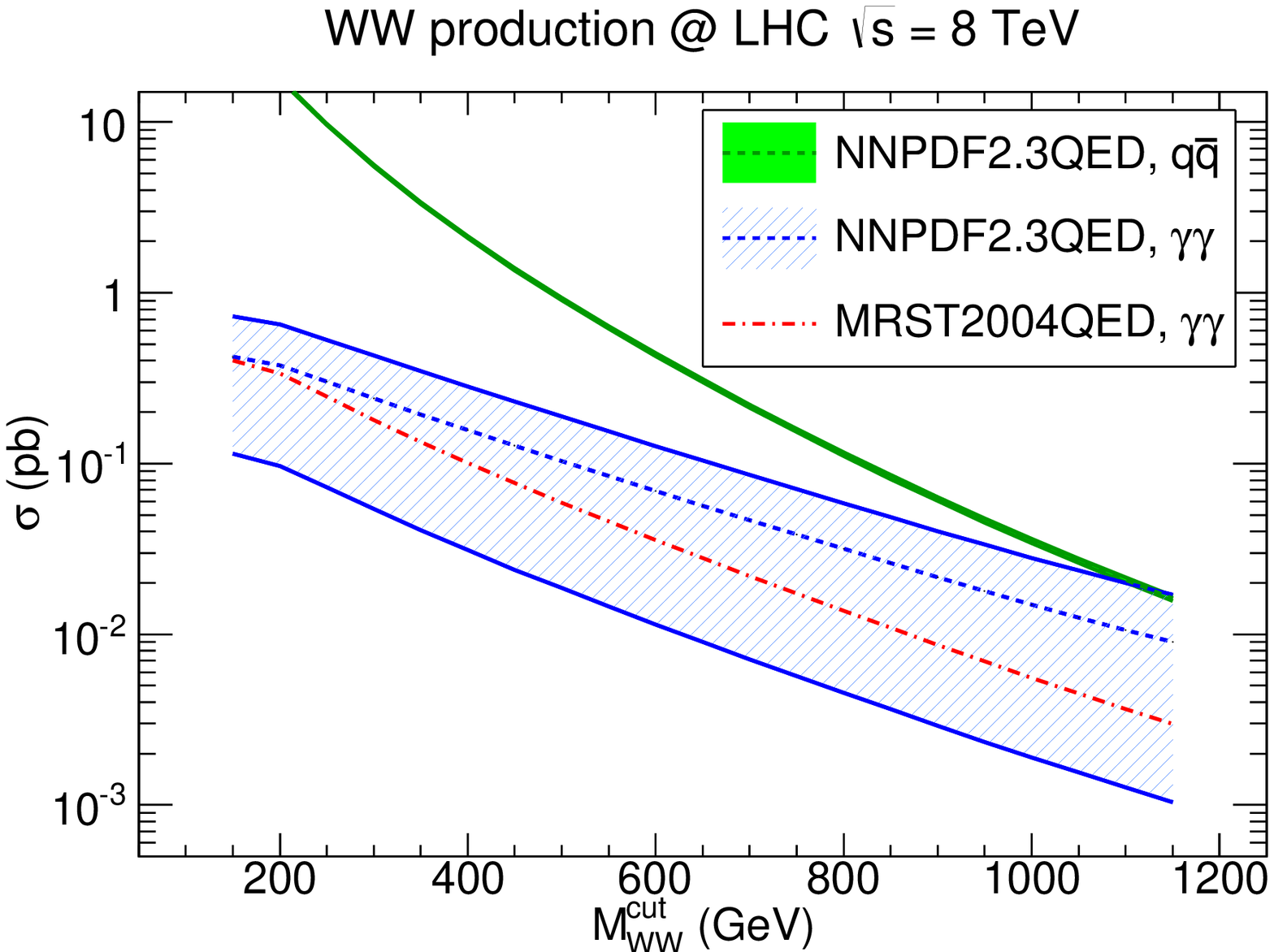}
\includegraphics[scale=0.37]{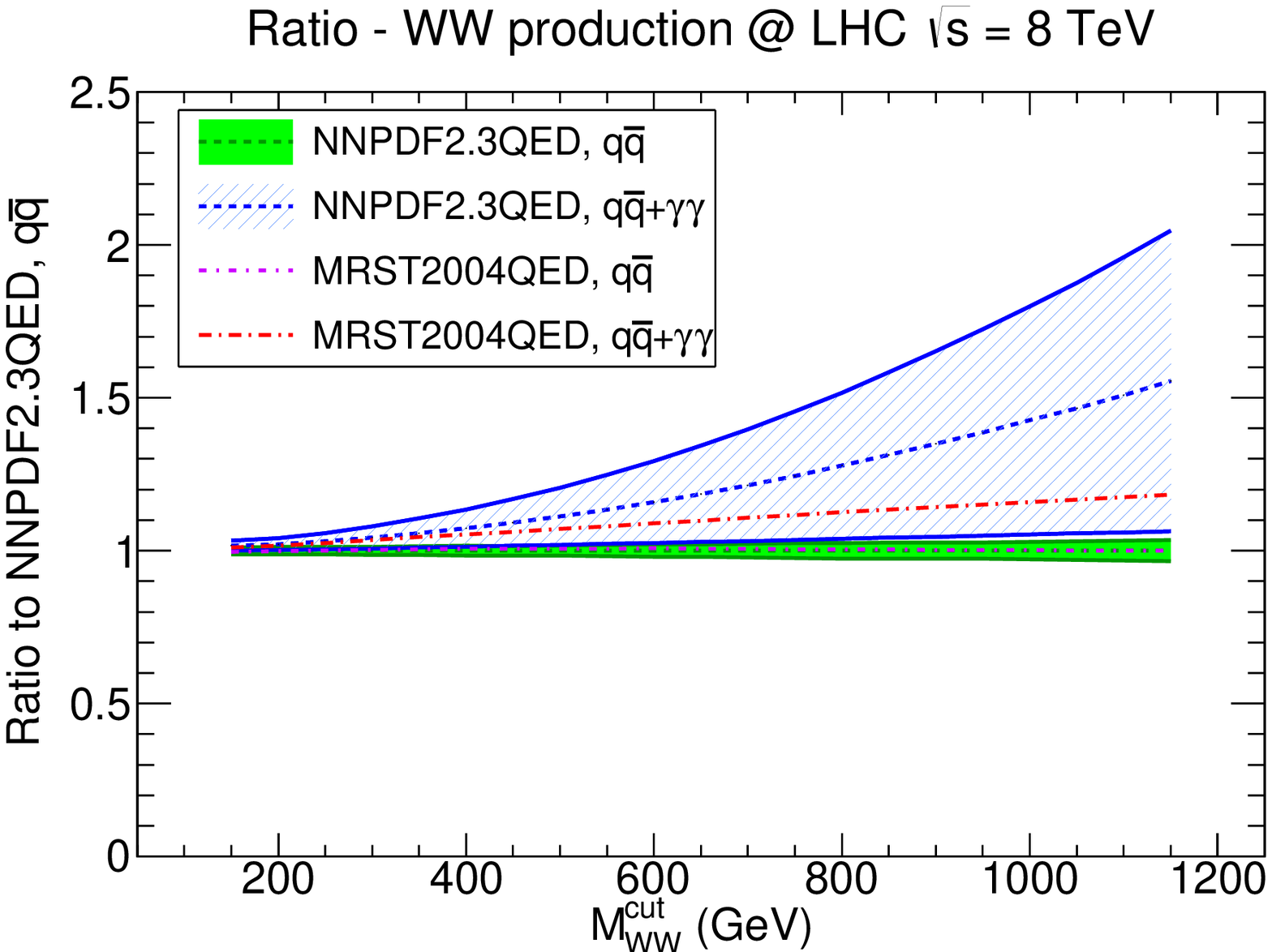}
\includegraphics[scale=0.37]{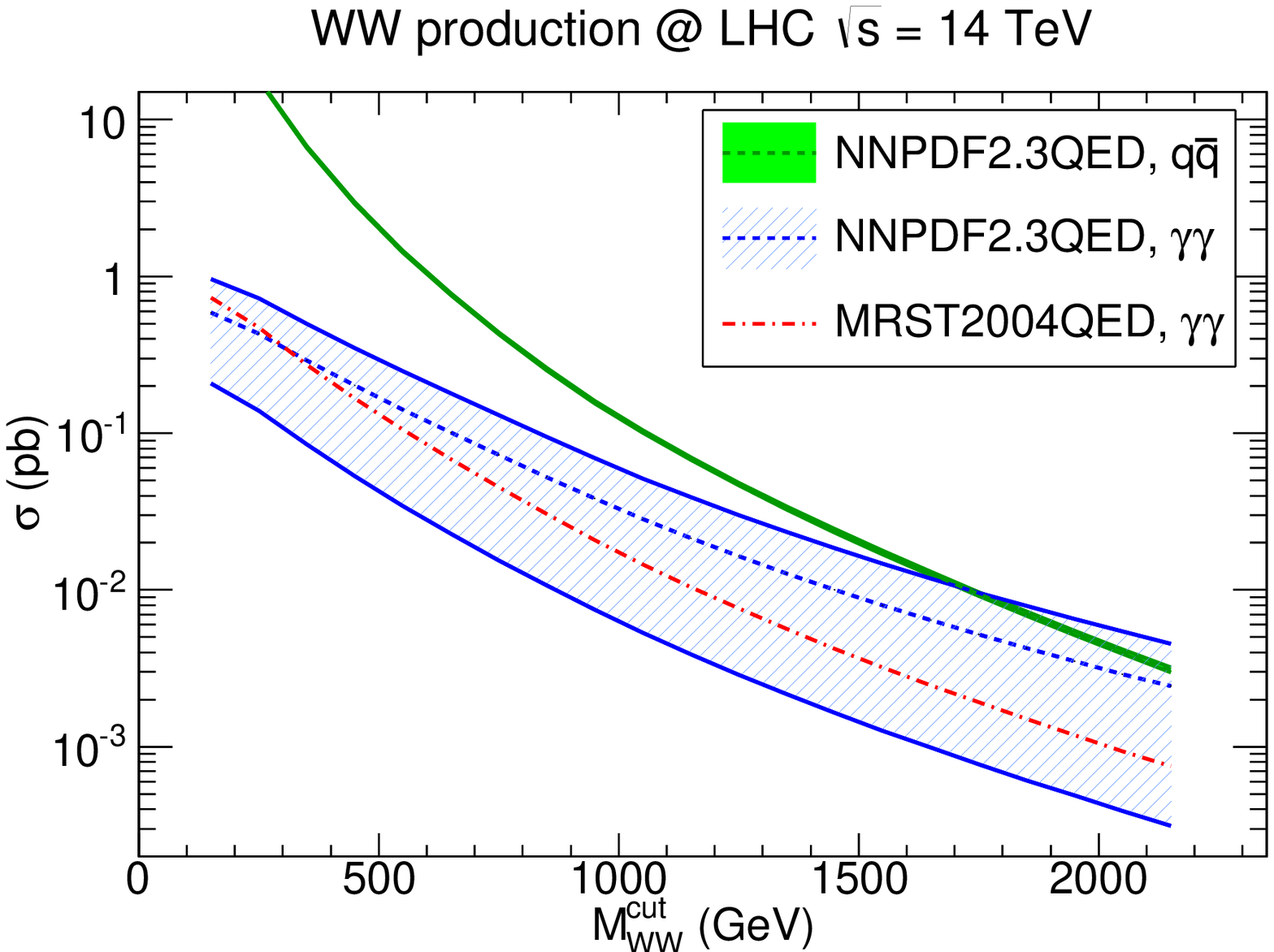}
\includegraphics[scale=0.37]{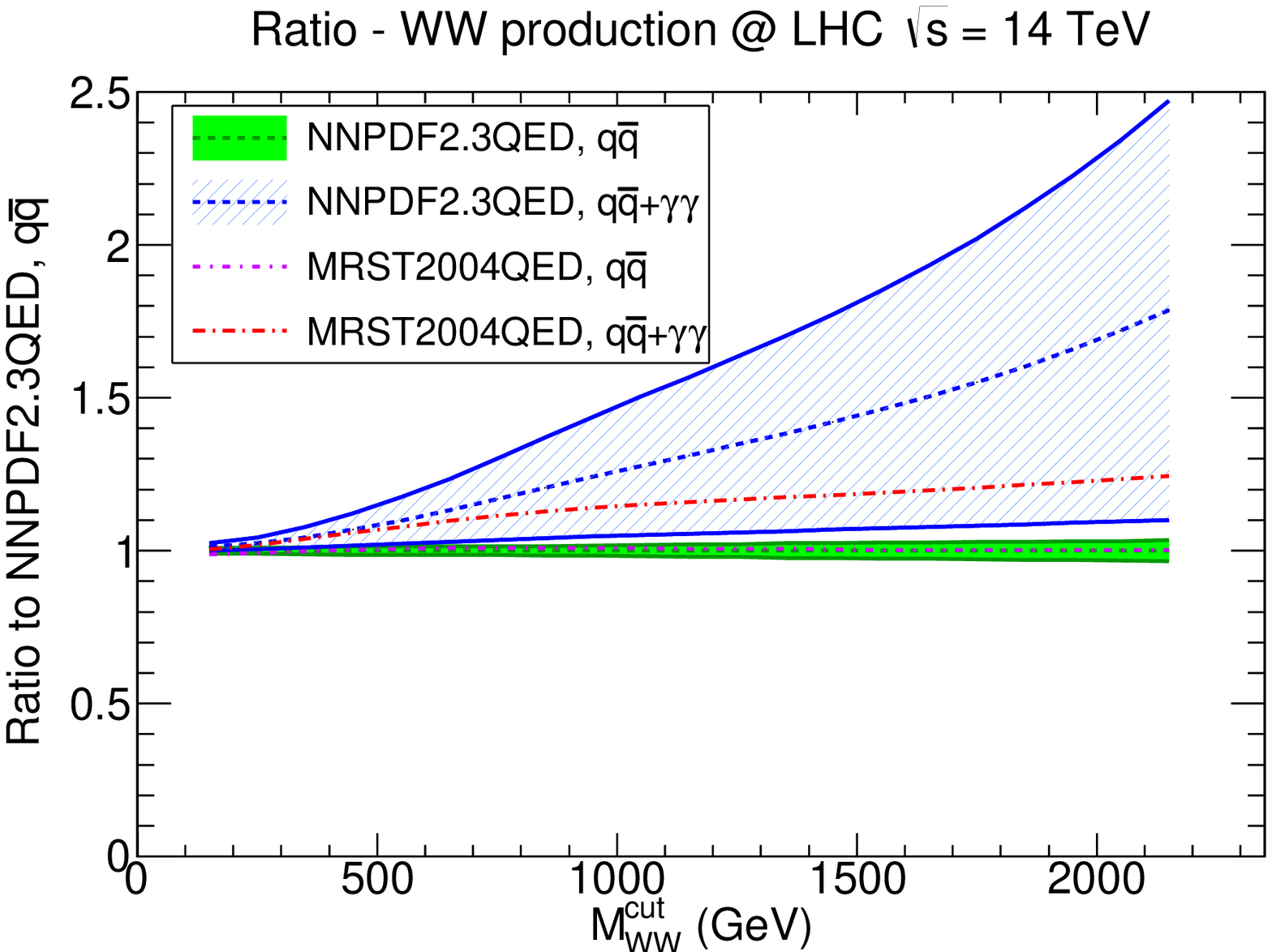}
\par\end{centering}
\caption{\small \label{fig:WWproduction} 
Photon-induced and quark-induced Born-level contributions to the
production of a $W$ pair with mass $M_{WW}>M_{WW}^{\rm cut}$ plotted
as a function of  $M_{WW}^{\rm cut}$ 
at the LHC 8 TeV (top) and LHC 14 TeV (bottom), computed
with the code of Ref.~\cite{Bierweiler:2012kw} and NNPDF2.3QED  NLO and
MRST2004QED PDFs.}
\end{figure}

To estimate the impact of
photon-induced contributions to $WW$ production, predictions have been
computed with either MRST2004QED or NNPDF2.3QED NLO PDFs.
They have been provided by the authors
of Ref.~\cite{Bierweiler:2012kw} 
using the code  and settings of Ref.~\cite{Bierweiler:2012kw}. 
In particular, the kinematical cuts in the transverse momentum and rapidity of the $W$ bosons
are 
\be
\label{eq:WWcuts}
p_{T,W} \ge 15~{\rm GeV} \, ,\quad |y_{W}|\le 2.5 \, .
\ee
%
 \begin{figure}[ht]
\begin{centering}
\includegraphics[scale=0.37]{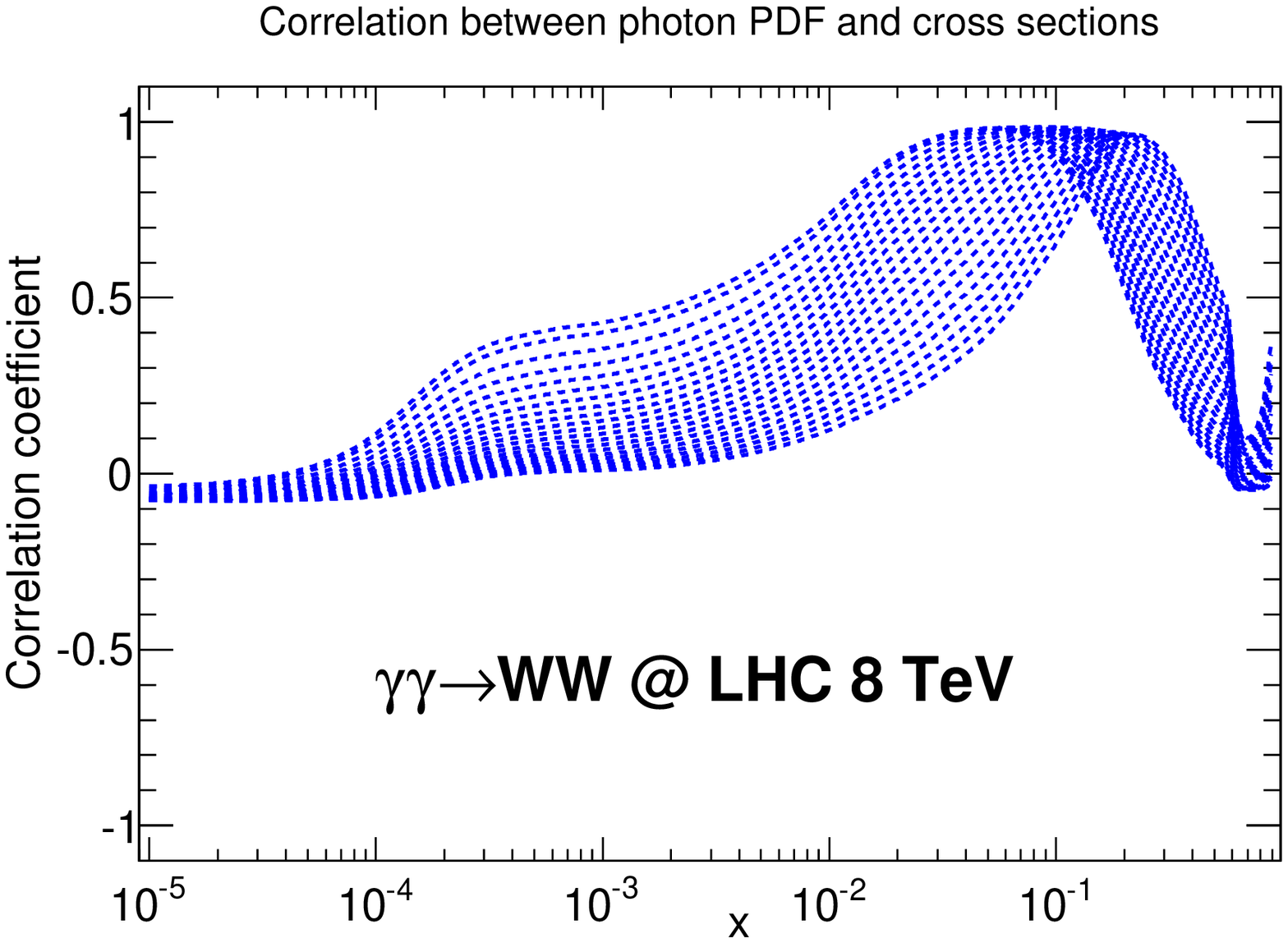}\includegraphics[scale=0.37]{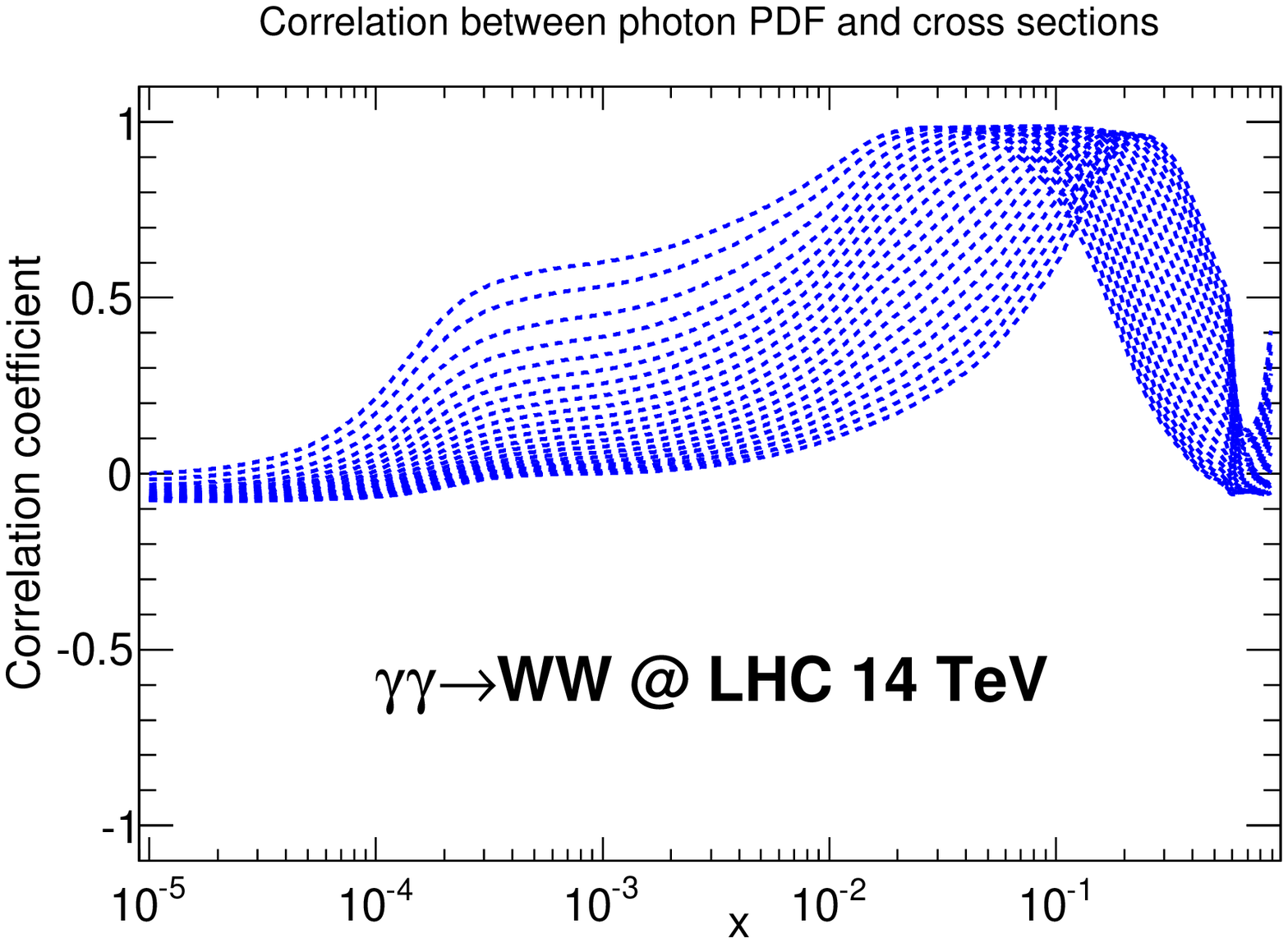}
\par\end{centering}
\caption{\small \label{fig:WWproductionc} Correlations between the $W$
  pair production cross section of Fig.~\ref{fig:WWproduction} and the  
photon PDF from the
  NNPDF2.3QED NLO set for $Q=10^4$~GeV$^2$. 
  Each curve corresponds to one of 40 equally
  spaced bins in which the $M_{WW}^{\rm cut}$ range of
  Fig.~\ref{fig:WWproduction} has been subdivided.
}
\end{figure}

In Fig.~\ref{fig:WWproduction} the cross-section for production of a
$W$ pair of mass $M_{WW}>M_{WW}^{\rm cut}$ is displayed as a function
of  $M_{WW}^{\rm cut}$,  at the  LHC 8  and 14 TeV. The 
Born $q\bar{q}$ and $\gamma\gamma$ initiated contributions are shown
(computed using LO QCD),
while we refer to Ref.~\cite{Bierweiler:2012kw} for the full
$O\lp \alpha \rp$ electroweak corrections, which depend only
weakly on the photon PDF.  It is clear that for large enough values of
the mass of the pair the photon-induced contribution becomes
increasingly important. Again, the relative size of the results
obtained using NNPDF2.3QED or MRST2004QED PDFs can be inferred from the
behavior of the luminosities shown in Fig.~\ref{fig:lumimrst}.

As in the case of Fig.~\ref{fig:Zproduction}, the large uncertainties
found for large values of $M_{WW}^{\rm cut}$  reflect the lack of
knowledge on the photon PDF at large $x\gsim0.1$. 
Indeed, in Fig.~\ref{fig:WWproductionc} we display the correlation between the
cross section of Fig.~\ref{fig:WWproduction} and the photon PDF at
$Q^2=10^4$~GeV$^2$ as a function of $x$, obtained
subdividing the range of  $M_{WW}^{\rm cut}$ of
Fig.~\ref{fig:WWproduction} into 40 bins of equal width, and then
computing the correlation for each bin. It is clear that this process
is sensitive to the photon PDF at large $x$, 
where the data of
Tab.~\ref{tab:expdata}  provide little or no constraint (recall
Fig.~\ref{fig:correlations}). Hence, predictions for
$W$ pair production obtained using MRST2004QED or NNPDF2.3QED should be
taken  with care: NNPDF2.3QED provides a more conservative
estimate of the uncertainties
involved, but perhaps overestimates the range of reasonable photon
PDF shapes.  However, future measurements of this
process could be used to pin down the photon PDF at large
$x$, and thus in turn improve the accuracy of the prediction for very high mass
Drell-Yan production discussed in Sect.~\ref{sec:wzprimesearch} and
Fig.~\ref{fig:Zproduction}, and conversely. Of course, in using
either, or both of these channels for new physics searches, care should
be taken that the sought-for new physics effects are not being hidden
in the PDFs themselves, which could be done by introducing suitable
kinematic cuts.


\section{Conclusions}
\label{sec:conclusions}
We have presented a first determination of PDFs with QED corrections
and including a photon PDF using NNPDF methodology.
The photon PDF is determined by
deep-inelastic scattering and neutral- and charged-current Drell-Yan production
data from the LHC. 
The LHC data  constrain the photon PDF in
the range $10^{-5}\lsim x\lsim 10^{-1}$ at the initial scale
$Q^2=2$~GeV$^2$. In comparison to the previous MRST2004QED set, we
also provide  PDF
uncertainties, and we determine the photon from data, rather than
from a model. We find that our central photon PDF is in good
agreement with the MRST2004QED result for $x\gsim 0.03$, but is rather
smaller for lower values of $x$.

The main shortcoming of our determination of the photon PDF is the
lack of experimental information for $x\gsim  0.1$. This induces
substantial uncertainties related to electroweak corrections in
processes which are relevant for new physics searches at the LHC, such
as high mass gauge boson production and double gauge boson
production. This latter process could in turn be used to constrain the
large $x$ photon PDF once accurate LHC data become available.

Now that a first determination of the photon PDF based on data is
available, it will be interesting to give a more precise assessment of
the impact of QED corrections on precision electroweak measurements at
the LHC, such as the determination of the $W$ mass~\cite{Bozzi:2011ww}.
A more systematic investigation of processes which
may be used to constrain the photon PDF is now in order.

\bigskip
\bigskip

The NNPDF2.3QED PDF sets, proton and neutron, NLO and NNLO,
are available from the NNPDF web site,
\begin{center}
{\bf \url{http://nnpdf.hepforge.org/}~}
\end{center}
and through the LHAPDF interface~\cite{Bourilkov:2006cj}. The list of
available  new grids is the following:
\begin{itemize}
\item {\tt NNPDF23\_nlo\_as\_0117\_qed.LHgrid}
\item {\tt NNPDF23\_nlo\_as\_0117\_qed\_neutron.LHgrid}
\item {\tt NNPDF23\_nnlo\_as\_0117\_qed.LHgrid}
\item {\tt NNPDF23\_nnlo\_as\_0117\_qed\_neutron.LHgrid}
\item {\tt NNPDF23\_nlo\_as\_0118\_qed.LHgrid}
\item {\tt NNPDF23\_nlo\_as\_0118\_qed\_neutron.LHgrid}
\item {\tt NNPDF23\_nnlo\_as\_0118\_qed.LHgrid}
\item {\tt NNPDF23\_nnlo\_as\_0118\_qed\_neutron.LHgrid}
\item {\tt NNPDF23\_nlo\_as\_0119\_qed.LHgrid}
\item {\tt NNPDF23\_nlo\_as\_0119\_qed\_neutron.LHgrid}
\item {\tt NNPDF23\_nnlo\_as\_0119\_qed.LHgrid}
\item {\tt NNPDF23\_nnlo\_as\_0119\_qed\_neutron.LHgrid}
\end{itemize}
Note that unlabelled (default) grids refer to the proton.

The NNPDF2.3QED sets are included in LHAPDF {\tt 5.9.0}
and subsequent releases.

\bigskip
\bigskip
\begin{center}
\rule{5cm}{.1pt}
\end{center}
\bigskip
\bigskip

{\bf\noindent  Acknowledgments \\}
We are grateful to Alessandro Vicini for discussions about
QED corrections and the {\tt HORACE} generator, and Giancarlo Ferrera for
assistance with {\tt DYNNLO}; to
James Stirling and Robert Thorne for information on
MRST2004QED set and for providing us with their code for direct photon
production in HERA; 
to Tobias Kasprzik for providing the predictions for $WW$ production
with electroweak corrections using NNPDF2.3QED; to 
Sasha Glazov and Uta Klein for providing information on the ATLAS
Drell-Yan measurements; and  to Katharina M\"uller
for details on the the LHCb data. We would also like to thank 
Stefan Dittmaier for discussions about QED and electroweak corrections 
at the LHC, and for emphasizing the
importance of scheme choices in the treatment of QED corrections.
S.C. and S.F. are supported by an Italian PRIN 2010 and by a European
EIBURS grant. J.~R. is supported by a Marie Curie 
Intra--European Fellowship of the European Community's 7th Framework 
Programme under contract number PIEF-GA-2010-272515.


\end{document}